\documentclass[preprint,12pt]{elsarticle}
\usepackage{graphicx} 
\usepackage[margin = 0.8in]{geometry}
\usepackage{amsmath}

\usepackage{amsthm}
\theoremstyle{plain}
\newtheorem{lemma}{Lemma}[section]

\theoremstyle{definition}

\usepackage{amssymb}

\usepackage{hyperref} 
\usepackage[capitalise]{cleveref}

\usepackage{arydshln}
\usepackage[dvipsnames]{xcolor}

\usepackage{booktabs}
\numberwithin{equation}{section}
\usepackage{threeparttable}
\usepackage{multirow} 

\newcommand{\M}{\textit{Medea}~}
\newcommand{\Mns}{\textit{Medea}}

\usepackage{tikz}
\DeclareRobustCommand*\circled[1]{\tikz[baseline=(char.base)]{
            \node[shape=circle,draw,inner sep=1pt] (char) {#1};}}

\begin{document}

\begin{frontmatter}
\title{Modeling \textit{Medea} gene-drive population replacement: thresholds and release strategies}
\author[1]{Zhuolin Qu\corref{cor1}}
\ead{zhuolin.qu@utsa.edu}
\cortext[cor1]{Corresponding author}
\affiliation[1]{organization={Department of Mathematics, University of Texas at San Antonio},
addressline={One UTSA Circle},
city={San Antonio},
state={TX},
postcode={78249},
country={USA}}

\begin{abstract}
Mosquito-borne diseases such as dengue, Zika, and yellow fever impose a substantial global health burden, motivating genetic control strategies that replace wild mosquito populations with disease-refractory ones. Maternal-effect dominant embryonic arrest (\textit{Medea}) is a gene drive in which the offspring of a \textit{Medea}-carrying mother die unless they inherit the \textit{Medea} allele, producing biased inheritance capable of driving a linked refractory trait to high prevalence. We develop and analyze a continuous-time compartmental model of \textit{Medea} dynamics in \textit{Aedes aegypti} that tracks mosquito abundance by life stage and genotype, and that generalizes the drive mechanism to allow both imperfect \textit{Medea}-killing and imperfect rescue. We characterize the biologically relevant equilibria and derive their local stability conditions, together with genotype-specific reproduction numbers and the basic reproduction number ($\mathcal{R}_0$) governing invasion from low frequency. Bifurcation analysis reveals bistability, and hence a release threshold that must be exceeded for \textit{Medea} to take over, even though $\mathcal{R}_0<1$ at baseline. Imperfect killing and imperfect rescue reshape this bifurcation structure in distinct ways: killing leakage governs the invasion threshold, while rescue efficiency governs the composition of the resulting population. Sensitivity analysis identifies the fitness coefficient and killing leakage as the dominant drivers of both threshold and coverage outcomes, with mosquito demographic parameters affecting absolute abundances but not genotype proportions. Simulations of batched release programs inform efficient field deployment and show that reversing an established drive is substantially more costly than achieving forward invasion, though co-releasing both sexes accelerates reversal considerably more than it accelerates invasion. 
\end{abstract}

\begin{keyword} 
gene drive; \textit{Medea}; population replacement; bifurcation analysis; threshold; release strategy
\end{keyword}

\end{frontmatter}

\section{Introduction}
Mosquito-borne diseases, such as dengue fever, malaria, chikungunya, and Zika, continue to impose an enormous burden on global public health. Together, these diseases account for over one million deaths annually \cite{AMCA}. Traditional approaches, including larviciding and residual insecticide spraying, are limited by high implementation cost and the accelerating spread of insecticide resistance \cite{Achee2019}.
These challenges have renewed interest in biological control strategies that target mosquito populations directly.

Several biological approaches have reached field deployment or large-scale trials. The Sterile Insect Technique (SIT) releases irradiated sterile males to suppress wild populations, though radiation-induced fitness penalties reduce mating competitiveness and constrain its efficacy \cite{dyck2005sterile, alphey2010sterileinsect}. \textit{Wolbachia}-based methods exploit cytoplasmic incompatibility, whereby matings between \textit{Wolbachia}-infected males and uninfected females produce no viable offspring, while simultaneously reducing the mosquitoes' ability to transmit pathogens \cite{HoffmannTurelli1997, bian2010endosymbiotic, dutra2016wolbachia}. Field programs deploying \textit{Wolbachia}-infected \textit{Aedes aegypti} have achieved sustained establishment and measurable reductions in dengue incidence in Australia, Indonesia, and elsewhere \cite{ryan2020establishment, indriani2020reduced, hoffmann2011successful}. Advances in CRISPR-based gene editing have more recently introduced a new class of interventions: gene-drive systems that can propagate engineered traits through wild populations with super-Mendelian inheritance \cite{champer2016cheating, burt2014heritable, alphey2014genetica}. Such systems hold transformative potential for mosquito control, but their capacity for self-propagating spread also raises significant ecological and regulatory concerns about unintended persistence or
transboundary dispersal \cite{oye2014regulating, james2018pathwaya, marshall2010cartagena}.

Among the various gene-drive architectures under development, threshold-dependent systems have attracted particular interest for their potential reversibility and spatial confinement \cite{buchman2018engineered, akbari2013synthetic}. Unlike self-propagating drives, these systems require that the engineered element exceed a critical introduction frequency before it can persist. Below this threshold, natural selection favors the wild-type, and the drive is eliminated; above it, the drive can establish and spread. This bistable behavior makes threshold-dependent systems more controllable and better suited to localized deployment, properties that are increasingly emphasized by regulatory agencies and the public in the context of field trials \cite{wang2022symbionts}. The \M (Maternal-Effect Dominant Embryonic Arrest) element is a prominent example of such a system. In \Mns-carrying females, the drive deposits a toxin during oogenesis that kills offspring who do not inherit the linked antidote, creating a maternal-effect inheritance bias that promotes invasion above a threshold but remains self-limiting below it \cite{beeman1992maternaleffect, chen2007synthetic, ward2011medea}. \M elements were first documented in flour beetles \cite{beeman1992maternaleffect} and have since been synthetically engineered in \textit{Drosophila} \cite{chen2007synthetic, buchman2018synthetically}, establishing a strong empirical basis for theoretical investigation.

Mathematical modeling has played a central role in characterizing the population dynamics of \M and related threshold-dependent gene drive systems. Early theoretical work established the foundational result that \M spreads to fixation under broad conditions in the absence of fitness costs, with the rate of spread determined by the strength of maternal-effect lethality \cite{wade1994population}. Building on this foundation, Ward et al.\ \cite{ward2011medea} developed a comprehensive discrete-time, genotype-frequency-based framework that tracks male and female genotype frequencies separately, characterizes four biologically relevant equilibria, and derives the unstable internal equilibrium allele frequency as the invasion threshold. Their analysis covers a range of maternal-effect killing and zygotic rescue efficiencies, and remains the primary theoretical reference for \M population dynamics. Parallel theoretical work has examined a broader class of threshold-dependent toxin-antidote architectures, such as inverse \M and two-locus underdominance systems, characterizing their equilibrium structure, invasion thresholds, and confinement properties in single- and two-population settings \cite{marshall2011inverse, marshall2012confinement, marshall2012general}. These analyses are grounded in discrete-generation, genotype-frequency frameworks and collectively establish a detailed picture of how threshold-dependent drives behave in well-mixed populations, with selected extensions to two-population migration settings to characterize spatial confinement properties \cite{marshall2012confinement}. More recently, the MGDrivE modeling platform \cite{SanchezMGDrivE2020, wu2021mgdrive} has extended this line of work into spatially explicit, stochastic simulation frameworks that incorporate full mosquito life stage structure, density-dependent larval competition, seasonal environmental variation, and metapopulation dynamics, providing a flexible testbed for evaluating gene drive deployment strategies across ecologically realistic settings. Together, these contributions span a modeling hierarchy from analytically tractable single-population genotype frequency recursions to high-dimensional spatially explicit simulations.

Despite this progress, a gap remains at the level of continuous-time structured population models that can leverage a dynamical systems approach to characterize \M population dynamics. Existing population genetic frameworks track the relative proportions of genotypes across generations but do not explicitly represent population size, life stage structure, or the density-dependent and environmental processes that govern mosquito abundance. These features are essential for connecting threshold conditions in genotype frequencies to the absolute numbers of individuals that must be released in the field, and for assessing how ecological factors such as carrying capacity and seasonal variation shape the population replacement dynamics.

In this paper, we develop and analyze a compartmental ordinary differential equation (ODE) model that explicitly tracks population abundances of mosquitoes by life stage (egg, female, and male) across three genotypes, which permits the incorporation of life-stage-specific and genotype-specific demographic parameters, density-dependent competition, and fitness costs across genotypes. The two-sex structure allows the maternal-effect mechanism to be expressed directly through the mating interactions between male and female genotypes, capturing the sex-asymmetric nature of \M inheritance. Leveraging the ODE framework, we derive dimensionless population reproduction numbers that govern the existence and stability of biologically relevant equilibria. We further employ bifurcation analysis to characterize how generalized \Mns-killing and rescue parameters, considered jointly, alter the qualitative structure of the system and the threshold conditions governing successful \M establishment.

We first present our mathematical model (\cref{sec:model}). Then we analyze equilibrium behavior (\cref{sec:analysis}) and present the bifurcation diagrams (\cref{sec:bifur}). We then assess the sensitivity to various parameters (\cref{sec:SA}) and examine control strategies through numerical simulation (\cref{sec:release}). Finally, we discuss the relevance of our results to control efforts (\cref{sec:discussion}).

\section{Model Formulation}\label{sec:model}
\begin{figure}[ht]
\centering
\includegraphics[width=\textwidth]{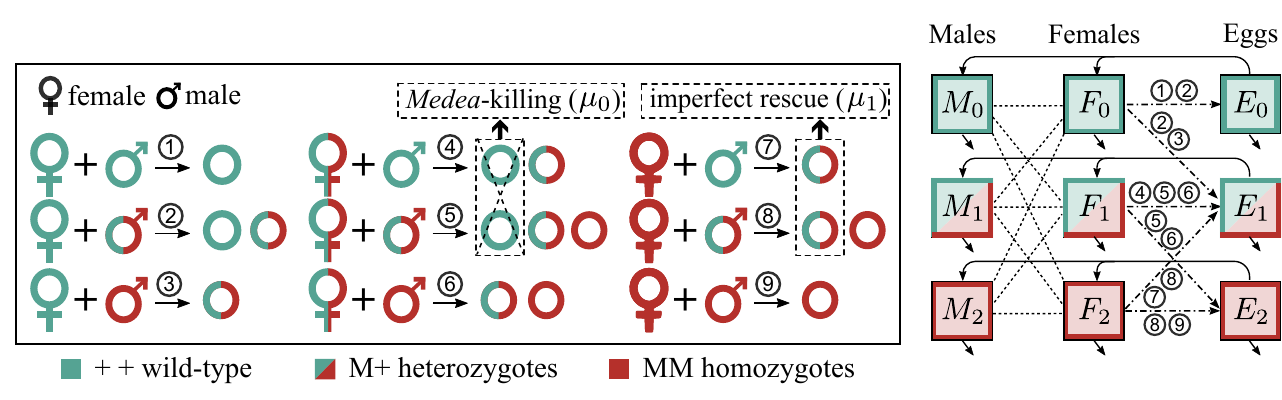} 
\caption{\textbf{Left:} Transmission of \M gene drive depends on parental genotypes. Nine possible crosses between female and male genotypes, numbered \circled{1}--\circled{9}. The maternal-effect \M killing happens in crosses \circled{4} and \circled{5} (a fraction of $\mu_0$ offspring survive), unless they inherit the \M element. For \Mns-bearing homozygous females, imperfect rescue happens in cases \circled{7} and \circled{8} (a fraction of $\mu_1$ offspring is rescued) among heterozygous offspring. All other crosses follow ordinary Mendelian inheritance. \textbf{Right:} Flowchart for the compartmental model capturing complex transmission routes of the \M gene drive across life stages and genotypes. Subscript 0 = wild-type (++); 1 = heterozygous (M+); 2 = \M homozygotes (MM).}
\label{fig:model_diagram}
\end{figure}

We formulate a continuous-time compartmental model that tracks the mosquito population by life stage (juveniles, combining eggs, larvae, and pupae; adult females; and adult males) and by genotype at the \M locus (wild-type, ++; heterozygous, M+; and \M homozygous, MM), yielding the nine state variables summarized in \cref{tab:variable} and illustrated in \cref{fig:model_diagram} (right panel).

Reproduction is modeled explicitly as a set of crosses between females and males of each genotype combination, numbered \circled{1}--\circled{9} in \cref{fig:model_diagram} (left) in increasing order of the female genotype index $i$ followed by the male genotype index $j$. The offspring genotype frequencies determined by standard Mendelian inheritance except where modified by the \M drive mechanism described below. We assume homogeneous mixing and equal mating competence among males regardless of genotype, so that the rate at which genotype-$i$ females mate with genotype-$j$ males is proportional to the frequency of genotype-$j$ males in the population; the resulting reproduction rate for the cross $F_i \times M_j$ is $F_i \frac{M_j}{N_M}\phi_{ij}$, where $\phi_{ij}$ denotes the per capita egg-laying rate for that cross and $N_M$ is the total male population, so that $M_j/N_M$ gives the mating probability with a genotype-$j$ male.

The \M drive alters offspring genotype frequencies through a maternal-effect toxin-antidote mechanism, which introduces the biased inheritance responsible for the selfish spread of \M through the population. Females carrying at least one copy of the drive (M+ or MM, crosses \circled{4}--\circled{9} in \cref{fig:model_diagram}) deposit a toxin during egg development; offspring that do not inherit a copy of the \M element from either parent are not rescued by the drive-linked antidote and are consequently nonviable. 
We refer to this as \Mns-killing: in crosses \circled{4} and \circled{5} (\cref{fig:model_diagram}, dashed box with cross out), where a heterozygous M+ female mates with a wild-type or heterozygous male, only a fraction $\mu_0$ of the wild-type offspring survive, which is reflected in the birth rate of $E_0$ with $\mu_0$ coefficient multiplied to the $F_1\times M_0$ and $F_1\times M_1$ terms. Experimental implementations in \textit{Drosophila melanogaster} have demonstrated near-complete killing \cite{chen2007synthetic, buchman2018synthetically}; however, resistance alleles in natural populations can reduce killing efficiency, yielding leakage values up to approximately $\mu_0 \approx 0.07$ \cite{buchman2018synthetically}. 
We set $\mu_0=0$ at baseline, corresponding to perfect \Mns-killing (no such offspring survive), but in general $0\le\mu_0\le1$ to allow for incomplete killing.

Additionally, during the rescue stage, offspring of \Mns-homozygous (MM) females experience double doses of toxin, and heterozygous offspring carrying only a single copy of the drive-linked antidote may fail to fully neutralize it, resulting in imperfect rescue. This phenomenon has been observed experimentally in \textit{Tribolium castaneum} and synthetic \textit{Drosophila} \M elements \cite{beeman1992maternaleffect, chen2007synthetic}. We refer to this as \M imperfect rescue: in crosses \circled{7} and \circled{8} (\cref{fig:model_diagram}, dashed box), where an MM female mates with a wild-type or heterozygous male, only a fraction $\mu_1$ of the resulting heterozygous offspring survive. This affects the birth rate of $E_1$ with the $\mu_1$ coefficient multiplied to the $F_2\times M_0$ and $F_2\times M_1$ terms. At baseline, we set $\mu_1=1$, corresponding to perfect rescue (all such offspring survive), but in general $0\le\mu_1\le1$ to allow for imperfect rescue.

Density dependence is incorporated through a logistic factor $\left(1-N_E/K\right)$ applied to the juvenile birth terms, where $N_E=E_0+E_1+E_2$ is the total juvenile population and $K$ is the carrying capacity, capturing resource competition among juveniles.

Beyond the toxin-antidote mechanism, \M carriage may impose an additional fitness cost, which we incorporate through genotype-dependent reductions in the reproduction rate $\phi_{ij}$ and through genotype-specific mortality rates at each life stage. The analytical results in the sections that follow hold for these general, genotype-dependent formulations, and specific assumptions are made for the numerical results, which are introduced separately in \cref{sec:numerical}. The remaining life-history parameters are taken to be genotype-independent, and the sex ratio at birth is assumed to be one-to-one.

The full system of ODEs is given in the system \eqref{eq:model}, with state variables and parameters summarized in \cref{tab:variable,tab:paramter}. We defer the discussion of additional modeling assumptions and their limitations to \cref{sec:discussion}.

\begin{table}[htbp]
\centering
\begin{tabular}{c l} 
\toprule
Var.& Description \\
\midrule
$E_0$ & Juvenile (eggs, larvae, pupae) population, wild-type (++)\\
$E_1$ & Juvenile (eggs, larvae, pupae) population, heterozygous (M+)\\
$E_2$ & Juvenile (eggs, larvae, pupae) population, \M homozygous (MM)\\
\midrule
$F_0$ & Female population, wild-type (++)\\
$F_1$ & Female population, heterozygous (M+)\\
$F_2$ & Female population, \M homozygous (MM)\\
\midrule
$M_0$ & Male population, wild-type (++)\\
$M_1$ & Male population, heterozygous (M+)\\
$M_2$ & Male population, \M homozygous (MM)\\
\midrule
$N_M$ & $=M_0+M_1+M_2$, Total male population\\
$N_E$ & $=E_0+E_1+E_2$, Total juvenile population\\
\bottomrule
\end{tabular}
\caption{Table of state variables. Subscript 0 = wild-type (++);  1 = heterozygous (M+); 2 = \M homozygous (MM); \label{tab:variable}}
\end{table}

\begin{align}
\frac{d E_0 }{dt} &= \left(
F_0 \frac{M_0}{N_M} \phi_{00} +  
\frac{1}{2} F_0 \frac{M_1}{N_M} \phi_{01} +
\frac{1}{2} F_1 \frac{M_0}{N_M} \phi_{10} \mu_{0}  +
\frac{1}{4} F_1 \frac{M_1}{N_M} \phi_{11} \mu_{0} \right)\left(1-\frac{N_E}{K}\right) - \psi E_0 - \mu_{e0} E_0,\nonumber\\
\frac{d E_1 }{dt} &= \left(
\frac{1}{2} F_0 \frac{M_1}{N_M} \phi_{01} +
F_0 \frac{M_2}{N_M} \phi_{02} +
\frac{1}{2} F_1 \frac{M_0}{N_M} \phi_{10} +
\frac{1}{2} F_1 \frac{M_1}{N_M} \phi_{11} + 
\frac{1}{2} F_1 \frac{M_2}{N_M} \phi_{12} +\right.\nonumber\\ 
& \left.\hspace{5cm} 
F_2 \frac{M_0}{N_M} \phi_{20} \mu_{1} +
\frac{1}{2} F_2 \frac{M_1}{N_M} \phi_{21} \mu_{1} \right)
\left(1-\frac{N_E}{K}\right)- \psi E_1 - \mu_{e1} E_1, \nonumber\\
\frac{d E_2 }{dt} &= \left( 
\frac{1}{4} F_1 \frac{M_1}{N_M} \phi_{11}+
\frac{1}{2} F_1 \frac{M_2}{N_M} \phi_{12} +
\frac{1}{2} F_2 \frac{M_1}{N_M} \phi_{21} + 
F_2 \frac{M_2}{N_M} \phi_{22} 
\right)\!\!\left(1-\frac{N_E}{K}\right) - \psi E_2 - \mu_{e2} E_2, \nonumber\\
\frac{d F_0 }{dt} &= \frac{1}{2} \psi E_0 - \mu_{f0} F_0, \qquad\qquad \frac{d M_0 }{dt} = \frac{1}{2} \psi E_0 - \mu_{m0} M_0, \label{eq:model}\\
\frac{d F_1 }{dt} &= \frac{1}{2} \psi E_1 - \mu_{f1} F_1, \qquad \qquad\frac{d M_1 }{dt} = \frac{1}{2} \psi E_1 - \mu_{m1} M_1, \nonumber\\
\frac{d F_2 }{dt} &= \frac{1}{2} \psi E_2 - \mu_{f2} F_2,
\qquad\qquad \frac{d M_2 }{dt} = \frac{1}{2} \psi E_2 - \mu_{m2} M_2.\nonumber
\end{align}

\begin{table}[htbp]
\centering
\begin{tabular}{lp{9cm}ccl}
\toprule
& Description & Baseline & Range & Ref.\\
\midrule
$\phi_0$ & Reproduction rate for wild-type females (day$^{-1})$& 3.7 & 1 – 8 & \cite{styer2007mortality,yang2011follow,qu2026multistage}\\
$\phi_{i,j}$ &  Reproduction rate for genotype-$i$ female mates with genotype-$j$ male $(=\phi_0 \times c^{i+j}, ~~i,j =0, 1, 2)$  (day$^{-1})$ & & &\\
$c$& Fitness coefficient & 0.9 & 0.45 – 1 & \cite{ward2011medea,chen2007synthetic,buchman2018synthetically} \\
\midrule 
$\psi$ & Juvenile development rate (day$^{-1})$ & 1/12 & 1/34 – 1/9 & \cite{soares-pinheiro2016eggs,foster2002mosquitoes}\\ 
\midrule
$\mu_{fi}$  & Female mortality rate for genotype $i ~(i=0, 1, 2)$  (day$^{-1})$ & 1/10 & 1/17.5 – 1/5 & \cite{nationalenvironmentagency2020aedes,zettel2009yellow}\\
$\mu_{mi}$ & Male mortality rate for genotype $i ~(i=0, 1, 2)$ (day$^{-1})$ & 1/7 & 1/10 – 1/4 &\cite{styer2007mortality} \\
$\mu_{ei}$ & Juvenile mortality rate for genotype $i ~(i=0, 1, 2)$ (day$^{-1})$ & 0.132 & 0.11 – 0.41& \cite{yang2011follow,soares-pinheiro2016eggs,qu2026multistage}\\
\midrule
$\mu_{0}$ & \Mns-killing leakage (crosses \circled{4}, \circled{5}): fraction of wild-type offspring of \M heterozygous females that survive & $0$ & 0 – 0.2 & \cite{chen2007synthetic,buchman2018synthetically,akbari2014novel}\\
$\mu_{1}$ & Rescue coefficient (crosses \circled{7}, \circled{8}): fraction of heterozygous offspring of \M homozygous females that survive & $1$ & 0.7 – 1 &\cite{beeman1992maternaleffect,chen2007synthetic,ward2011medea} \\
\midrule
$K$ & Carrying capacity & $10,000^\dag$ & & \cite{scott2000longitudinala}\\
\bottomrule
\end{tabular}
\caption{Model parameters, baseline values, and ranges. $^\dag$ The baseline value for $K$ serves only to set the scale of numerical simulations.\label{tab:paramter}}
\end{table}

\section{Equilibria and Stability Analysis} \label{sec:analysis}
We investigate the qualitative behavior of the proposed model, system \eqref{eq:model}, including the existence and local stability of its equilibria. The system admits four biologically relevant equilibria: (1) the \Mns-free equilibrium (MFE), consisting entirely of wild-type mosquitoes; (2) the \Mns-complete equilibrium (MCE), consisting entirely of \M homozygous (MM) mosquitoes; (3) the \Mns-carrying equilibrium, in which wild-type mosquitoes have been eliminated but heterozygous and homozygous genotypes coexist; and (4) the coexistence equilibrium, in which all three genotypes are simultaneously present. We further derive a set of dimensionless population reproduction numbers $\mathcal{G}_{ijk}$ that admit direct biological interpretation, and they contribute to the basic reproduction number $\mathcal{R}_0$, governing invasion of the \M element into a wild-type population.

Local stability of equilibria is established via the Metzler matrix theory below, which we include for convenience:
\begin{lemma}[Metzler Stability, \cite{kamgang2008computation}]\label{lem:metzler}
A real square matrix is called Metzler if all of its off-diagonal entries are non-negative, and Metzler stable if, in addition, all of its eigenvalues have negative real part.
Suppose $\mathbf{M}$ is a Metzler matrix that admits the block decomposition
\[
\mathbf{M} = 
\begin{bmatrix}
A & B  \\
C & D  \\
\end{bmatrix},
\]
where $A$ and $D$ are square. Then $\mathbf{M}$ is Metzler stable if and only if $A$ and $D - CA^{-1}B$ are both Metzler stable.
\end{lemma}

\subsection{\Mns-free Equilibrium (MFE)} \label{sec:MFE}
This equilibrium represents a population composed entirely of wild-type mosquitoes. Setting the right-hand side of system \eqref{eq:model} to zero and imposing $E_1 = E_2 = F_1 = F_2 = M_1 = M_2 = 0$, we obtain the MFE,
\begin{align*}
& E_0^{MFE} = \left(1-\frac{1}{\mathcal{G}_{00}}\right) K, \quad  F_0^{MFE}  = \frac{\psi}{2\mu_{f0}}E_0^{MFE}, \quad  M_0^{MFE}  = \frac{\psi}{2\mu_{m0}}E_0^{MFE},\\[10pt]
& E_1^{MFE} = E_2^{MFE} =F_1^{MFE} = F_2^{MFE} = M_1^{MFE} = M_2^{MFE} = 0,
\end{align*}
where $\mathcal{G}_{00}$ is defined below in \cref{eq:G00}.

\subsubsection{Population Reproduction Numbers near \Mns-free Equilibrium}
We define three dimensionless population reproduction numbers that characterize the per-capita production of female offspring from each relevant mating type. 

\paragraph{Population Reproduction Number for Wild-type Matings, $\mathcal{G}_{00}$}
\begin{equation}\label{eq:G00}
\mathcal{G}_{00} = \frac{1}{2} \frac{\psi}{\psi + \mu_{e0}}\frac{\phi_{00}}{\mu_{f0}}
\end{equation}
is the population reproduction number for the resident wild-type mosquito population. This quantity can be interpreted as follows: given a wild-type female $F_0$ mating with a wild-type male $M_0$ (the cross $F_0\times M_0$ corresponding to the subscript in $\mathcal{G}_{00}$), on average she reproduces $\phi_{00}/\mu_{f0} $ offspring over her expected lifetime of $1/\mu_{f0}$ days, a fraction $\psi/(\psi + \mu_{e0})$ of these offspring survive the egg stage to reach adulthood, and $1/2$ of surviving adults are females.

\paragraph{Population Reproduction Numbers for ++ and M+ Matings, $\mathcal{G}_{011}$ and $\mathcal{G}_{101}$}
Two further population reproduction numbers describe the heterozygous crosses that produce \Mns-carrying (M+) offspring,
\begin{equation}\label{eq:G011}
\mathcal{G}_{011} = \frac{1}{2} \cdot \frac{1}{2} \frac{\psi}{\psi + \mu_{e1}}\frac{\phi_{01}}{\mu_{f0}}
\end{equation}
and
\begin{equation}\label{eq:G101}
\mathcal{G}_{101} = \frac{1}{2} \cdot \frac{1}{2} \frac{\psi}{\psi + \mu_{e1}}\frac{\phi_{10}}{\mu_{f1}},
\end{equation}
corresponding to the crosses $F_0\times M_1 \to F_1$ (cross \circled{2}) and $F_1\times M_0 \to F_1$ (cross \circled{4}), respectively (the subscripts in each case indicating the female genotype, male genotype, and tracked offspring genotype).
Both quantities follow the same structural interpretation as $\mathcal{G}_{00}$ -- the mating female's expected lifetime egg output ($\phi_{01}/\mu_{f0}$ or $\phi_{10}/\mu_{f1}$), scaled by the fraction of eggs surviving to adulthood at the heterozygote-specific rate $\psi/(\psi+\mu_{e1})$, and by the $1/2$ female sex ratio. One structural difference is an additional factor of $1/2$, reflecting that only half the offspring for the cross is M+ genotype under Mendelian inheritance, rather than the entire brood as in the homozygous scenario for $\mathcal{G}_{00}$.
For $\mathcal{G}_{101}$, we further note that the cross $F_1\times M_0$ is precisely one of the \Mns-killing crosses (cross \circled{4} in \cref{fig:model_diagram}): the wild-type offspring for this cross is subject to maternal-killing effect, but this does not enter $\mathcal{G}_{101}$ itself, since it tracks only the surviving M+ offspring, which are unaffected by the killing mechanism.

\paragraph{Biological Significance}
These population reproduction numbers are dimensionless quantities that measure genotype-specific reproductive success relative to the generation time scale. When \Mns-induced fitness cost is absent ($\phi_{01} = \phi_{10} = \phi_{00}$, $\mu_{e1} = \mu_{e0}$, $\mu_{f1} = \mu_{f0}$), $\mathcal{G}_{011} = \mathcal{G}_{101} = \frac{1}{2}\mathcal{G}_{00}$, reflecting that heterozygous crosses \circled{2} and \circled{4} produce heterozygous offspring at exactly half the rate that wild-type cross \circled{1} produce wild-type offspring. Under biologically-relevant parameter regime, we have 
\begin{equation}\label{eq:bio_cond1}
\mathcal{G}_{00}>1    
\end{equation}
ensuring a self-sustaining wild-type population in the absence of \Mns. With \Mns-included fitness cost,  $\phi_{01}<\phi_{00}, ~ \phi_{10} <\phi_{00}$, $\mu_{e1} > \mu_{e0}$, and $\mu_{f1} > \mu_{f0}$, it implies
\begin{equation}\label{eq:bio_cond2}
\mathcal{G}_{011}<\frac{1}{2}\mathcal{G}_{00}, \quad \mathcal{G}_{101}<\frac{1}{2}\mathcal{G}_{00},
\end{equation}
which implies the reproductive disadvantage of \Mns-carrying genotypes; the maternal-effect killing mechanism, however, generates biased inheritance that is in favor of \M cohorts and can offset this disadvantage sufficiently to permit invasion.

The interplay of $\mathcal{G}_{00}$, $\mathcal{G}_{011}$, and $\mathcal{G}_{101}$ determines an invasion threshold, formalized through the basic reproduction number $\mathcal{R}_0$ in \cref{sec:R0}.

\subsubsection{Stability of \Mns-free Equilibrium} For convenience, we reorder the state variables by genotypes, $\mathbf{X}=\{E_0, F_0, M_0, E_1, F_1, M_1, E_2, F_2, M_2\}$, so that the Jacobian matrix of system \eqref{eq:model}, evaluated at the MFE, takes the block form
\begin{align*}
&\hspace{-1.5cm}\mathcal{J}_0\!\!= \!\!
\resizebox{1.15\textwidth}{!}{\ensuremath{
\left[
\begin{array}{@{}ccc:ccc:ccc@{}}
-\frac{\phi_{00} \psi}{2 \mu_{f0}} & \frac{2 \mu_{f0} (\mu_{e0} + \psi)}{\psi} & 0  & \mu_{e0} + \psi - \frac{\phi_{00}  \psi}{2 \mu_{f0}} & \frac{\mu_0 \mu_{f0} \phi_{10} (\mu_{e0} + \psi)}{\phi_{00} \psi} & -\frac{\mu_{m0} (2 \phi_{00} - \phi_{01}) (\mu_{e0} + \psi)}{\phi_{00} \psi} & \mu_{e0} + \psi - \frac{\phi_{00}  \psi}{2 \mu_{f0}} & 0 & -\frac{2 \mu_{m0} (\mu_{e0} + \psi)}{\psi} \\ 
\frac{\psi}{2} & -\mu_{f0} & 0  & 0 & 0 & 0 & 0 & 0 & 0 \\ 
\frac{\psi}{2} & 0 & -\mu_{m0} & 0 & 0 & 0 & 0 & 0 & 0\\ 
\hdashline
0 & 0 & 0 & -\mu_{e1}- \psi & \frac{ \mu_{f0} \phi_{10}(\mu_{e0} + \psi)}{\phi_{00}\psi}  & \frac{ \mu_{m0} \phi_{01}(\mu_{e0} + \psi)}{\phi_{00}\psi}  & 0 & \frac{2\mu_1 \mu_{f0} \phi_{20} (\mu_{e0} + \psi)}{\phi_{00} \psi} & \frac{2 \mu_{m0} \phi_{02} (\mu_{e0} + \psi)}{\phi_{00} \psi} \\ 
0 & 0 & 0 & \frac{\psi}{2} & -\mu_{f1} & 0  & 0 & 0& 0 \\ 
0 & 0 & 0 & \frac{\psi}{2} & 0 & -\mu_{m1} & 0 & 0 & 0 \\ 
\hdashline
0 & 0 & 0 & 0 & 0 & 0 & -\mu_{e2} - \psi & 0 & 0  \\ 
0 & 0 & 0 & 0 & 0 & 0 & \frac{\psi}{2} & -\mu_{f2} & 0  \\ 
0 & 0 & 0 & 0 & 0 & 0 & \frac{\psi}{2} & 0 & -\mu_{m2}
\end{array}
\right]}}\\ 
& = \begin{bmatrix}
A_{1,1} & A_{1,2} & A_{1,3}\\
\mathcal{O}_{3\times 3} & A_{2,2} & A_{2,3}\\
\mathcal{O}_{3\times 3} & \mathcal{O}_{3\times 3} & A_{3,3}
\end{bmatrix}
\end{align*}
which is a block upper triangular matrix, so the eigenvalues of $\mathcal{J}_0$ are given by the union of the eigenvalues of the diagonal blocks, $A_{1,1}, ~ A_{2,2}$, and $A_{3,3}$. 

The block $A_{3,3}$ is lower triangular with negative diagonal entries, so all of its eigenvalues are negative. The block $A_{1,1}$ 
$$
A_{1,1} = \left[\begin{array}{cc:c}
-\dfrac{\phi_{00} \psi}{2\mu_{f_0}} & \dfrac{2 \mu_{f_0} (\mu_{e0} + \psi)}{\psi} & 0 \\
\dfrac{\psi}{2} & -\mu_{f_0} & 0 \\
\hdashline
\dfrac{\psi}{2} & 0 & -\mu_{m0}
\end{array}\right] = \begin{bmatrix}
B_{1,1} & \mathcal{O}_{2\times 1} \\
B_{2,1} & B_{2,2} 
\end{bmatrix}.
$$
is Metzler, since all off-diagonal entries are non-negative. The submatrix $B_{1,1}$ has negative eigenvalues as 
\begin{equation}\label{eq:MFE_cond1}
tr(B_{1,1})<0, \quad det(B_{1,1}) = \mu_{f0}(\mu_{e0}+\psi) \left(\mathcal{G}_{00}-1\right)>0, \quad \text{provided} \quad \mathcal{G}_{00}>1.
\end{equation}
The last condition is satisfied at the biologically relevant parameter regime for the persistence of wild-type mosquitoes. Thus, $B_{1,1}$ is Metzler stable. Additionally, $B_{2,2}-B_{2,1}B_{1,1}^{-1} \mathcal{O}_{2\times 1} = -\mu_{m0}<0$ is a trivially negative, so by the Metzler Stability lemma (\cref{lem:metzler}), $A_{1,1}$ is Metzler stable and all of its eigenvalues are negative.

The block $A_{2,2}$ can be decomposed similarly,
$$
A_{2,2} = \left[\begin{array}{cc:c}
-\mu_{e1} - \psi & \frac{ \mu_{f0} \phi_{10}(\mu_{e0} + \psi)}{\phi_{00}\psi}  & \dfrac{\mu_{m0} \phi_{01} (\mu_{e0} + \psi)}{\phi_{00} \psi} \\
\dfrac{\psi}{2} & -\mu_{f1} & 0 \\
\hdashline
\dfrac{\psi}{2} & 0 & -\mu_{m1}
\end{array} \right] = \begin{bmatrix}
C_{1,1} & C_{1,2} \\
C_{2,1} & C_{2,2} 
\end{bmatrix}.
$$
The submatrix $C_{1,1}$ is Metzler stable, as 
\begin{equation}\label{eq:MFE_cond2}
tr(C_{1,1}) <0,\quad det(C_{1,1}) =  \mu_{f1} (\psi + \mu_{e1})
- \frac{\phi_{10}\,\mu_{f0}}{2\phi_{00}} (\psi + \mu_{e0})> 0,  \quad \text{provided} \quad \mathcal{G}_{00}>\mathcal{G}_{101}.
\end{equation}
Furthermore, 
$$
C_{2,2}-C_{2,1}C_{1,1}^{-1}C_{1,2} = -\mu_{m1}
+ \frac{\mu_{f1}\,\mu_{m0}\,\phi_{01}\,(\mu_{e0} + \psi)}
{2\phi_{00}\mu_{f1}(\mu_{e1} + \psi)
- \mu_{f0}\phi_{10}(\mu_{e0} + \psi)}<0 
$$
provided
\begin{equation}\label{eq:MFE_cond3}
\mathcal{G}_{00}>\mathcal{G}_{101}, \quad \text{and}\quad \frac{\mathcal{G}_{011}\frac{1}{\mu_{m1}} + \mathcal{G}_{101}\frac{1}{\mu_{m0}}}{\mathcal{G}_{00}\frac{1}{\mu_{m0}}}<1.
\end{equation}
Notice that the second inequality is precisely the condition $\mathcal{R}_0<1$ for the basic reproduction number derived in \cref{sec:R0}.

Combining the three blocks, the MFE is locally asymptotically stable if $\mathcal{G}_{00}>1$, $\mathcal{G}_{00}>\mathcal{G}_{101}$, and $\mathcal{R}_0<1$ (\cref{eq:MFE_cond1,eq:MFE_cond2,eq:MFE_cond3}); each of these conditions holds under the biologically relevant regime specified in \cref{eq:bio_cond1,eq:bio_cond2}, so the MFE is stable throughout the parameter regime of interest.

\subsection{Basic Reproduction Number}\label{sec:R0}
Following the next-generation matrix approach \cite{van2002reproduction} and considering the subsystem for $\mathbf{X} = [E_1, F_1, M_1, E_2, F_2, M_2]^T$ tracking the \Mns-carrying compartments, we obtain the basic reproduction number
\begin{equation}\label{eq:R0}
\mathcal{R}_0= 
\frac{(\mu_{f1}\mu_{m0}\phi_{01}
+ \mu_{f0}\mu_{m1}\phi_{10})
(\mu_{e0} + \psi)}{2\mu_{f1}\mu_{m1}\phi_{00}
(\mu_{e1} + \psi)}
=
\frac{
\mathcal{G}_{011}\,\frac{1}{\mu_{m1}}
\;+\;
\mathcal{G}_{101}\,\frac{1}{\mu_{m0}}
}{
\mathcal{G}_{00}\,\frac{1}{\mu_{m0}}
}
\end{equation}
where $\mathcal{G}_{00}$, $\mathcal{G}_{011}$, and $\mathcal{G}_{101}$ are the population reproduction numbers given in \cref{eq:G00,eq:G011,eq:G101}.

\paragraph{Biological Interpretation} 
$\mathcal{R}_0$ carries the classical interpretation to measure the invasiveness of an infection. Here, infection corresponds to \Mns-carrying status, and $\mathcal{R}_0$ gives the expected number of \Mns-carrying offspring produced per \Mns-carrying individual, normalized by the wild-type population reproduction, when a small introduction is made into a field population at the MFE. As in analyses of the \textit{Wolbachia} population-replacement strategy \cite{qu2018modeling}, $\mathcal{R}_0$ in \cref{eq:R0} takes the form of a ratio of the invasion fitness of \Mns-carrying individuals to the fitness of the resident wild-type population at the MFE, directly revealing the underlying competition dynamics between the two.

Specifically, the numerator sums the contributions of the two reciprocal heterozygous crosses that produce M+ offspring, $\mathcal{G}_{011}$ (from $F_0\times M_1$) and $\mathcal{G}_{101}$ (from $F_1\times M_0$), each weighted by the corresponding male lifespan, $1/\mu_{m1}$ for M+ males and $1/\mu_{m0}$ for wild-type males; when male lifespan carries no fitness cost ($\mu_{m1}=\mu_{m0}$), the two crosses contribute equally. The denominator is the reproductive output of the resident wild-type population, $\mathcal{G}_{00}$, weighted by the same wild-type male lifespan $1/\mu_{m0}$, and serves as the reference baseline.

The ratio, $\mathcal{R}_0$, therefore compares the combined reproductive success of the two heterozygous crosses directly against the wild-type cross:
\begin{itemize}
\item When $\mathcal{R}_0 > 1$: \Mns-carrying individuals produce more offspring per capita than wild-type individuals near the MFE, and \M can invade from an arbitrarily low frequency, with no threshold introduction required. 
\item When $\mathcal{R}_0 = 1$: This happens when there is no \Mns-induced fitness cost, and the $M+$ and wild-type cohorts have equal per-capita reproductive rates, so the \M frequency neither grows nor declines. 
\item When $\mathcal{R}_0 < 1$ (more biologically-relevant case): Wild-type individuals outcompete $M+$ individuals near MFE due to fitness costs (increased embryonic mortality $\mu_{e1} > \mu_{e0}$, reduced fecundity $\phi_{10} < \phi_{00}$ and $\phi_{01} < \phi_{00}$, or reduced female survival $\mu_{f1} > \mu_{f0}$). The MFE is locally stable as shown in \cref{sec:MFE}, and \M cannot invade from low frequency. A threshold introduction is required for \M to persist.
\end{itemize}

\subsection{\Mns-complete Equilibrium (MCE)}\label{sec:MCE}
This equilibrium represents a population entirely composed of \M homozygous (MM) mosquitoes, with no heterozygous or wild-type cohorts present.
Setting the right-hand side of system \eqref{eq:model} to zero and imposing $E_0 = E_1 = F_0 = F_1 = M_0 = M_1 = 0$, we obtain the \Mns-complete equilibrium (MCE),
\begin{align*}
& E_2^{MCE} = \left(1-\frac{1}{\mathcal{G}_{22}}\right) K, \quad  F_2^{MCE}  = \frac{\psi}{2\mu_{f2}}E_2^{MCE}, \quad  M_2^{MCE}  = \frac{\psi}{2\mu_{m2}}E_2^{MCE},\\[10pt]
& E_0^{MCE} = E_1^{MCE} =F_0^{MCE} = F_1^{MCE} = M_0^{MCE} = M_1^{MCE} = 0,
\end{align*}
where $\mathcal{G}_{22}$ is the population reproduction number for \Mns-homozygous mosquitoes, defined below in \cref{eq:G22}.

\subsubsection{Population Reproduction Numbers near \Mns-complete Equilibrium}
Similar to the population reproduction numbers near the MFE, we define population reproduction numbers that characterize the dynamics near the MCE.

\paragraph{Population Reproduction Number for \Mns-homozygous Matings, $\mathcal{G}_{22}$}
\begin{equation}\label{eq:G22}
\mathcal{G}_{22} = \frac{1}{2} \frac{\psi}{\psi + \mu_{e2}}\frac{\phi_{22}}{\mu_{f2}}
\end{equation}
is the population reproduction number for the resident \Mns-homozygous population, representing the expected number of MM female offspring produced by an MM female mated with an MM male; its structure and interpretation are identical to $\mathcal{G}_{00}$ (\cref{eq:G00}), with all parameters replaced by their MM-genotype counterparts.

\paragraph{Population Reproduction Numbers for M+ and MM Matings, $\mathcal{G}_{211}$ and $\mathcal{G}_{121}$}
Two further population reproduction numbers describe the crosses between heterozygous and homozygous individuals that produce M+ offspring,
\begin{equation*}
\mathcal{G}_{211} = \frac{1}{2} \cdot \frac{1}{2} \frac{\psi}{\psi + \mu_{e1}}\frac{\phi_{21}}{\mu_{f2}}\mu_1 \quad \text{and}\quad \mathcal{G}_{121} = \frac{1}{2} \cdot \frac{1}{2} \frac{\psi}{\psi + \mu_{e1}}\frac{\phi_{12}}{\mu_{f1}}
\end{equation*}
corresponding to the crosses $F_2\times M_1 \to F_1$ (cross \circled{8} in \cref{fig:model_diagram}) and $F_1\times M_2 \to F_1$ (cross \circled{6}), respectively.
Both follow the same structural interpretation as $\mathcal{G}_{101}$ near the MFE (\cref{eq:G101}) but differ in one respect: the cross $F_2\times M_1$ (cross \circled{8}) is one of the potential imperfect \Mns-rescue crosses, where only a fraction $\mu_1$ of its heterozygous offspring survive, and this factor appears explicitly in $\mathcal{G}_{211}$.

\paragraph{Biological Constraints}
We require the \Mns-homozygous population to be self-sustaining in isolation,
\begin{equation}\label{eq:bio_cond3}
\mathcal{G}_{22} > 1.    
\end{equation}
Assuming \M imposes a fitness cost such that heterozygous (M+) individuals are fitter than homozygous (MM) individuals, that is, $\phi_{22}\le \phi_{21}$, $\phi_{22}\le\phi_{12}$, $\mu_{e2}\ge \mu_{e1}$, and $\mu_{f2}\ge \mu_{f1}$, we have the following constraints
\begin{equation}\label{eq:bio_cond4}
\mathcal{G}_{121} > \dfrac{1}{2}\mathcal{G}_{22},\quad 
\mathcal{G}_{211} > \dfrac{1}{2}\mathcal{G}_{22}\mu_1.
\end{equation} 

\subsubsection{Stability of MCE} 
The Jacobian matrix of the system \eqref{eq:model}, evaluated at the MCE, is given by
\begin{align*}
\hspace{-1.5cm}\mathcal{J}_2\!\!&= \!\!
\resizebox{1.05\textwidth}{!}{\ensuremath{
\left[
\begin{array}{ccc:ccc:ccc}
-\mu_{e0} - \psi & 0 & 0 & 0 & 0 & 0 & 0 & 0 & 0 \\
\frac{\psi}{2} & -\mu_{f0} & 0 & 0 & 0 & 0 & 0 & 0 & 0 \\
\frac{\psi}{2} & 0 & -\mu_{m0} & 0 & 0 & 0 & 0 & 0 & 0 \\
\hdashline
0 & \frac{2\mu_{f2} \phi_{02} (\mu_{e2} + \psi)}{\phi_{22} \psi} & \frac{2 \mu_1 \mu_{m2} \phi_{20} (\mu_{e2} + \psi)}{\phi_{22} \psi} & -\mu_{e1} - \psi & \frac{\mu_{f2} \phi_{12} (\mu_{e2} + \psi)}{\phi_{22} \psi} & \frac{\mu_1 \mu_{m2} \phi_{21} (\mu_{e2} + \psi)}{\phi_{22} \psi} & 0 & 0 & 0 \\
0 & 0 & 0 & \frac{\psi}{2} & -\mu_{f1} & 0 & 0 & 0 & 0 \\
0 & 0 & 0 & \frac{\psi}{2} & 0 & -\mu_{m1} & 0 & 0 & 0 \\
\hdashline
\mu_{e2} + \psi-\frac{\phi_{22}\psi}{2\mu_{f2}} & 0 & -\frac{2\mu_{m2}(\mu_{e2}+\psi)}{\psi} & \mu_{e2} + \psi -\frac{\phi_{22} \psi}{2 \mu_{f2}}  & \frac{\mu_{f2} \phi_{12} (\mu_{e2} + \psi)}{\phi_{22} \psi} & \frac{\mu_{m2} (\phi_{21}-2\phi_{22}) (\mu_{e2} + \psi)}{\phi_{22} \psi} &  -\frac{\phi_{22}\psi}{2\mu_{f2}} & \frac{2 \mu_{f2} (\mu_{e2} + \psi)}{\psi} & 0 \\
0 & 0 & 0 & 0 & 0 & 0 & \frac{\psi}{2} & -\mu_{f2} & 0 \\
0 & 0 & 0 & 0 & 0 & 0 & \frac{\psi}{2} & 0 & -\mu_{m2}
\end{array}
\right]}}\\
& = \begin{bmatrix}
A_{1,1} &  \mathcal{O}_{3\times 3} & \mathcal{O}_{3\times 3}\\
A_{2,1} & A_{2,2} & \mathcal{O}_{3\times 3}\\
A_{3,1} & A_{3,2} & A_{3,3}
\end{bmatrix},
\end{align*}
which is block lower triangular, so the eigenvalues of $\mathcal{J}_2$ are given by the union of the eigenvalues of the diagonal blocks $A_{1,1}$, $A_{2,2}$, and $A_{3,3}$. We derive the conditions under which each block is stable.

The block $A_{1,1}$ is lower triangular with strictly negative diagonal entries, so all of its eigenvalues are negative, and $A_{1,1}$ is unconditionally stable.

The block
$$
A_{3,3} = \left[\begin{array}{cc:c}
-\dfrac{\phi_{22} \psi}{2\mu_{f_2}} & \dfrac{2 \mu_{f_2} (\mu_{e2} + \psi)}{\psi} & 0 \\[10pt]
\dfrac{\psi}{2} & -\mu_{f_2} & 0 \\
\hdashline
\dfrac{\psi}{2} & 0 & -\mu_{m_2}
\end{array}\right] = \begin{bmatrix}
B_{1,1} & \mathcal{O}_{2\times 1} \\
B_{2,1} & B_{2,2}
\end{bmatrix}
$$
is Metzler. The submatrix $B_{1,1}$ is a Metzler matrix with
\begin{equation}\label{eq:MCE_cond1}
tr(B_{1,1})<0, \quad det(B_{1,1}) = -\mu_{e2}\mu_{f2} - \mu_{f2}\psi + \dfrac{\phi_{22}\psi}{2}>0 \quad \text{provided} \quad \mathcal{G}_{22}>1
\end{equation}
Thus, $B_{1,1}$ is Metzler stable if $\mathcal{G}_{22}>1$. The matrix
$$
B_{2,2}-B_{2,1}B_{1,1}^{-1}B_{1,2} = -\mu_{m2}<0,
$$
which is Metzler stable. Therefore, by \cref{lem:metzler}, matrix $A_{3,3}$ is Metzler stable under the condition \cref{eq:MCE_cond1}.

The block 
$$
A_{2,2}=\left[
\begin{array}{cc:c}
- \mu_{e1} - \psi & \frac{\mu_{f2} \phi_{12}(\mu_{e2} + \psi)}{\phi_{22} \psi} & \frac{\mu_1 \mu_{m2} \phi_{21}(\mu_{e2} + \psi)}{\phi_{22} \psi} \\
\frac{\psi}{2} & - \mu_{f1} & 0 \\ 
\hdashline
\frac{\psi}{2} & 0 & - \mu_{m1}
\end{array}
\right]= \begin{bmatrix}
C_{1,1} & C_{1,2} \\
C_{2,1} & C_{2,2} 
\end{bmatrix}
$$
decomposes similarly. The submatrix $C_{1,1}$ has negative trace and a positive determinant,
\begin{equation}\label{eq:MCE_cond2}
det(C_{1,1})=\mu_{e1}\mu_{f1} + \mu_{f1}\psi - \frac{\mu_{e2}\mu_{f2}\phi_{12}}{2\phi_{22}} - \frac{\mu_{f2}\phi_{12}\psi}{2\phi_{22}} > 0, \quad\text{provided}\quad \mathcal{G}_{22} > \mathcal{G}_{121}.
\end{equation}
Thus, $C_{1,1}$ is Metzler stable under condition \cref{eq:MCE_cond2}. Moreover, 
$$
C_{2,2}-C_{2,1}C_{1,1}^{-1}C_{1,2} = -\mu_{m1} + \mu_1 \frac{ \mu_{f1} \mu_{m2} \phi_{21} (\mu_{e2} + \psi)}{2\mu_{f1}\phi_{22}(\mu_{e1} + \psi) - \mu_{f2}\phi_{12}(\mu_{e2} + \psi)}<0,
$$
is Metzler stable provided
\begin{equation}\label{eq:MCE_cond3}
\mathcal{G}_{22} > \mathcal{G}_{121}, \quad \text{and} \quad \mathcal{R}_2=\frac{\mathcal{G}_{211} \frac{1}{\mu_{m1}} + \mathcal{G}_{121} \frac{1}{\mu_{m2}}}{\mathcal{G}_{22} \frac{1}{\mu_{m2}}} < 1.
\end{equation}
Analogous to $\mathcal{R}_0$ near the MFE (\cref{sec:R0}), $\mathcal{R}_2$ measures the expected number of \M heterozygous (M+) female offspring produced per MM female offspring when a small cohort of M+ individuals is introduced into a population at the MCE. When $\mathcal{R}_2<1$, it indicates that the resident MM population outcompetes a reintroduced M+ cohort near the MCE.
Hence $A_{2,2}$ is Metzler stable, and all of its eigenvalues are negative, whenever $\mathcal{G}_{22} > \mathcal{G}_{121}$ and $\mathcal{R}_2<1$.

Combining the three blocks, the MCE is locally asymptotically stable if $\mathcal{G}_{22}>1$, $\mathcal{G}_{22} > \mathcal{G}_{121}$, and $\mathcal{R}_2<1$ (\cref{eq:MCE_cond1,eq:MCE_cond2,eq:MCE_cond3}).

The first condition holds under the biologically relevant constraint \cref{eq:bio_cond3}, but the latter two are not guaranteed: 
\begin{itemize}
\item Under perfect rescue ($\mu_1=1$ in $\mathcal{G}_{211}$) and biological constraints on fitness cost (\cref{eq:bio_cond4} and reduced male life span $\mu_{m2}>\mu_{m1}$), $\mathcal{R}_2>1$, thus the MCE is unstable.
\item Under imperfect rescue ($\mu_1<1$) with the same biological constraints on fitness cost, 
$\mathcal{G}_{211}$ is reduced, and the latter two stability conditions are more easily satisfied, so the MCE is more likely to be stable for small $\mu_1$ and small fitness costs.
\end{itemize}

\subsection{Internal Equilibria}
Beyond the MFE and MCE, the system admits two additional biologically relevant equilibria in which more than one genotype persists.
The \textbf{\Mns-carrying equilibrium} represents a population in which wild-type individuals have been eliminated, but both \Mns-bearing genotypes persist ($E_0 = 0$, $E_1, E_2 > 0$), capturing the scenario in which the \M element successfully invades and drives the wild-type population to local extinction while heterozygotes and \Mns-homozygotes coexist. The \textbf{coexistence equilibrium} has all three genotypes simultaneously present ($E_0, E_1, E_2 > 0$), representing a balance among wild-type, heterozygous, and \Mns-homozygous cohorts.

Characterizing these internal equilibria analytically is considerably more involved than the boundary cases treated in \cref{sec:MFE,sec:MCE}: the equilibrium equations are nonlinear in the state variables, and their solution does not admit a closed form unless under highly restrictive parametric conditions. Rather than pursue this algebra in full generality, we adopt a simplified but biologically motivated parameterization for the remainder of the analysis and use numerical bifurcation methods to characterize the existence and stability of the internal equilibria as a function of key parameters (\cref{sec:bifur}).
This approach provides direct insight into the system's qualitative behavior across biologically relevant parameter regimes. It reveals bifurcation phenomena that would otherwise be difficult to anticipate from boundary analysis alone.

\section{Numerical Results}\label{sec:numerical}
For the numerical bifurcation, sensitivity, and simulation results that follow, we parameterize the model using life-history data for \textit{Aedes aegypti}, with \Mns-specific parameters drawn from synthetic \M constructs characterized in \textit{Drosophila}, since no functional synthetic \M construct has yet been demonstrated in \textit{Aedes aegypti}. Baseline values and ranges are summarized in \cref{tab:paramter}.
We additionally adopt two simplifying assumptions for this parameterization, used only in the numerical results that follow and not required for the analytical results of \cref{sec:analysis}: adult and juvenile mortality rates are assumed equal across genotypes (\cref{eq:assum}), and \Mns-induced fitness costs are encoded multiplicatively through a single coefficient $c$ acting on the reproduction rate, $\phi_{ij} = \phi_0\,c^{i+j}$ (\cref{eq:assum2}).

Full derivations, literature sources, and justification for the parameterization and assumptions are given in \ref{sec:app_param}. All codes for generating the numerical results are developed in MATLAB 2025a and will be made available upon acceptance. We use the MATLAB \texttt{ode45} solver to integrate the ODE model.

\subsection{Bifurcation Diagrams} \label{sec:bifur}
\begin{figure}[ht!]
\centering
\includegraphics[width=0.48\linewidth]{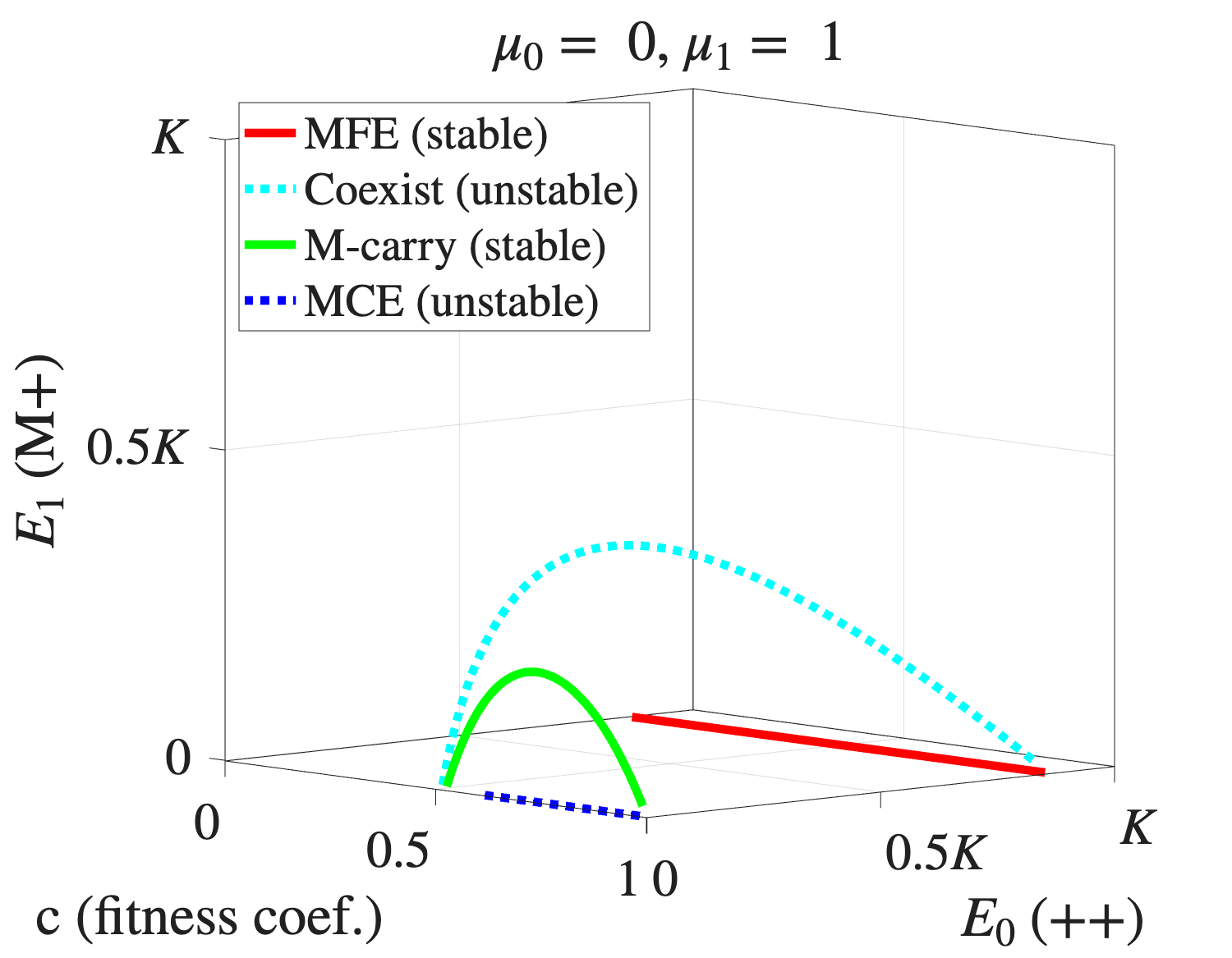}\hfill\includegraphics[width=0.48\linewidth]{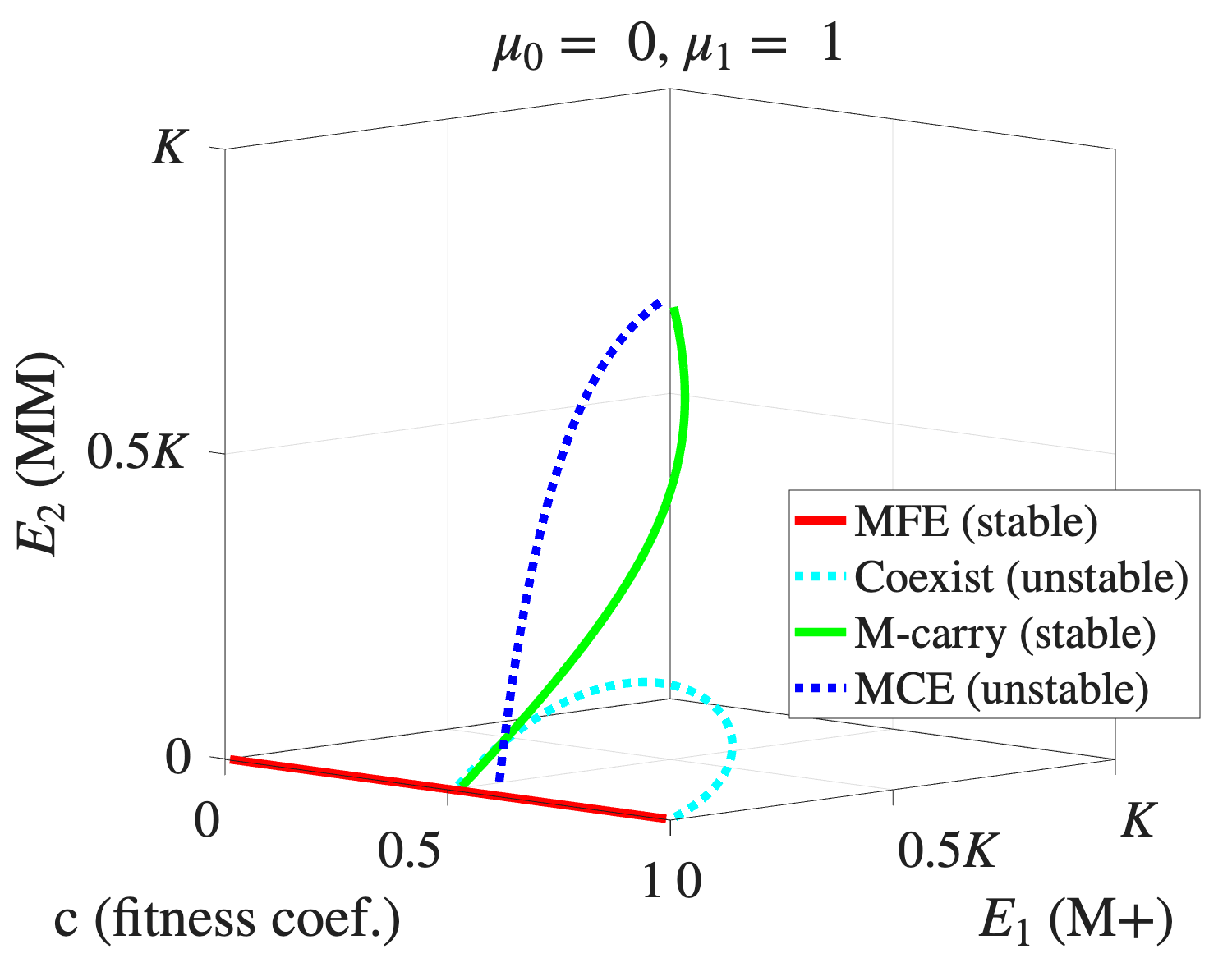}
\includegraphics[width=0.48\linewidth]{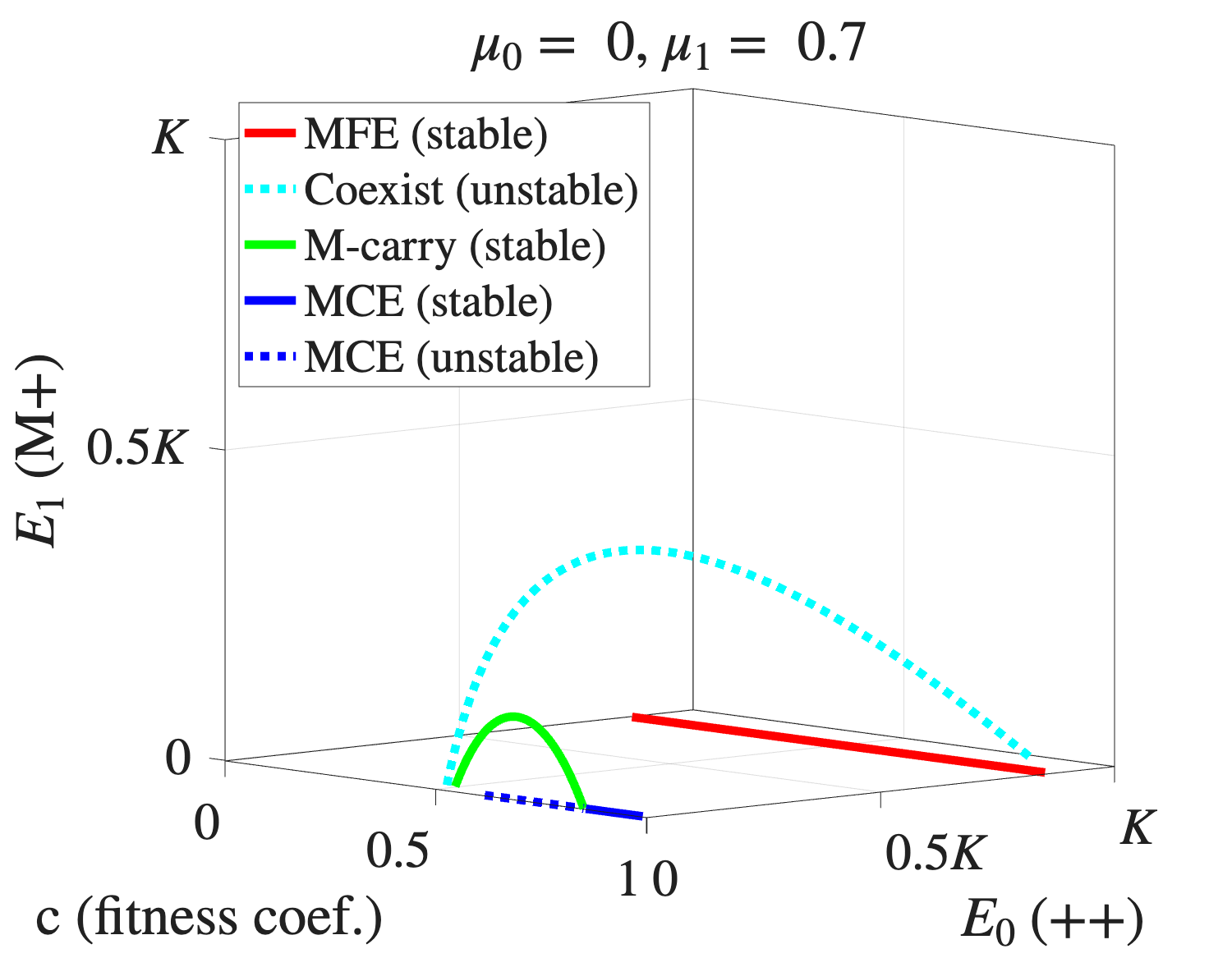}\hfill\includegraphics[width=0.48\linewidth]{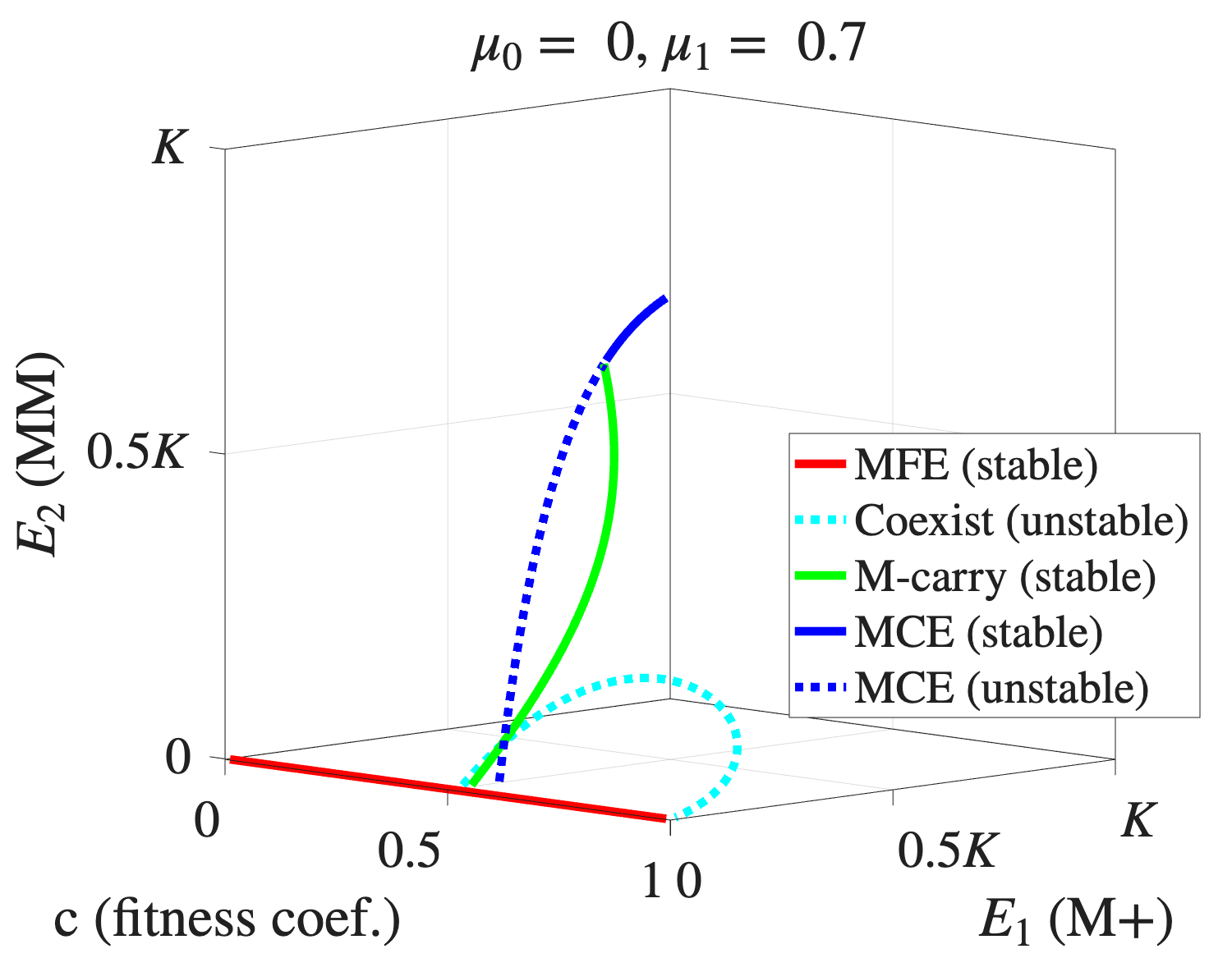}
\includegraphics[width=0.48\linewidth]{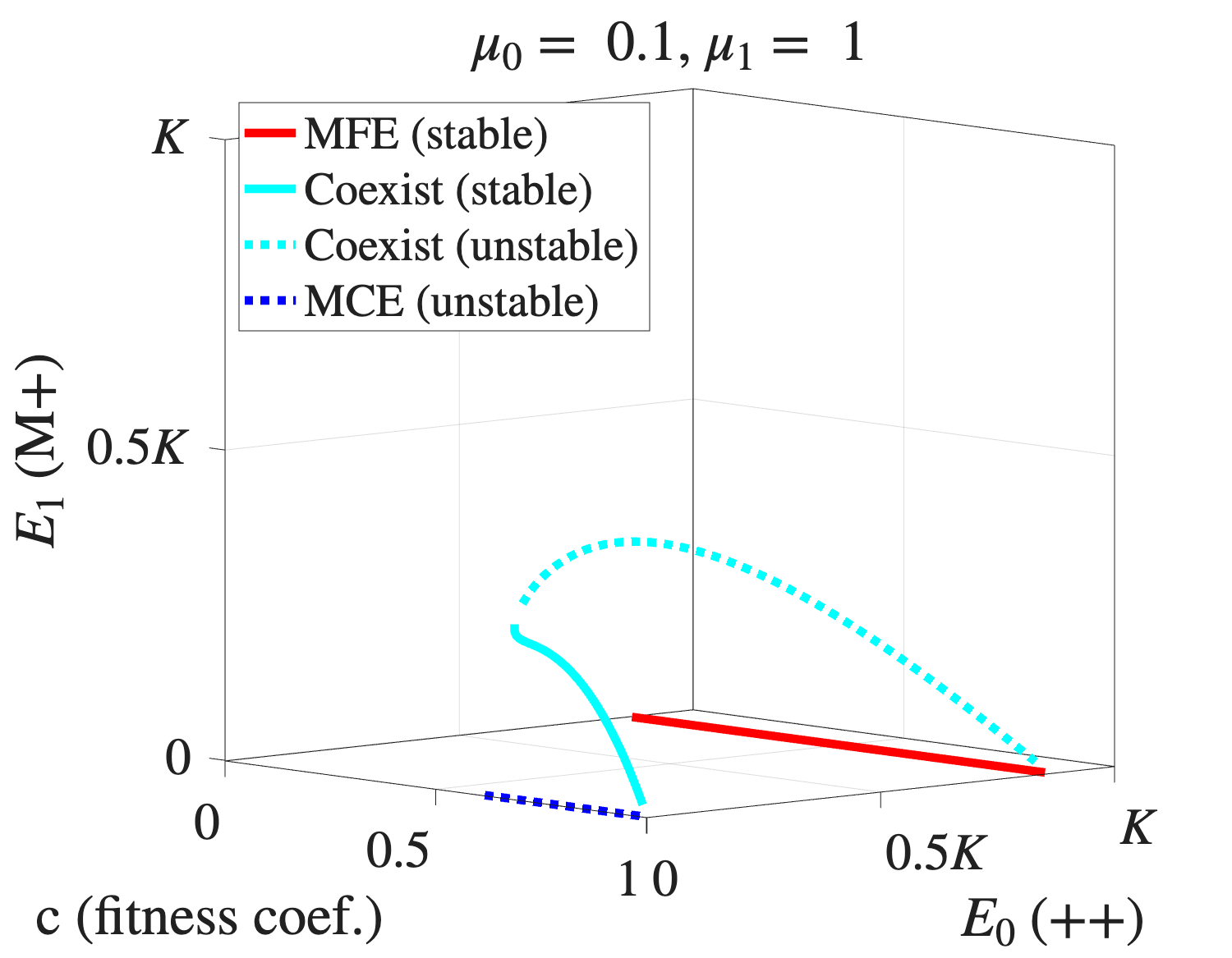}\hfill\includegraphics[width=0.48\linewidth]{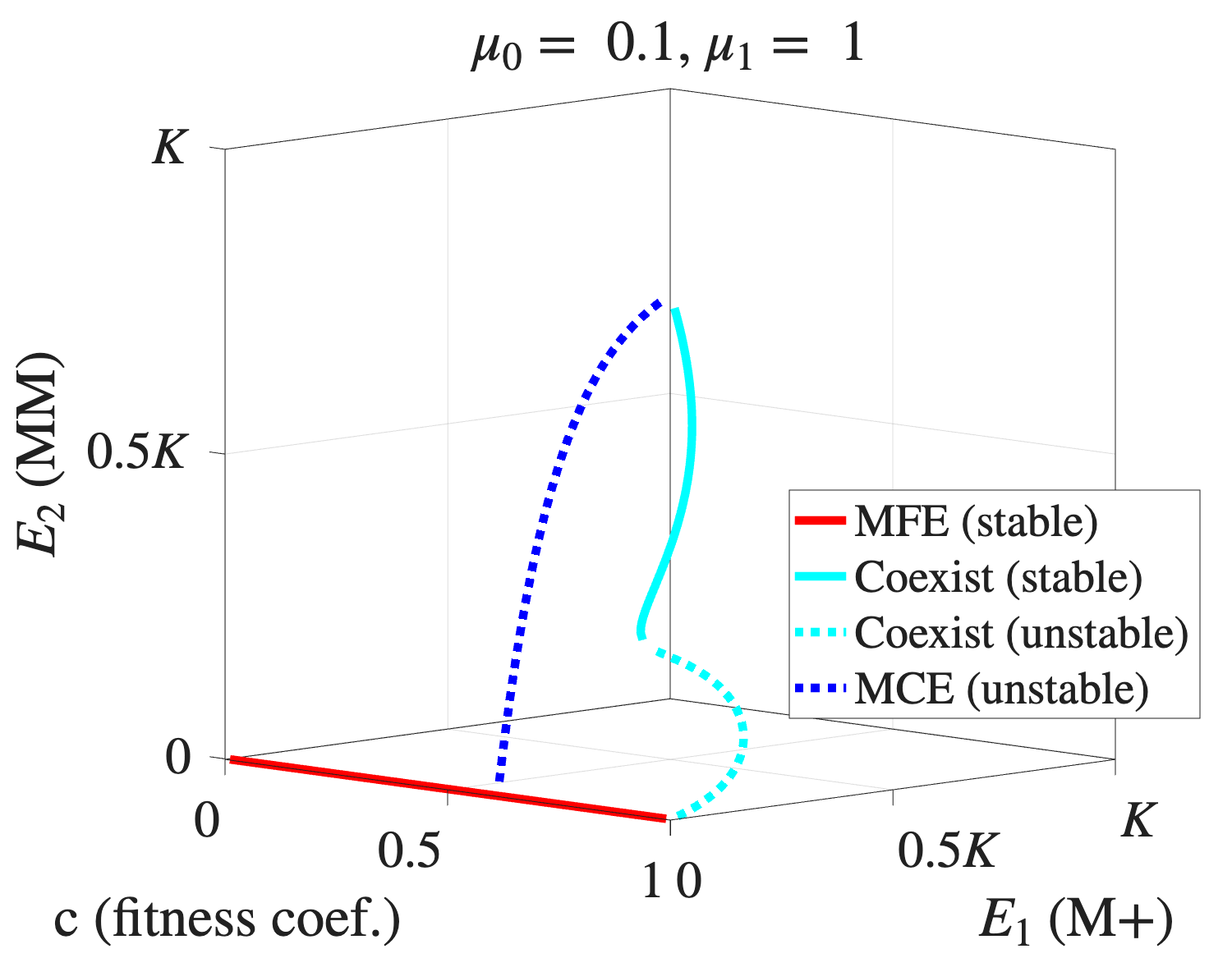}
\caption{Bifurcation diagrams with varying fitness coefficient $c$, \Mns-killing leakage parameter $\mu_0$, and rescue efficiency $\mu_1$. Solid lines indicate stable equilibria; dotted curves represent unstable branches. More parameter combinations of $\mu_0$ and $\mu_1$ are presented in \ref{sec:app_bifur}.} 
\label{fig:bifur}
\end{figure}

Building on \cref{sec:analysis}, we use numerical tools to trace the equilibria as functions of the fitness coefficient $c$. Three representative scenarios are presented for different values of the \Mns-killing leakage parameter, $\mu_0$, and rescue efficiency parameter, $\mu_1$. For each fixed $(\mu_0,\mu_1)$ pair, we solve for equilibria across a fine grid of $c$ values and classify their types and stability. This analysis reveals how imperfect killing and rescue mechanisms alter the qualitative dynamical structure of the system.

\paragraph{Perfect \M Killing and Rescue ($\mu_0 = 0, \mu_1 = 1$).} This is the baseline scenario, where maternal-effect killing is complete and zygotic rescue is perfect. Four equilibrium types appear: the MFE (wild-type only), the MCE (MM genotype only), the \Mns-carrying equilibrium (no wild-type), and the coexistence equilibrium (all three genotypes present). 

The MFE is stable and represents the field population in the absence of any \Mns-modified individuals; the MCE, by contrast, is unstable wherever it exists ($c\ge 0.62$), so complete replacement by \Mns-homozygous mosquitoes is sensitive to perturbations and cannot be stably sustained in practice. The \Mns-carrying branch instead exists over a wider fitness range ($c\ge 0.53$) and is stable throughout. This state corresponds to a more practical route to achieve population replacement, where \Mns-bearing individuals (M+ and MM genotypes) stably coexist and resist invasion by wild-type.  The unstable coexistence equilibrium acts as a separatrix, separating the stable MFE and \Mns-carrying equilibrium: trajectories initiated below this threshold converge to the MFE, while those above it converge to the \Mns-carrying equilibrium. 

This bifurcation structure reveals an invasion threshold for \Mns: successful establishment requires releasing a threshold number of \Mns-bearing individuals, above which the modified population persists. The threshold release numbers of M+ and MM individuals, given by the coexistence equilibrium, decrease in $c$ within the high fitness range ($c\ge 0.71$), permitting \M to establish from progressively smaller introductions. Once established, the population stabilizes at a wild-type-free state, comprising a mixed cohort of M+ and MM genotypes, whose composition shifts toward MM genotypes as the fitness coefficient increases.

\paragraph{Imperfect \M Rescue ($\mu_0 = 0, \mu_1 = 0.7<1$).} We now consider imperfect rescue, $\mu_1 < 1$, where only a fraction of the M+ offspring of MM mothers survive the maternal toxin. The bifurcation structure remains qualitatively similar to the perfect-rescue case: the MFE is stable, the \Mns-carrying branch exists and is stable throughout ($ 0.55 \le c\le 0.85$), and the coexistence branch continues to act as a separatrix. 
However, one critical change emerges: the MCE becomes stable in the high-fitness regime ($c \ge 0.85$ for $\mu_1=0.7$). This means that complete replacement by MM genotypes is now achievable when fitness costs are sufficiently low. 

The role of $\mu_1$ is therefore to govern the outcome of population replacement once \M invasion has occurred by shifting the genotype composition between M+ and MM: smaller $\mu_1$ (weaker rescue) enlarges the MCE stability region, since incomplete rescue penalizes M+ disproportionately relative to MM. As shown in \cref{fig:bifur_mu1}, the MCE stability range falls from $c\ge 0.85$ at $\mu_1 = 0.7$ to $c\ge 0.76$ at $\mu_1=0.5$. Notably, $\mu_1$ has little effect on the coexistence equilibrium, and thus leaves the corresponding invasion threshold largely unchanged. 

\paragraph{Imperfect \M Killing ($\mu_0 = 0.1>0, \mu_1 = 1$)} In contrast to $\mu_1$, which alters the outcome of \M invasion through the MCE, $\mu_0$ produces a qualitative shift in the bifurcation structure of the internal coexistence branches. When the \Mns-killing leakage coefficient $\mu_0>0$, the \Mns-carrying equilibrium disappears entirely: as seen in the $E_0$--$E_1$ diagram (\cref{fig:bifur}, last row), the stable \Mns-carrying equilibrium bends toward the $E_0$ direction and becomes a stable coexistence branch, meeting the unstable coexistence branch at a transcritical bifurcation point. Biologically, killing leakage maintains a sustained reservoir of wild-type offspring and prevents their complete elimination, so all three genotypes persist along this new stable branch. This outcome is not observed at $\mu_0 = 0$, where perfect killing always drives the wild-type to zero. 

The $\mu_0$ coefficient further reshapes the unstable coexistence branch, and thereby the invasion threshold for \M to establish in the population. A larger $\mu_0$ narrows the range of $c$ over which invasion is possible: the unstable coexistence branch exists for $c\ge 0.58$ at $\mu_0=0.02$, $c\ge 0.66$ at $\mu_0=0.1$, and $c\ge 0.73$ at $\mu_0=0.2$ (\cref{fig:bifur_mu0}). Within the range where invasion is possible, larger leakage also raises the invasion threshold itself: at $c=0.8$, the threshold $E_2$ value is 553 at $\mu_0=0.02$, 683 at $\mu_0=0.1$, and 935 at $\mu_0=0.2$. Overall, increased killing leakage makes population replacement progressively more difficult, requiring either higher fitness or larger introduction numbers to achieve invasion.





\subsection{Sensitivity Analysis} \label{sec:SA}
To assess the robustness of the bifurcation structure to parameter uncertainty, we perform a global sensitivity analysis using Latin Hypercube Sampling with Partial Rank Correlation Coefficient (LHS-PRCC) \cite{marino2008methodology}, drawing $N = 10{,}000$ samples from the parameter space. For most parameters, we sample uniformly over $[0.5X, 1.5X]$, where $X$ denotes the baseline value given in \cref{tab:paramter}. Applying a comparable proportional perturbation to every parameter ensures that the resulting rankings reflect the intrinsic sensitivity of the model to each parameter, rather than differences in the widths of the sampled intervals, which would otherwise confound the comparison. Three parameters are treated differently: the fitness coefficient $c$ and \Mns-specific parameters $\mu_0$ and $\mu_1$, governing \Mns-killing leakage and rescue efficiency. Their baseline values ($c=0.9, \mu_0 = 0$ and $\mu_1 = 1$) lie near, or at the boundaries of their feasible intervals, so a symmetric proportional range is not meaningful. Unless otherwise stated, for these three parameters, we instead sample over the biologically relevant ranges reported in \cref{tab:paramter}.
    
\subsubsection{Feasibility of \M Replacement and Reversibility} 
\begin{figure}[ht]
\centering
\includegraphics[width=\linewidth]{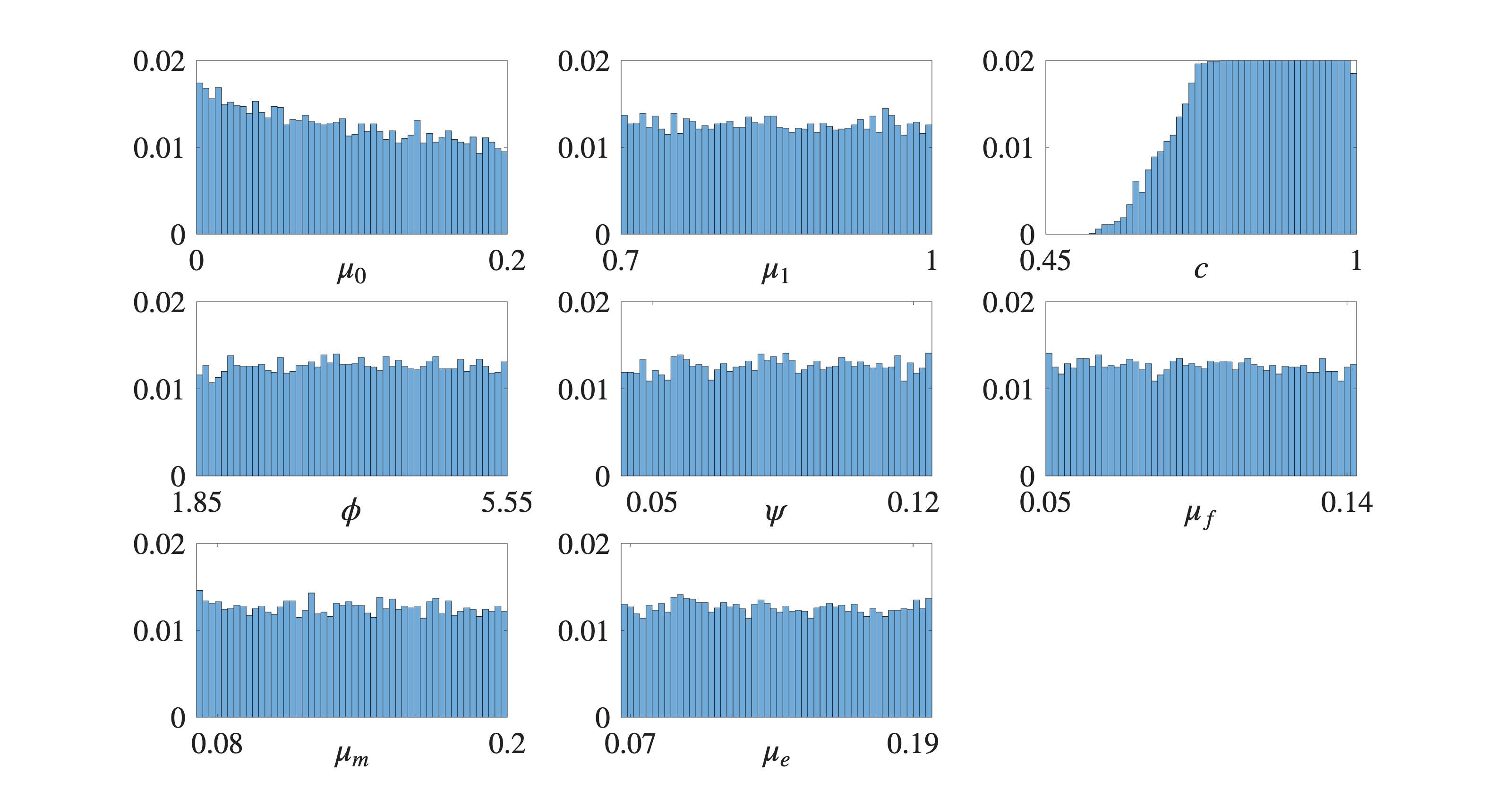}
\caption{Marginal histograms of the LHS samples for which the unstable coexistence equilibrium exists. A uniform histogram indicates that the parameter does not influence the feasibility of the bistable regime; a non-uniform histogram indicates that it does.}
\label{fig:SA_exist} 
\end{figure}

We first examine how the existence of the \M replacement threshold, represented by the unstable coexistence equilibrium, varies across parameter space. 
The same unstable coexistence equilibrium governs both the forward invasion threshold and the reversibility threshold, approached from opposite directions in initial conditions, so its existence determines the feasibility of replacement and of reversal alike. 
Because this quantity of interest (QOI) is binary (the threshold either exists or does not exist), it violates the monotonicity assumption required for PRCC; so we analyze the LHS output directly instead. For each parameter, we plot the marginal histogram of the samples for which the coexistence equilibrium exists (\cref{fig:SA_exist}): a uniform histogram indicates insensitivity, while a non-uniform histogram indicates that the parameter influences whether the bistable regime is feasible.

The fitness coefficient $c$ emerges as by far the most influential parameter. The coexistence equilibrium is essentially absent for values of $c$ near the lower bound $(c\approx 0.45)$, and its frequency rises sharply as $c$ approaches one, consistent with the \M biology: when the fitness of \Mns-carrying cohorts is too low, the drive cannot persist regardless of the initial conditions. 

The \Mns-killing leakage $\mu_0$ shows a modest effect, with the threshold occurring slightly less frequently at higher leakage values. This reflects the tendency of imperfect \Mns-killing to sustain wild-type reproduction, which lowers the likelihood that the \M element persists.

The remaining parameters, including the rescue efficiency parameter $\mu_1$ and the mosquito life-history parameters, produce approximately uniform histograms across their sampled ranges, indicating that the feasibility of the bistable regime is robust to uncertainty in these parameters.

\subsubsection{Threshold Level for Invasion and Reversal} 
\label{sec:SA_threshold}
\begin{figure}[ht]
\centering
\includegraphics[width=0.33\linewidth]{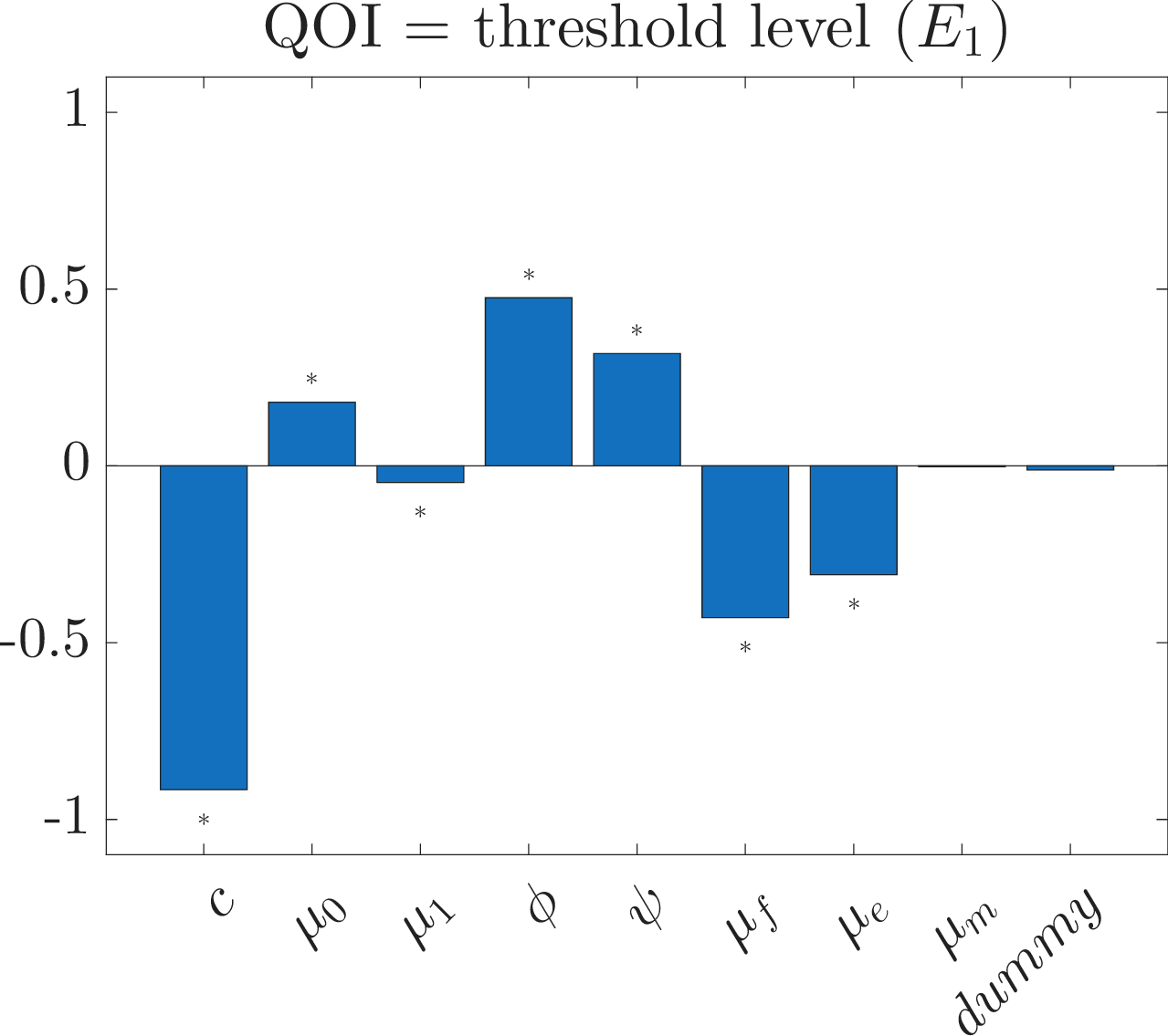}\hfill\includegraphics[width=0.33\linewidth]{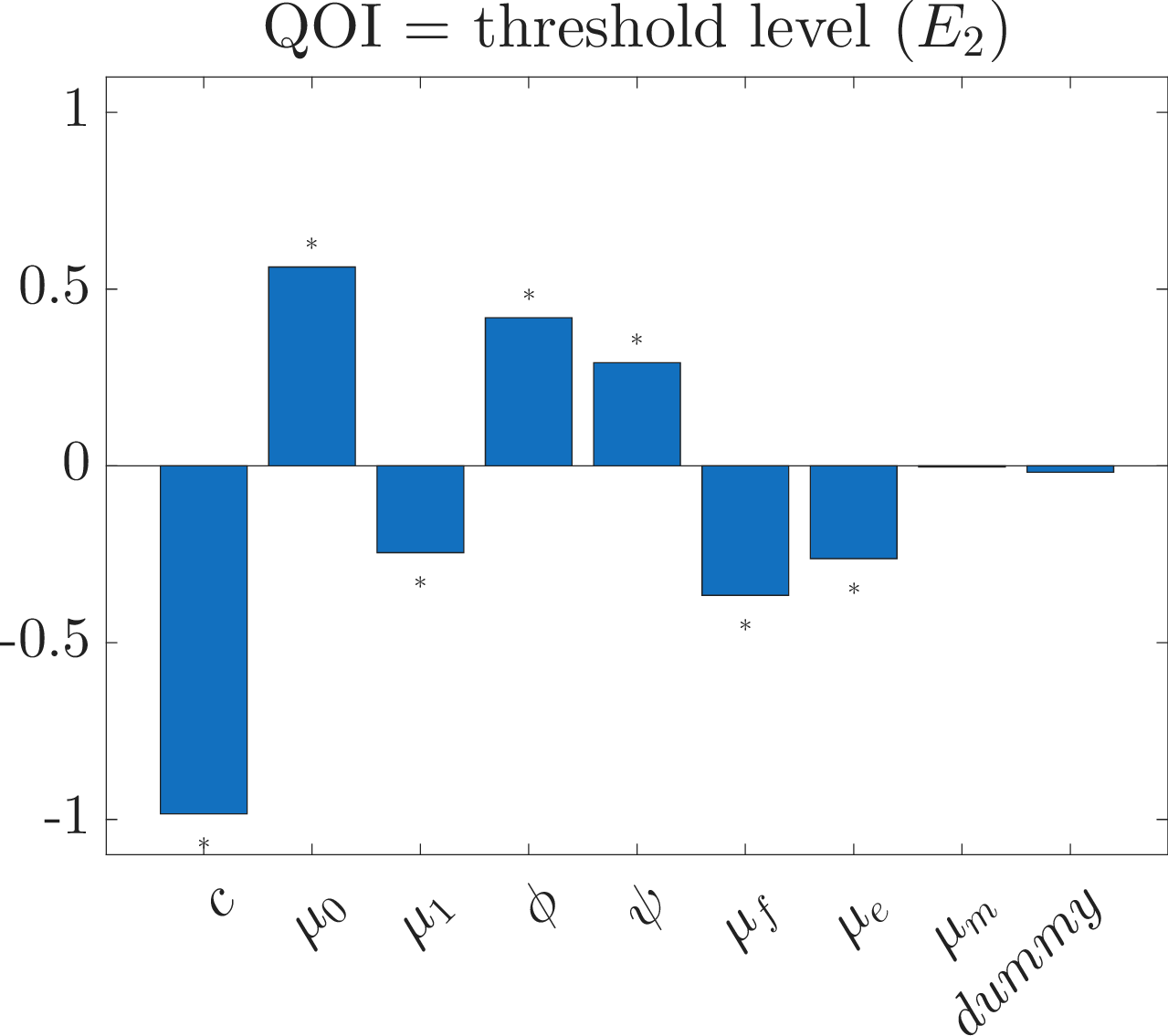}\hfill\includegraphics[width=0.33\linewidth]{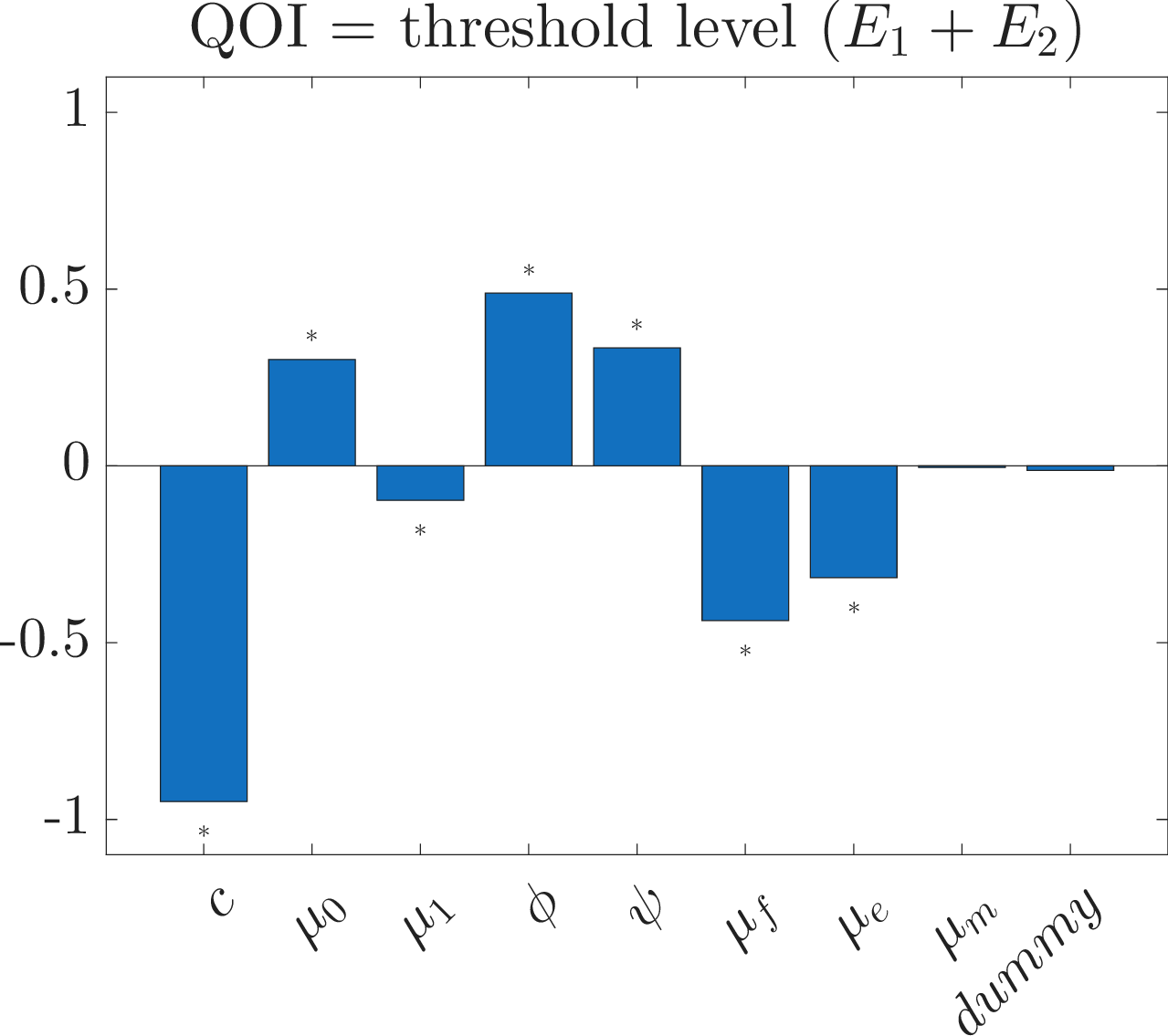}\\
\includegraphics[width=0.33\linewidth]{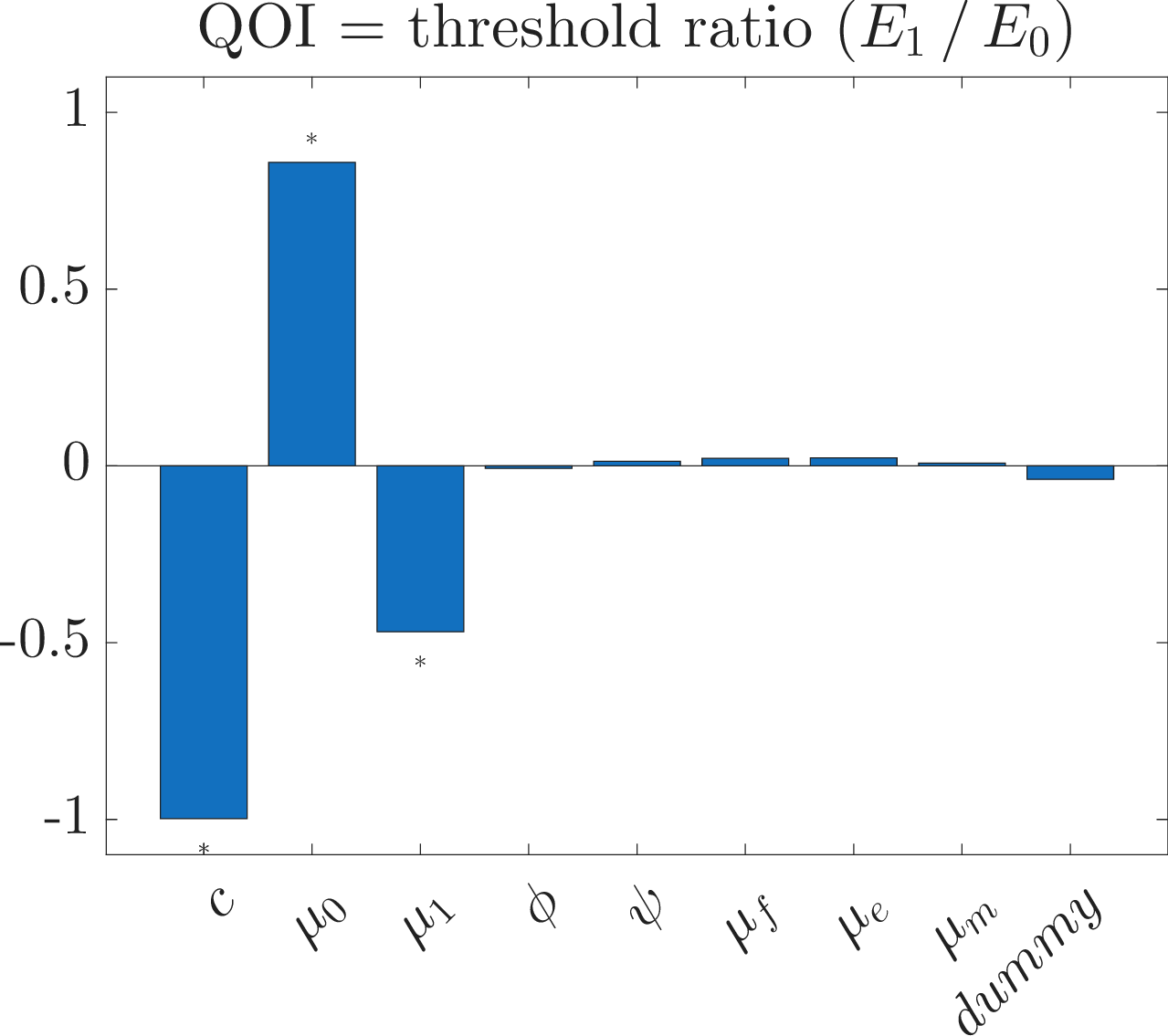}\hfill
\includegraphics[width=0.33\linewidth]{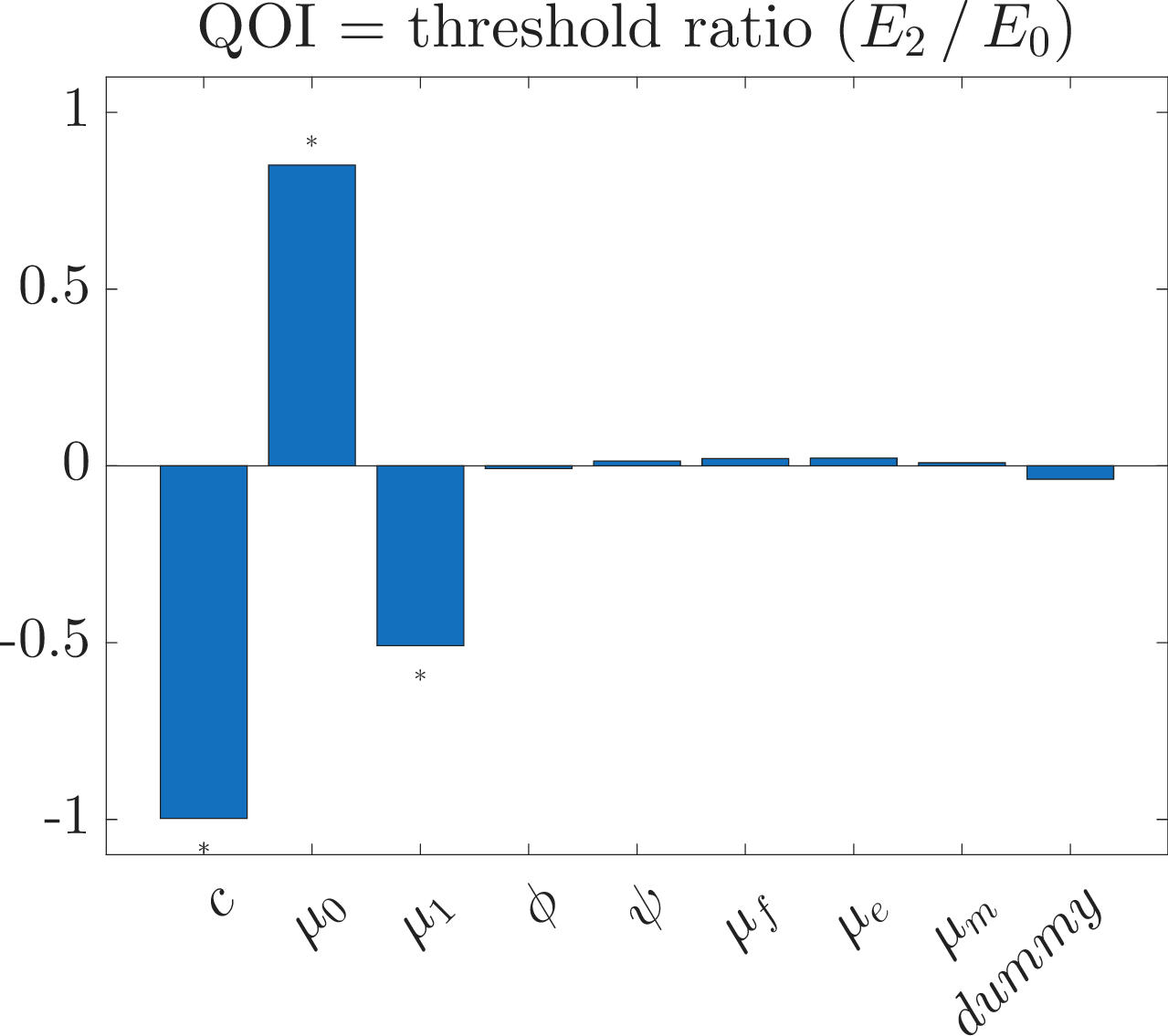}\hfill
\includegraphics[width=0.33\linewidth]{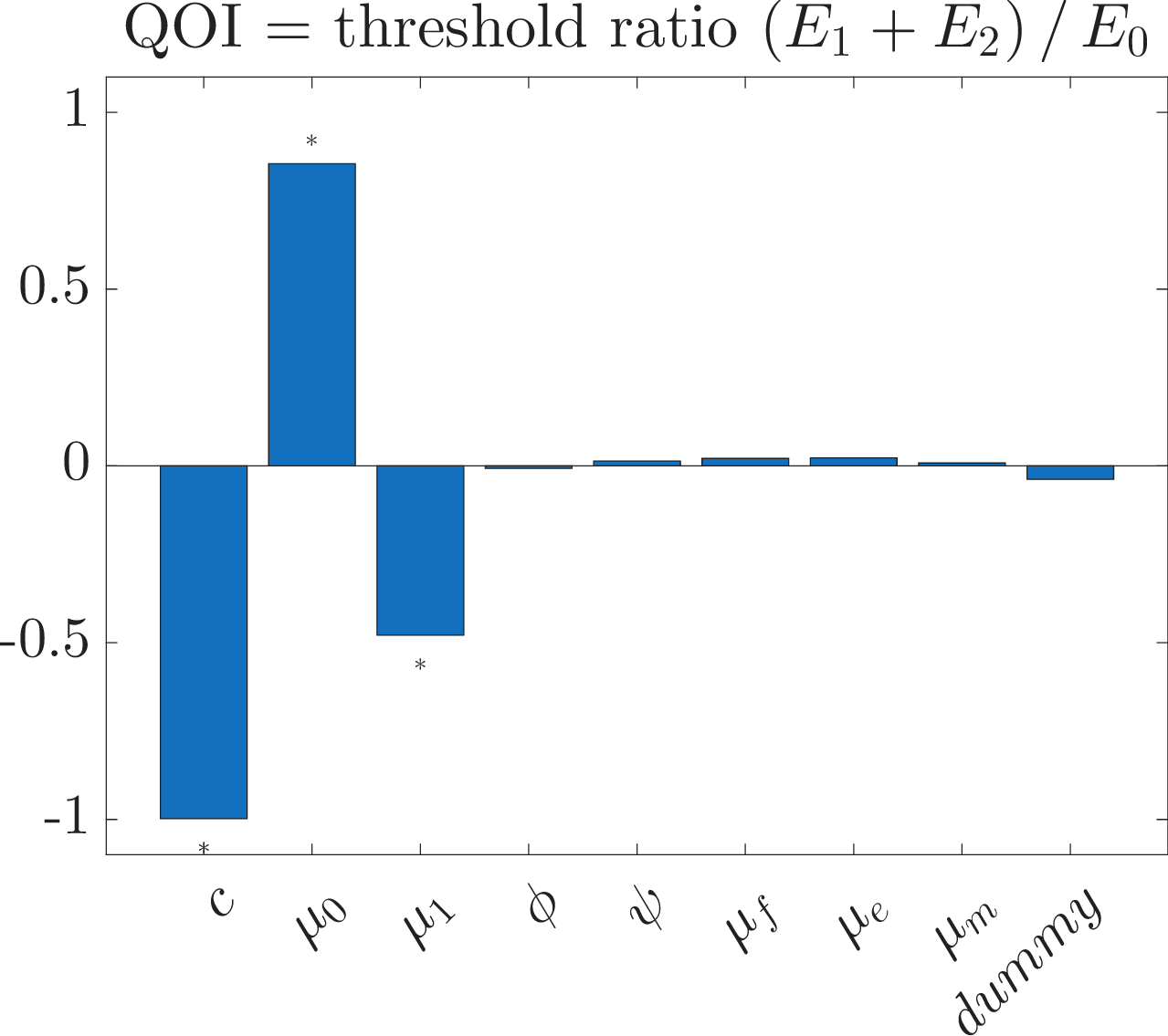}
\caption{PRCC sensitivity analysis on the threshold level for \M invasion and reversal, given by the unstable coexistence equilibrium. The QOIs are indicated in the figure titles.}
\label{fig:SA_threshold} 
\end{figure}

Conditional on the \M replacement being feasible, we assess the threshold level by examining the composition of the unstable coexistence steady state $E^* = (E_0^*, E_1^*, E_2^*)$, which determines the minimum introduction frequency required for both \M invasion and reversal. We consider six QOIs: the absolute abundances $E_1^*$, $E_2^*$, and $E_1^* + E_2^*$ (top row of \cref{fig:SA_threshold}), and the ratios relative to wild-type $E_1^*/E_0^*$, $E_2^*/E_0^*$, and $(E_1^*+E_2^*)/E_0^*$ (bottom row). To ensure that the QOI responses satisfy the monotonicity assumption required for PRCC, we restrict the sampling range for the fitness coefficient $c$ to $[0.7, 1]$ for this and all subsequent subsections; monotonicity checks are provided in \ref{sec:app_SA}. 

The fitness coefficient $c$ consistently dominates across all six metrics, yielding a strong negative PRCC of approximately $-1$ regardless of which genotype is tracked at the threshold. This confirms that a higher fitness for \Mns-carrying individuals substantially reduces the introduction frequency required for invasion or reversal.

For the absolute abundance metrics (top row), the demographic life-history parameters show a moderate secondary influence. Higher fecundity $\phi$ and maturation rate $\psi$ are positively correlated with the threshold level, while higher female and juvenile mortality rates $\mu_f$ and $\mu_e$ are negatively correlated; together, these reflect the fact that conditions favoring overall mosquito reproduction scale up all genotype abundances proportionally, so the absolute threshold level rises accordingly. The male mortality rate $\mu_m$ does not reach significance in any abundance metric.

In contrast, the demographic parameters lose their influence entirely when the threshold is expressed as a ratio relative to the wild-type abundance (bottom row). These ratios are governed exclusively by the \Mns-specific parameters $c$, $\mu_0$, and $\mu_1$, indicating that the population replacement process is intrinsically determined by the \M mechanism rather than by the background life-history of the mosquito.

Among the \Mns-specific parameters, $\mu_0$ is the second most influential after $c$, while $\mu_1$ shows a moderate effect. In terms of absolute abundance (top row), both exert a larger influence on the MM homozygous genotype ($E_2^*$) than on the M+ heterozygous genotype ($E_1^*$), consistent with the fact that an MM offspring requires an \Mns-allele from both parents and therefore arises only from crosses between two \Mns-carrying individuals, so that any effect on the \Mns-carrying cohort compounds when propagating to the MM class. Mechanistically, higher killing leakage $\mu_0$ raises the threshold by permitting greater wild-type reproduction, while higher rescue efficiency $\mu_1$ lowers the threshold by improving the survival of \Mns-carrying M$+$ offspring that would otherwise be lost to the maternal toxin.

\subsubsection{Stable Coverage of \Mns}\label{sec:SA_coverage}
\begin{figure}[ht]
\centering
\includegraphics[width=0.24\linewidth]{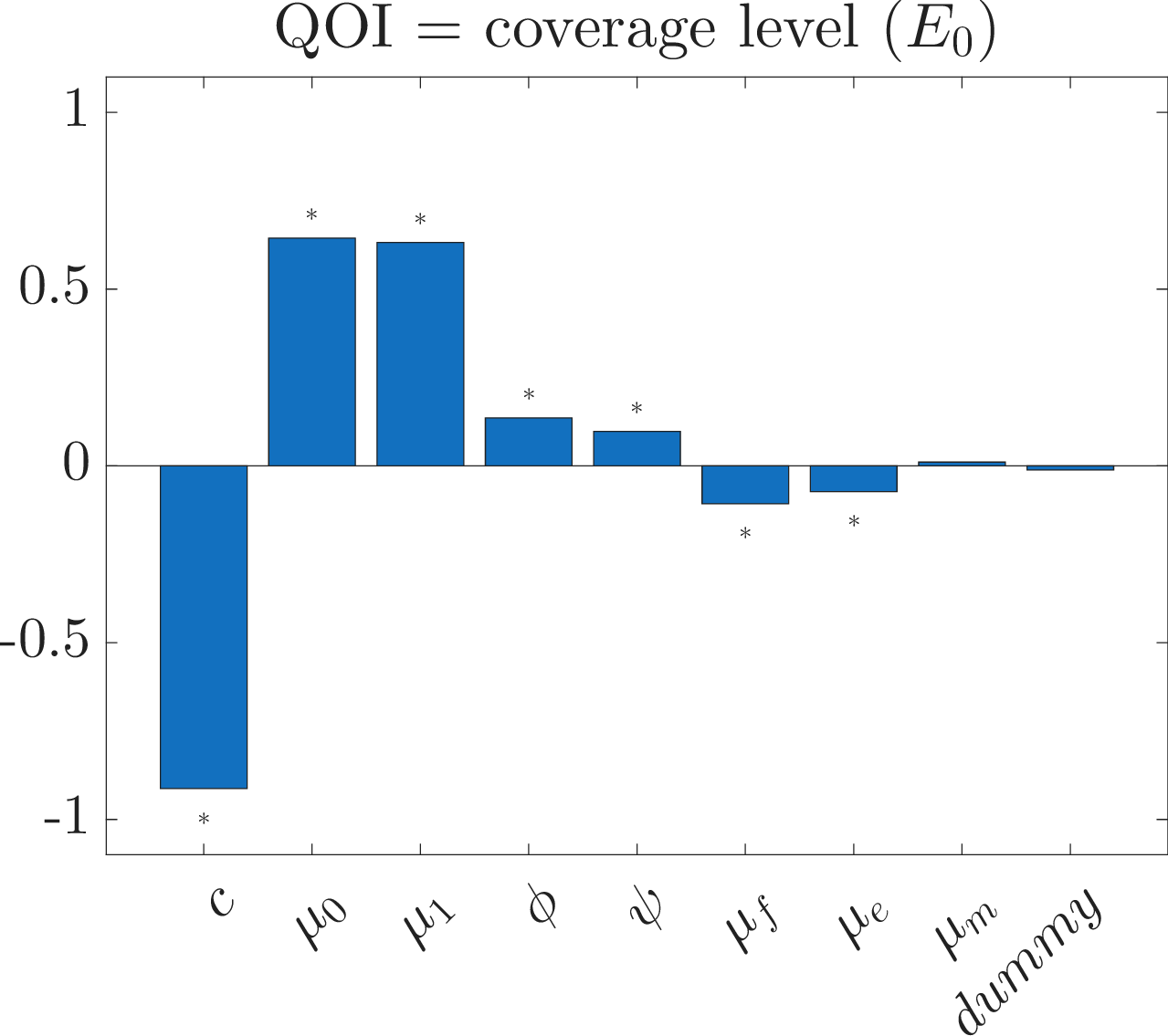}\hfill
\includegraphics[width=0.24\linewidth]{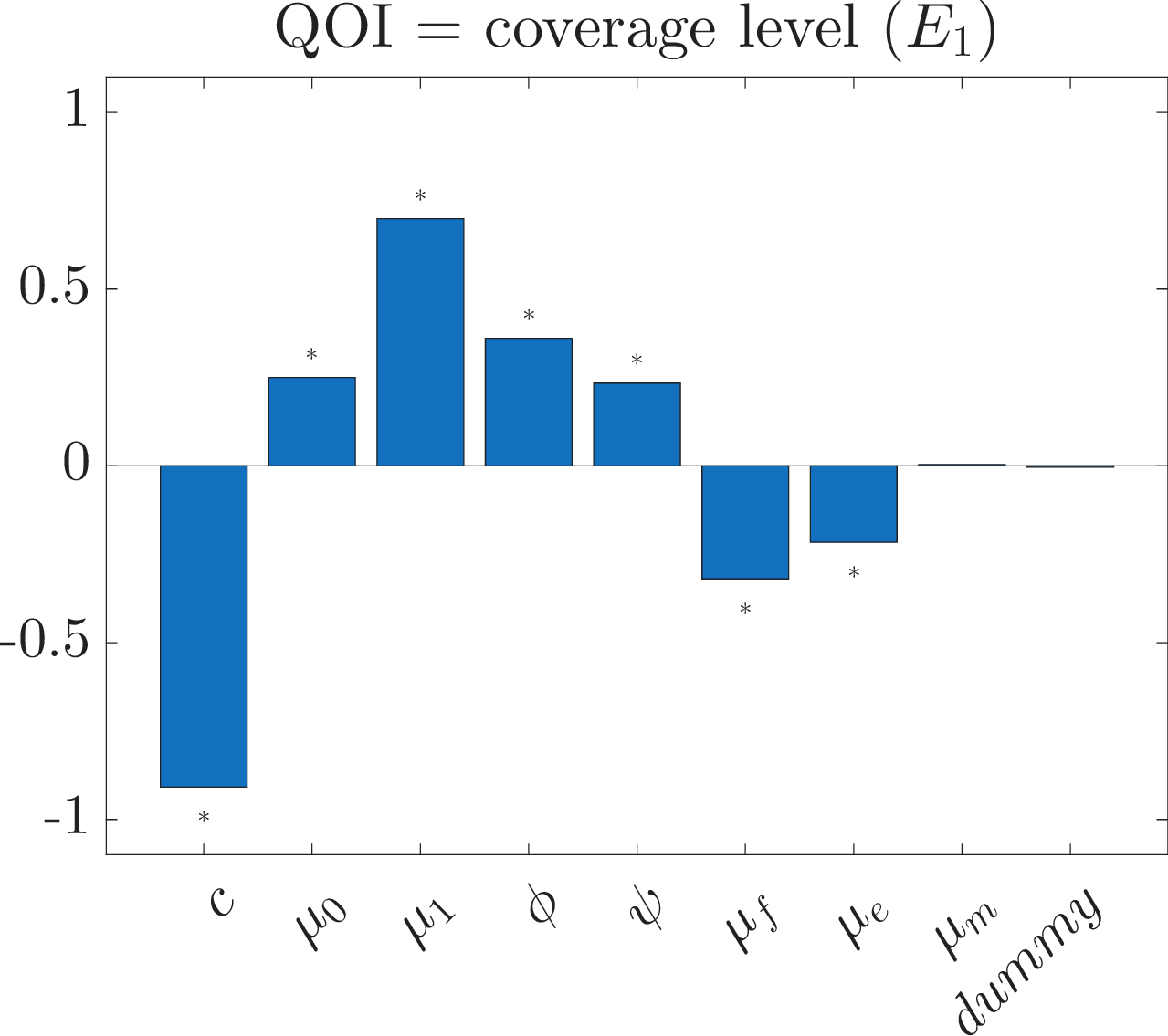}\hfill\includegraphics[width=0.24\linewidth]{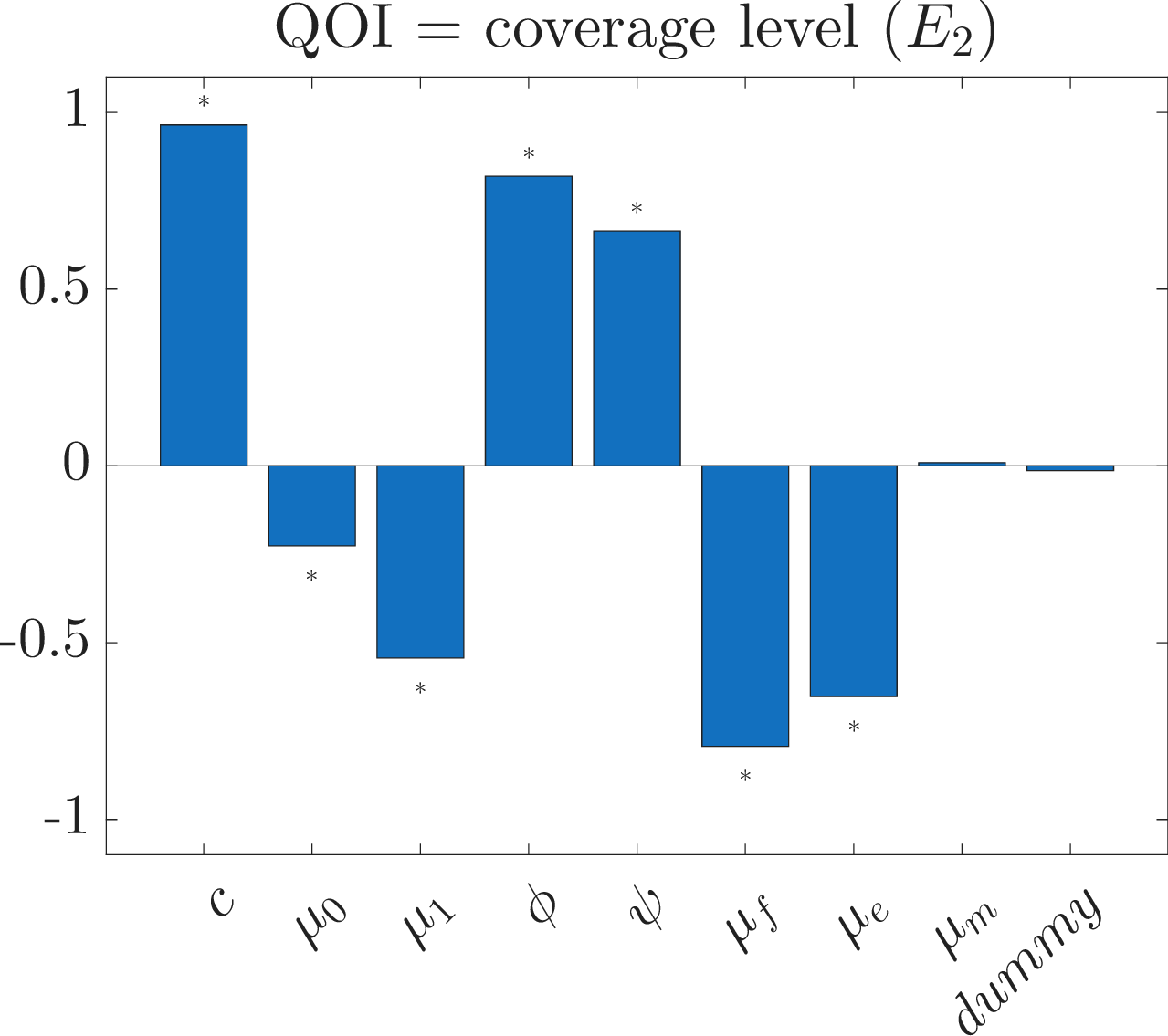}\hfill\includegraphics[width=0.24\linewidth]{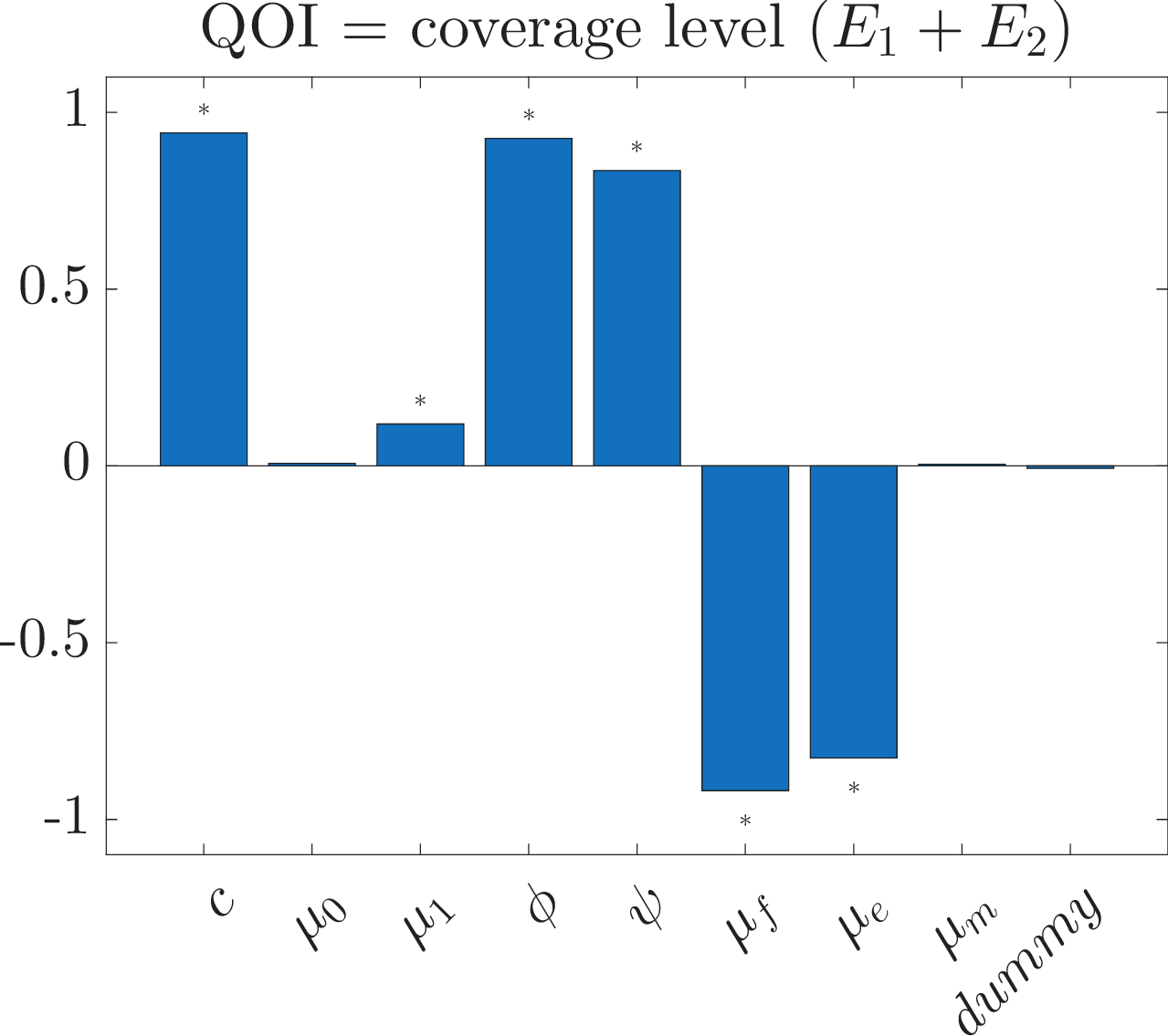}\\
\includegraphics[width=0.24\linewidth]{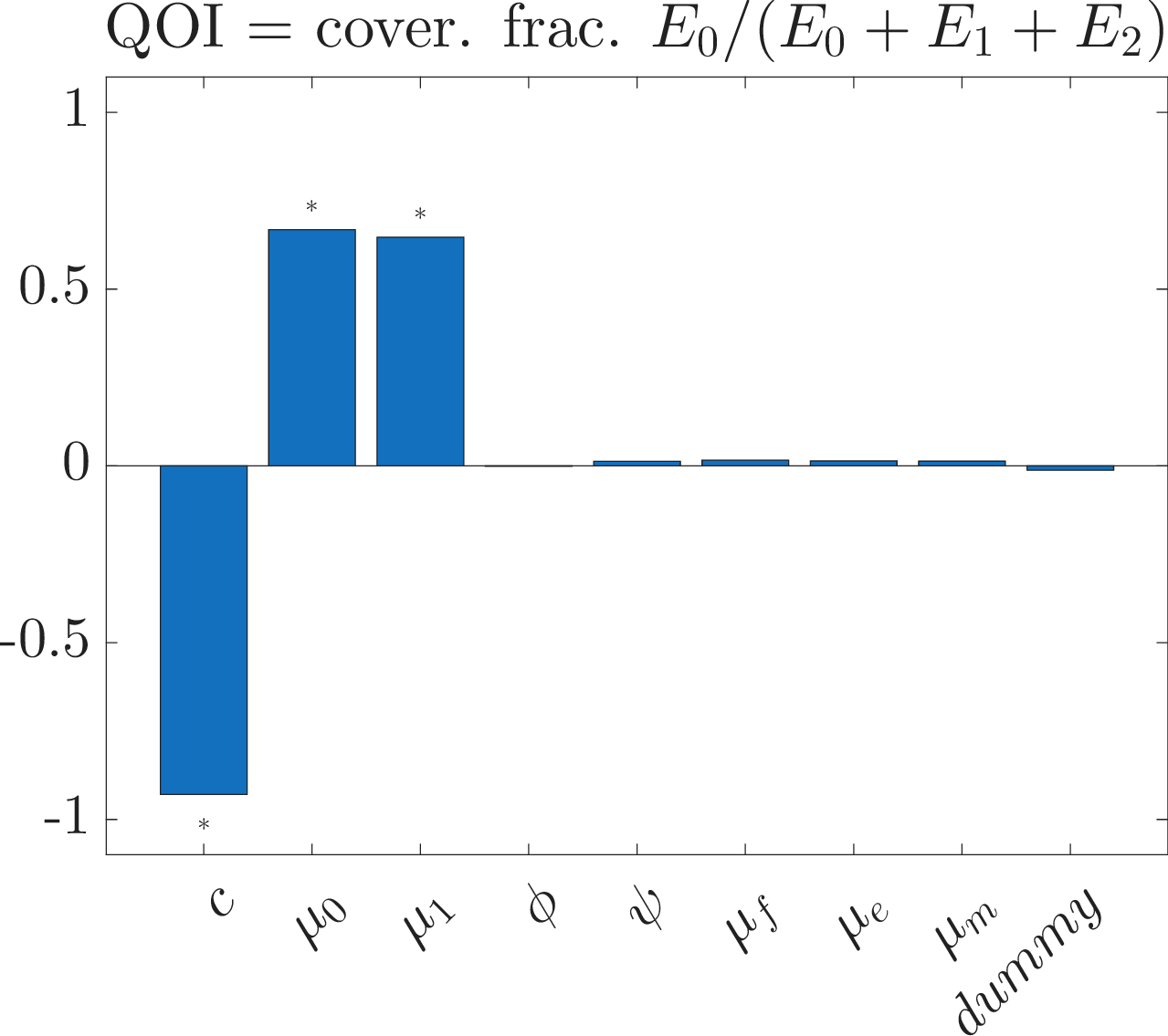}\hfill
\includegraphics[width=0.24\linewidth]{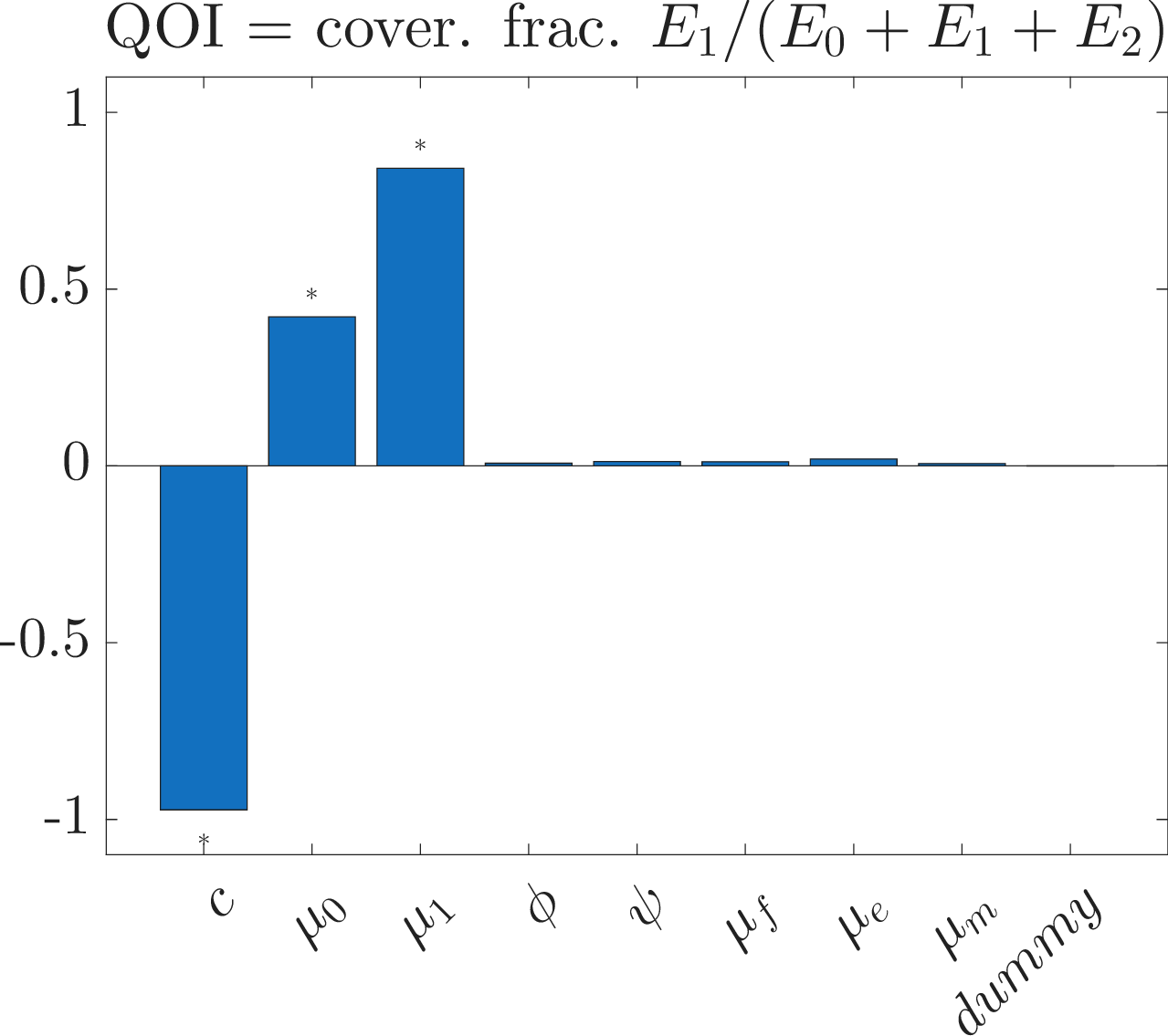}\hfill
\includegraphics[width=0.24\linewidth]{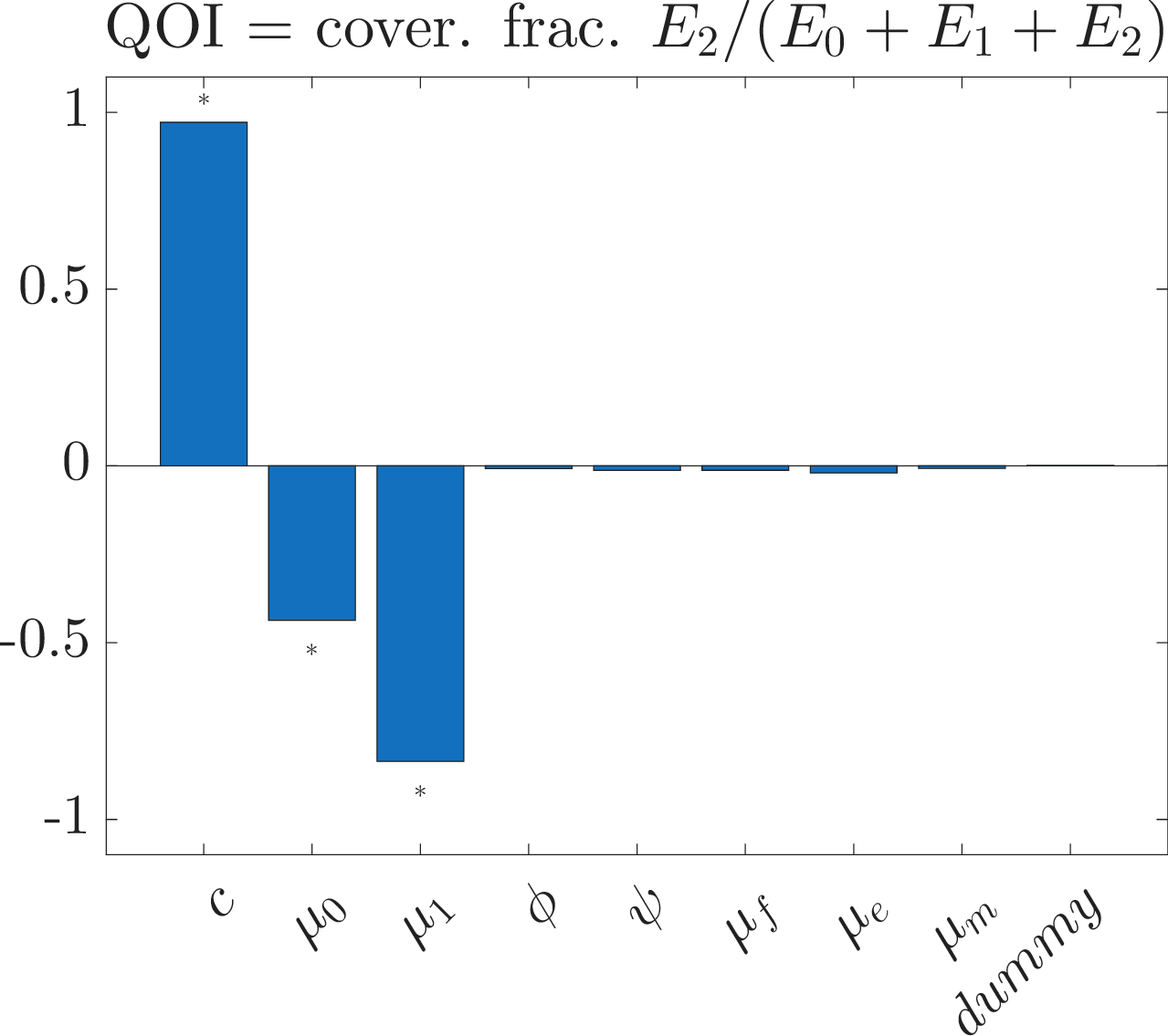}\hfill
\includegraphics[width=0.26\linewidth]{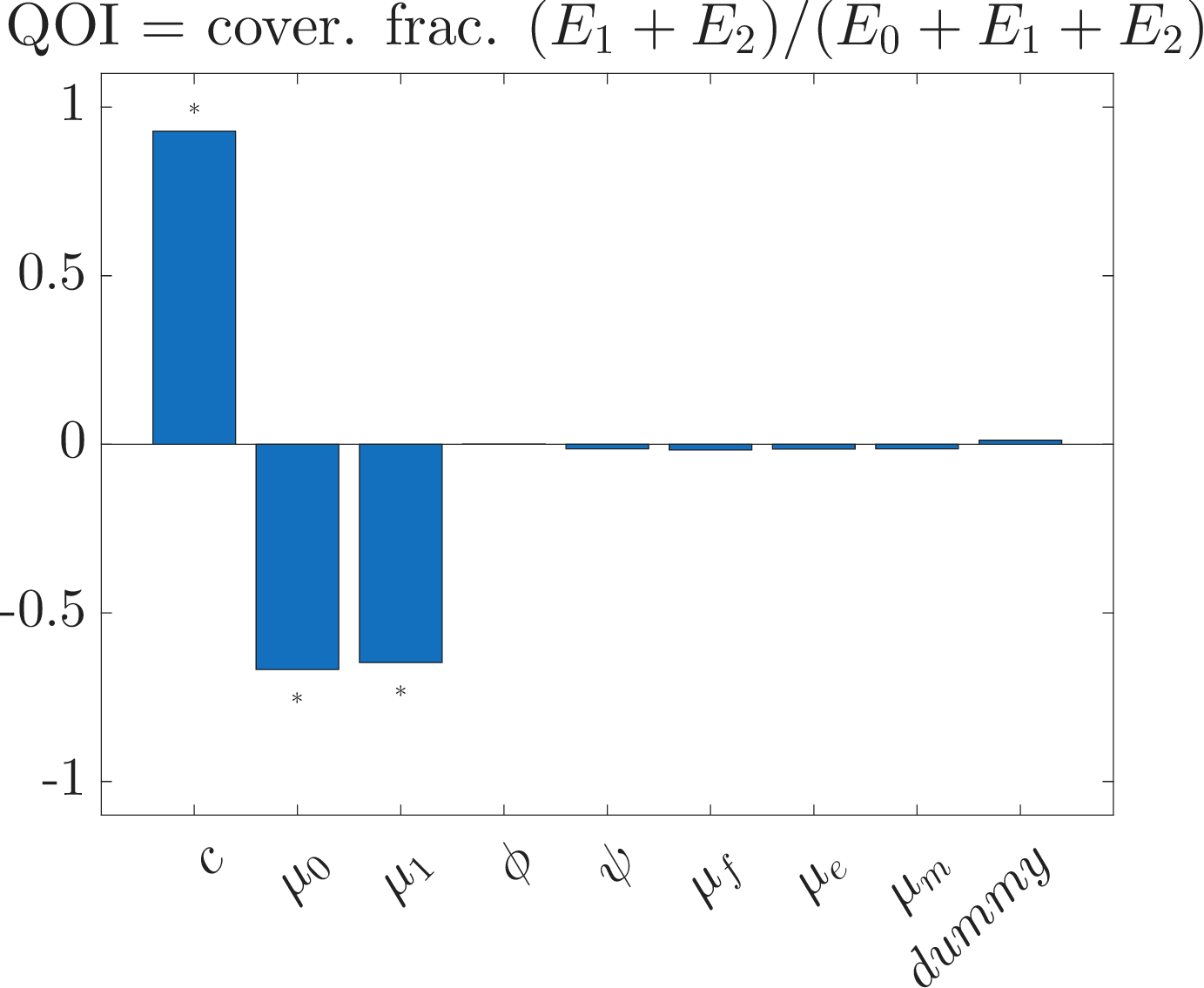}
\caption{PRCC sensitivity analysis on the stable \M coverage at the post-invasion equilibrium. This equilibrium may be the \Mns-carrying equilibrium, the MCE, or the stable coexistence equilibrium, depending on $\mu_0$ and $\mu_1$ (see \cref{sec:bifur}). The QOIs are indicated in the figure titles. \textit{cover. frac.} = coverage fraction.}
\label{fig:SA_coverage}
\end{figure}

Conditional on the population being above the invasion threshold, we examine the stable prevalence of \Mns. Depending on the values of $\mu_0$ and $\mu_1$, this stable coverage state may be the \Mns-carrying equilibrium, the MCE, or the stable coexistence equilibrium, as characterized in \cref{sec:bifur}. We consider eight QOIs: the absolute abundances $E_0^*$, $E_1^*$, $E_2^*$, and $E_1^*+E_2^*$ (top row of \cref{fig:SA_coverage}), and the corresponding coverage fractions $E_0/(E_0+E_1+E_2)$, $E_1/(E_0+E_1+E_2)$, $E_2/(E_0+E_1+E_2)$, and $(E_1+E_2)/(E_0+E_1+E_2)$ (bottom row).

The role of the demographic life-history parameters mirrors the pattern observed at the invasion threshold (\cref{sec:SA_threshold}): they exert a moderate influence on the absolute abundance metrics but become entirely non-significant when the QOI is expressed as a fraction, and $\mu_m$ does not reach significance under any metric. This reinforces the conclusion that, within the present model assumptions, the proportional composition of the stable population is governed by the \M mechanism alone, independently of background mosquito demography.

For the coverage fractions (bottom row), only the \Mns-specific parameters $c$, $\mu_0$, and $\mu_1$ are significant, with $c$ remaining the most influential across all fraction metrics. A higher fitness coefficient shifts the stable composition toward the MM homozygous genotype, increasing the $E_2$ fraction while decreasing both the $E_0$ and $E_1$ fractions.

Higher killing leakage $\mu_0$ acts in the opposite direction by permitting greater wild-type reproduction at the stable state, raising the $E_0$ fraction and, consequently, increasing heterozygous $E_1$ production while suppressing the MM fraction $E_2$. 

Higher rescue efficiency $\mu_1$ saves heterozygous M$+$ offspring that would otherwise be lost to the maternal toxin, increasing the $E_1$ fraction; however, these additional heterozygotes contribute more towards wild-type production than toward MM production, so the net effect is a reduction in the $E_2$ fraction.

\subsection{Release Strategies} \label{sec:release}
We further numerically explore various field release strategies to inform the principle for efficient \Mns-based population replacement and reversal. 

We focus on releasing homozygous \M males (MM) only. Male mosquitoes do not bite and therefore pose no direct pathogen-transmission risk, while MM homozygotes maximize the \M drive signal delivered to the next generation by ensuring that every mating event with a wild-type female produces heterozygous offspring. Alternative release compositions are also considered, and the implementation generalizes straightforwardly by incrementing the corresponding compartments.

Releases are implemented as impulsive perturbations to the ODE system: at each scheduled release time, the compartment $M_2$ is incremented by a prescribed batch size and integration resumes from the updated state, mimicking the discrete, periodic nature of field deployment without introducing continuous source terms.

A repetitive release program is characterized by two parameters. The batch size $\Delta M_2$ is the number of MM males released per event, expressed relative to the wild-type male abundance at the \Mns-free equilibrium, $M_0^{\mathrm{MFE}}$. The release frequency, in batches per week, determines the inter-release interval $\tau$ in days: a frequency of 1 corresponds to weekly releases ($\tau = 7$), while a frequency of 7 corresponds to daily releases ($\tau = 1$). For cross-strategy comparisons, we also report the weekly deployment, defined as $\Delta M_2 \times (\text{release frequency})$, which measures the total number of \M males deployed per week and serves as a proxy for the practical constraint imposed by factory output or field logistics.

For each release configuration, we determine the minimum number of batches $n^*$ needed to satisfy the establishment criterion: the \M prevalence among adult females, $(F_1+F_2)/(F_0+F_1+F_2)$, must reach 90\% within a time horizon of $T = 365$ days. To avoid discontinuities arising from integer-only batch counts, $n^*$ is treated as a real number: the first $\lfloor n^* \rfloor$ releases are full batches of size $\Delta M_2$, and the fractional remainder determines the size of a final partial batch. The minimum is identified via a bisection search. We track three QOIs: (i) the minimum batch count $n^*$; (ii) the total release cost $n^* \cdot \Delta M_2$, a proxy for cumulative resource expenditure expressed in units of $M_0^{\mathrm{MFE}}$; and (iii) the active release phase, the time elapsed from the first to the last batch, beyond which the population evolves without further intervention until settling at the replacement equilibrium.

\subsubsection{Forward Invasion}
\begin{figure}[ht]
\centering
\includegraphics[width=0.49\linewidth]{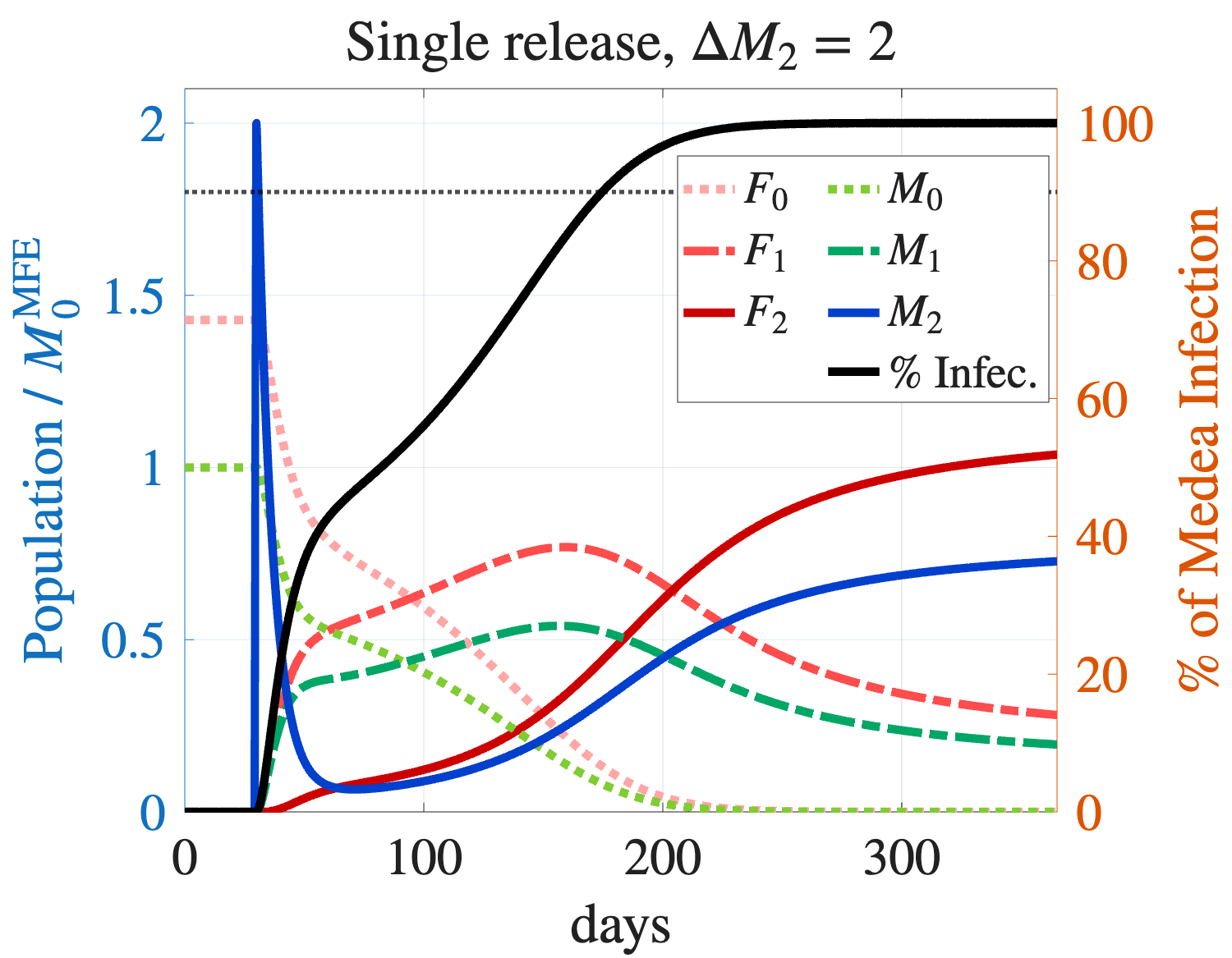}\hfill\includegraphics[width=0.49\linewidth]{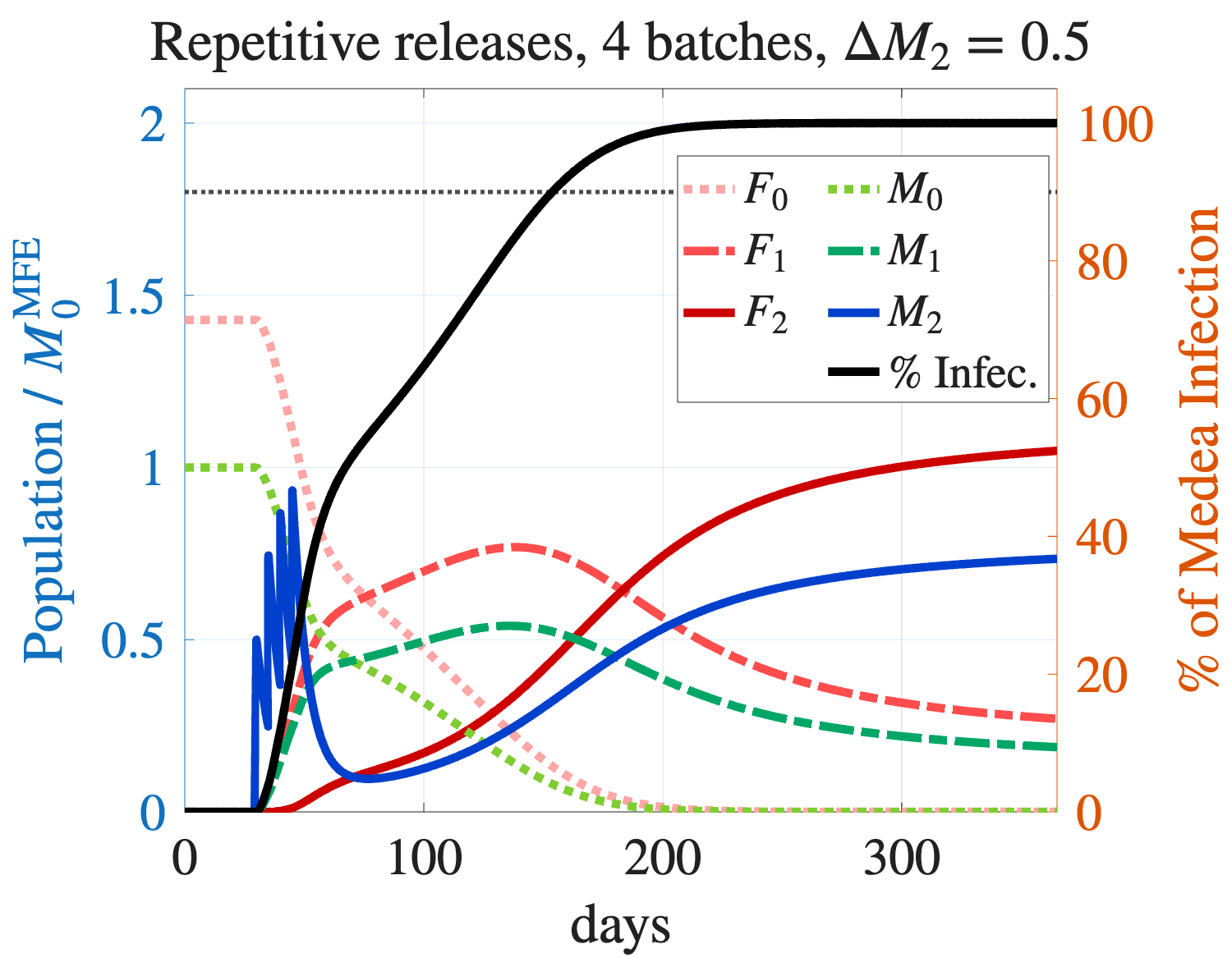}
\caption{Single vs.\ repetitive releases of equal total cost ($\Delta M_2^{\mathrm{total}} = 2$). \textbf{Left:} single release, achieving 90\% \M prevalence at $t = 175$ days. \textbf{Right:} four equal batches at five-day intervals, achieving the same threshold at $t = 154$ days, 21 days earlier.}
\label{fig:releases}
\end{figure}

We first consider forward invasion: starting from the MFE, batched releases are used to drive the population past the invasion threshold and establish stable \M coverage.
\paragraph{Trade-offs between batch size, release frequency, and total cost}
\Cref{fig:releases} illustrates the potential benefit of batch splitting under a fixed total release budget. A single release of $\Delta M_2 = 2$ achieves 90\% \M prevalence at $t = 175$ days, whereas distributing the same total into four equal batches of $\Delta M_2 = 0.5$ at five-day intervals achieves the same threshold at $t = 154$ days, 21 days earlier and at identical total cost. The acceleration arises because a large single bolus temporarily inflates total male abundance, increasing density-dependent competition among the immediate offspring and diluting the per-capita mating efficiency of released MM males, whereas staged releases allow each cohort to propagate the \M element through natural mating before the next batch augments the pool.

\begin{figure}[ht]
\centering
\includegraphics[width=0.45\linewidth]{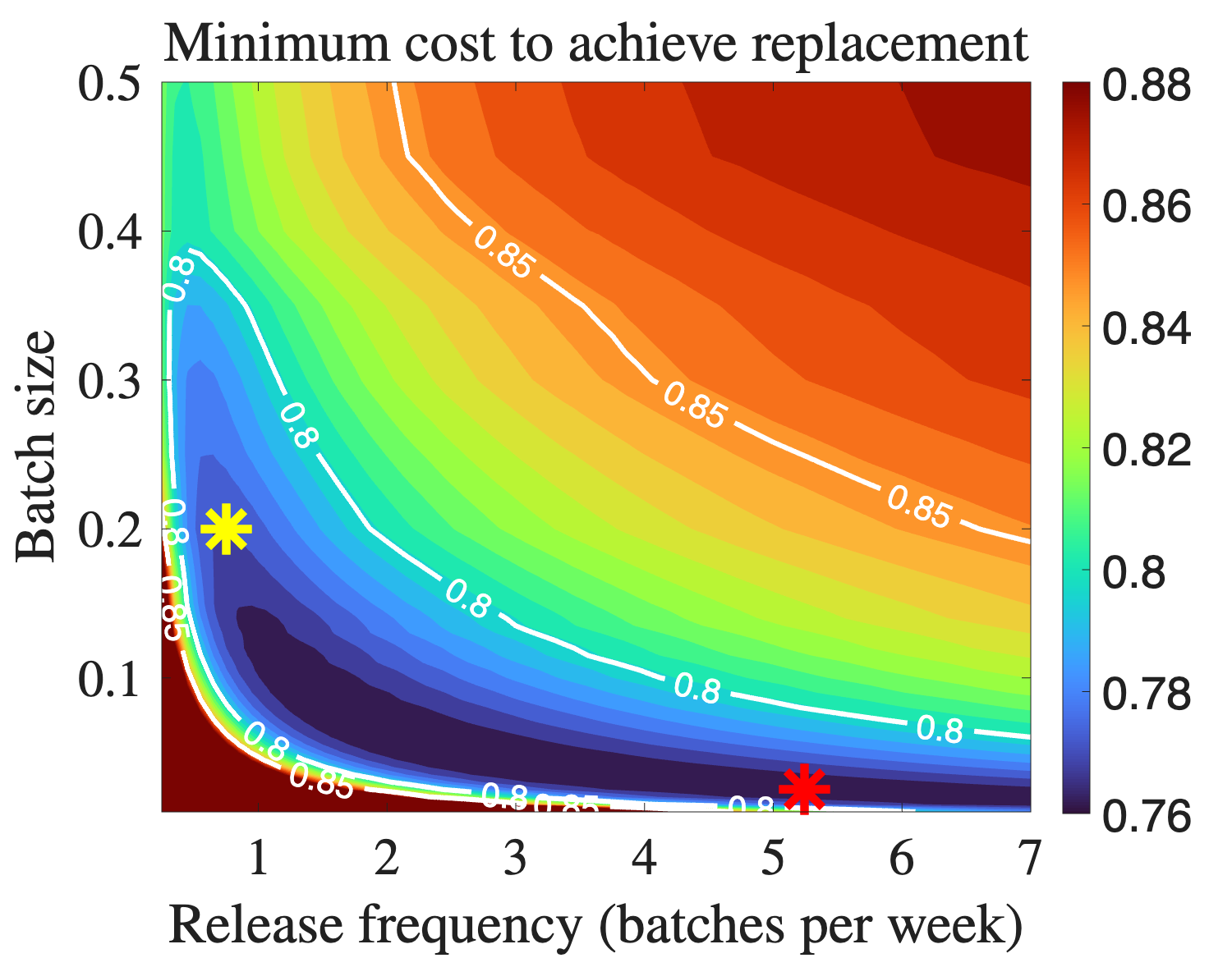}
\includegraphics[width=0.45\linewidth]{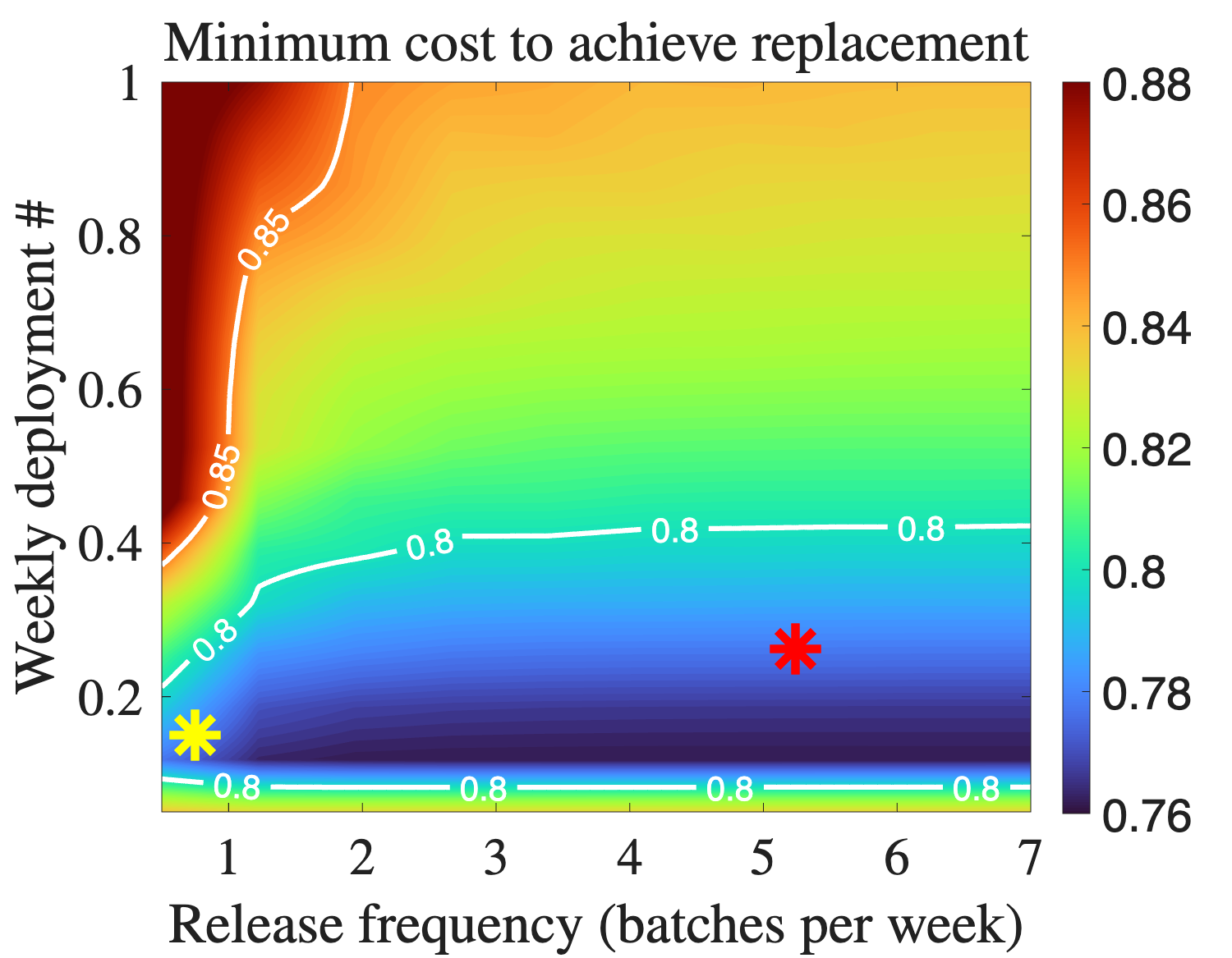}
\includegraphics[width=0.45\linewidth]{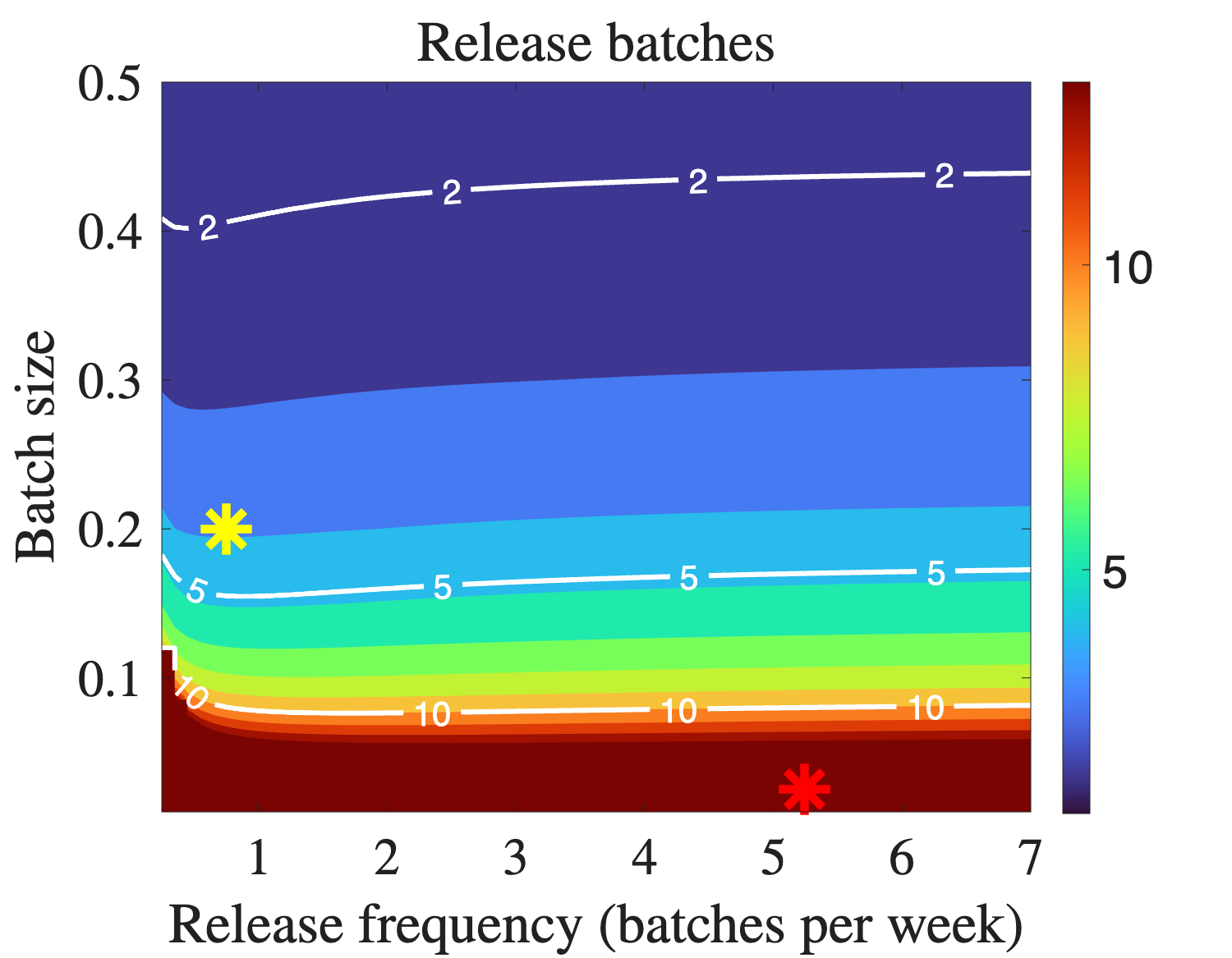}
\includegraphics[width=0.45\linewidth]{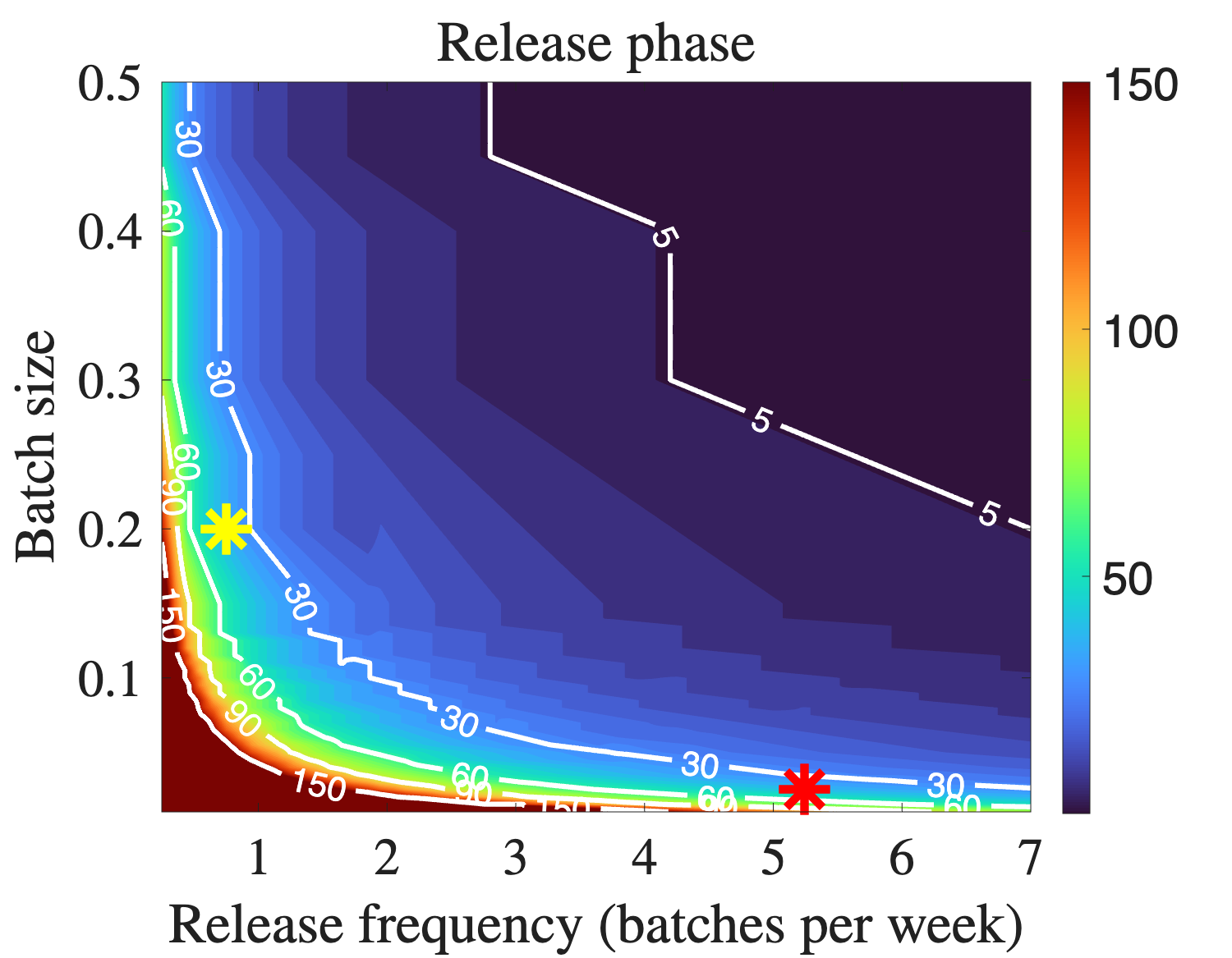}
\caption{Release optimization heatmaps. (A)~Total release cost $n^* \cdot \Delta M_2$ as a function of batch size and release frequency. (B)~The same cost plotted against weekly deployment and release frequency. (C)~Minimum batch count $n^*$. (D)~Active release phase, in days from first to last batch. Stars mark representative operating points: yellow star ($\Delta M_2 \approx 0.2$, $\approx 0.85$ batches/week) and red star ($\Delta M_2 \approx 0.05$, $\approx 3.5$ batches/week). All panels use the 90\% establishment criterion within $T = 365$ days.}
\label{fig:heatmaps}
\end{figure}

\Cref{fig:heatmaps} surveys a broad range of release configurations: batch sizes $\Delta M_2 \in [0.05, 0.5]$ and release frequencies from roughly biweekly to daily. 
For small batch sizes ($\Delta M_2 \lesssim 0.1$), frequent releases minimize total cost, with a minimum near $0.76$ attained at approximately $5.24$ batches per week (red star, $\Delta M_2 \approx 0.025$), requiring around $30$ batches over an active release phase of approximately $41$ days. 
For moderate batch sizes ($\Delta M_2 \approx 0.2$), considerably less frequent releases suffice: approximately $0.75$ batches per week (yellow star), at comparable total cost ($\approx 0.77$) and a similarly short active release phase of roughly $37$ days. Across both regimes, the optimal total cost and weekly deployment are similar, but larger batches require fewer and less frequent release events, which may be a logistical advantage in field operations. For large batch sizes ($\Delta M_2 \gtrsim 0.4$), only one or two batches suffice to cross the invasion threshold regardless of release frequency (\cref{fig:heatmaps}, panel~C), and wider inter-release spacing is more efficient (panel~A), since it allows the first cohort to propagate the \M element before additional males are introduced. The total cost in this regime nonetheless rises relative to the small-to-moderate range.

Across all batch sizes, shortening the active release phase requires increasing the release frequency (panel~D), which in turn tends to raise total cost, particularly for moderate-to-large batches (panel~A); this presents an explicit trade-off between deployment duration and resource expenditure.

The reparameterized view in panel~B clarifies how the dominant bottleneck shifts with weekly capacity. When weekly deployment is low ($\lesssim 0.1$), total cost and active release phase are governed almost entirely by deployment capacity: for a fixed weekly output at this level, adjusting the within-week release schedule yields negligible benefit, and the principal lever is simply to increase total weekly output. When weekly deployment is higher ($\gtrsim 0.4$), scheduling becomes the primary efficiency lever: for a fixed weekly output, distributing it into more frequent and smaller batches reduces total cost relative to fewer, larger deployments.

\begin{figure}[ht]
\centering
\includegraphics[width=0.5\linewidth]{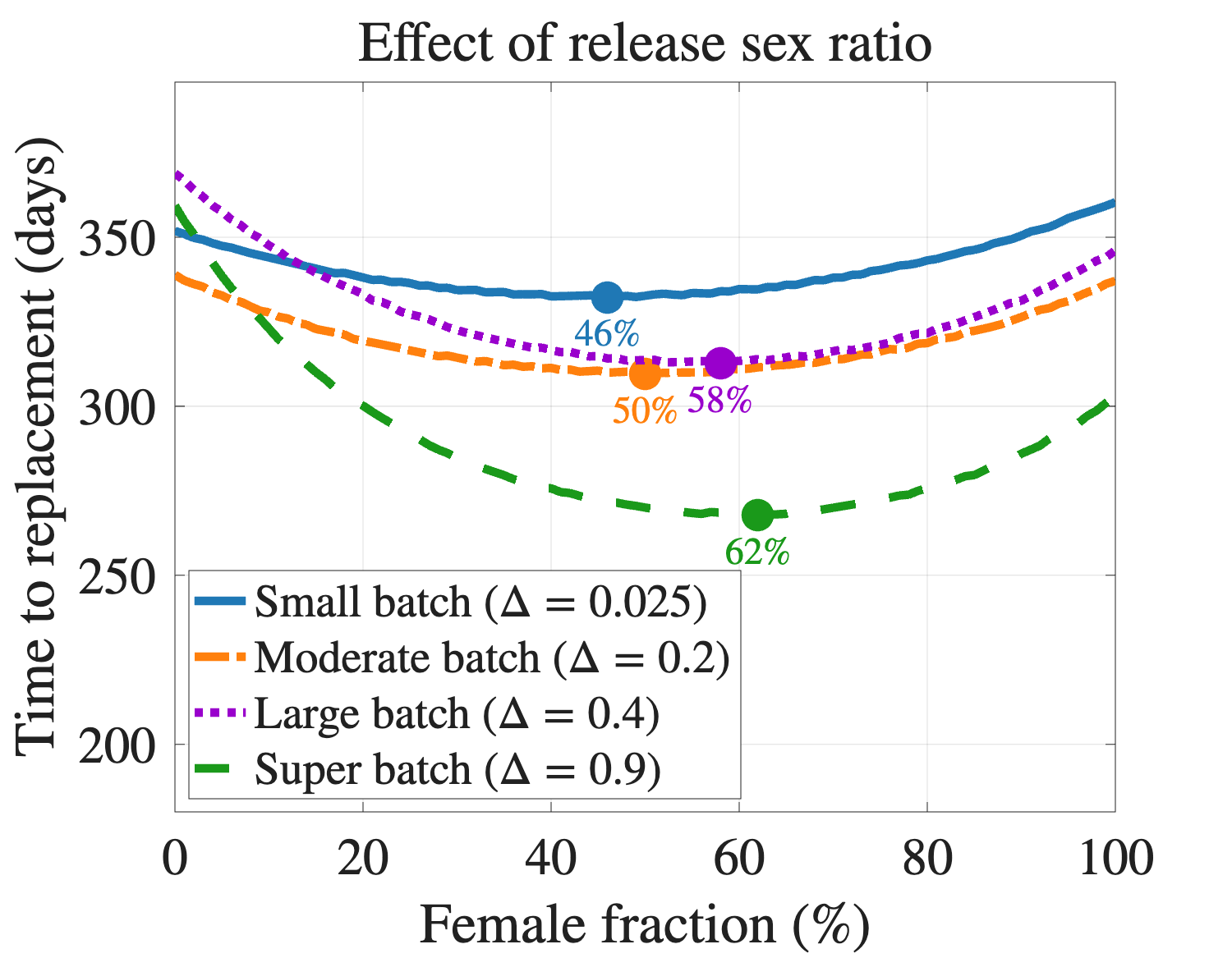}
\caption{Time to replacement for varying release sex ratio across four batch size regimes: small ($\Delta = 0.025$, $\tau = 7/5$ days, 31 batches), moderate ($\Delta = 0.2$, $\tau = 7/0.75$ days, 4 batches), large ($\Delta = 0.4$, $\tau = 7/0.45$ days, 2 batches), and a single super-batch ($\Delta = 0.9$). The optimal release sex ratio, giving the fastest replacement, is marked with a dot in each regime, with the corresponding female fraction indicated.}
\label{fig:release_ratio}
\end{figure}

\paragraph{Release composition: Mixed MM males and MM females}
We examine the effect of release sex composition on invasion speed by comparing mixed MM male and MM female releases ($\Delta = \Delta M_2+\Delta F_2$) of varying sex ratio against MM male-only releases across four batch size regimes, ranging from small repetitive releases ($\Delta  = 0.025$) to a single super-batch ($\Delta = 0.9$). The results are presented in \cref{fig:release_ratio}.

The optimal female fraction of the total release budget increases with batch size, rising from 46\% for small batches to 50\%, 58\%, and 62\% for the moderate, large, and super-batch regimes, respectively. The magnitude of the resulting benefit increases correspondingly, with mixed-sex releases achieving replacement 19 days (5\%), 27 days (8\%), 57 days (15\%), and 91 days (25\%) earlier than MM male-only releases across the four regimes. For small batch sizes, the effect is modest over a one-year horizon, so the male-only strategy remains effectively optimal given its simpler implementation; for larger batch sizes, the benefit becomes pronounced enough to justify the added logistical complexity of releasing females.

The acceleration conferred by mixed-sex releases does not arise from any mechanistic advantage unique to female releases. Both released MM males and MM females mate predominantly with wild-type partners and produce heterozygous (M+) daughters at the same per-mating rate (under $\mu_0 = 0$, $\mu_1 = 1$), via cross~\circled{3} (++ female $\times$ MM male) and cross~\circled{7} (MM female $\times$ ++ male) respectively (\cref{fig:model_diagram}). Those M+ daughters then express the \M maternal-killing mechanism in the subsequent generation, eliminating wild-type offspring and driving population replacement. The two release pathways are therefore mechanistically equivalent.

One contributing factor is the demographic asymmetry between sexes: adult females have a longer lifespan than adult males (10 days vs.\ 7 days at baseline), so released MM females remain reproductively active longer and contribute more M+ offspring per individual over their lifetime. When adult lifespans are set equal, the benefit of mixed-sex releases is accordingly reduced (results not shown).

A second, more fundamental factor is the reduction of male mating competition. Under male-only releases of $N$ individuals, all released MM males compete among themselves for matings with wild-type females, and the per-male mating rate scales inversely with total male abundance $M_0 + N$. Under a mixed half-and-half release, however, only $N/2$ males enter the mating pool, reducing intra-male competition, while the $N/2$ released females contribute equivalent M+ offspring production through cross~\circled{7} without inflating the male pool denominator. This effect is negligible when $N \ll M_0$ but becomes substantial when $N \sim M_0$, explaining why the optimal female fraction and the associated speedup both increase with batch size.

As a side note, we also considered releasing mixed genotypes. As expected, the optimal scenario is to exclusively release MM genotypes, regardless of males or females (see \ref{sec:app_release}). Although releasing M+ females directly leads to \Mns-killing in the first generation (cross~\circled{4}), it produces only half as many M+ offspring as MM females. Maintaining a higher abundance of M+ females is the key to invasion speed and threshold.

\subsubsection{Reversal Releases}
Having established the forward invasion dynamics in the preceding section, we next examine the reverse process: recovering a wild-type-dominant population from the stable \Mns-dominant equilibrium through batched releases of wild-type (++) mosquitoes. We use the same impulsive-release framework as before, with successful reversal as the wild-type fraction among adult females, $F_0/(F_0+F_1+F_2)$, exceeding 90\% within the same time horizon. Unlike forward invasion, reversal cannot be achieved by releasing males alone. Released ++ males with the resident \Mns-carrying females (M+ or MM), and because of the maternal-killing effect (crosses \circled{4} and \circled{5}), only the \Mns-carrying offspring can survive. Wild-type restoration therefore requires releasing ++ females, whose offspring include wild-type offspring. 

\begin{figure}[ht]
\centering
\includegraphics[width=0.33\linewidth]{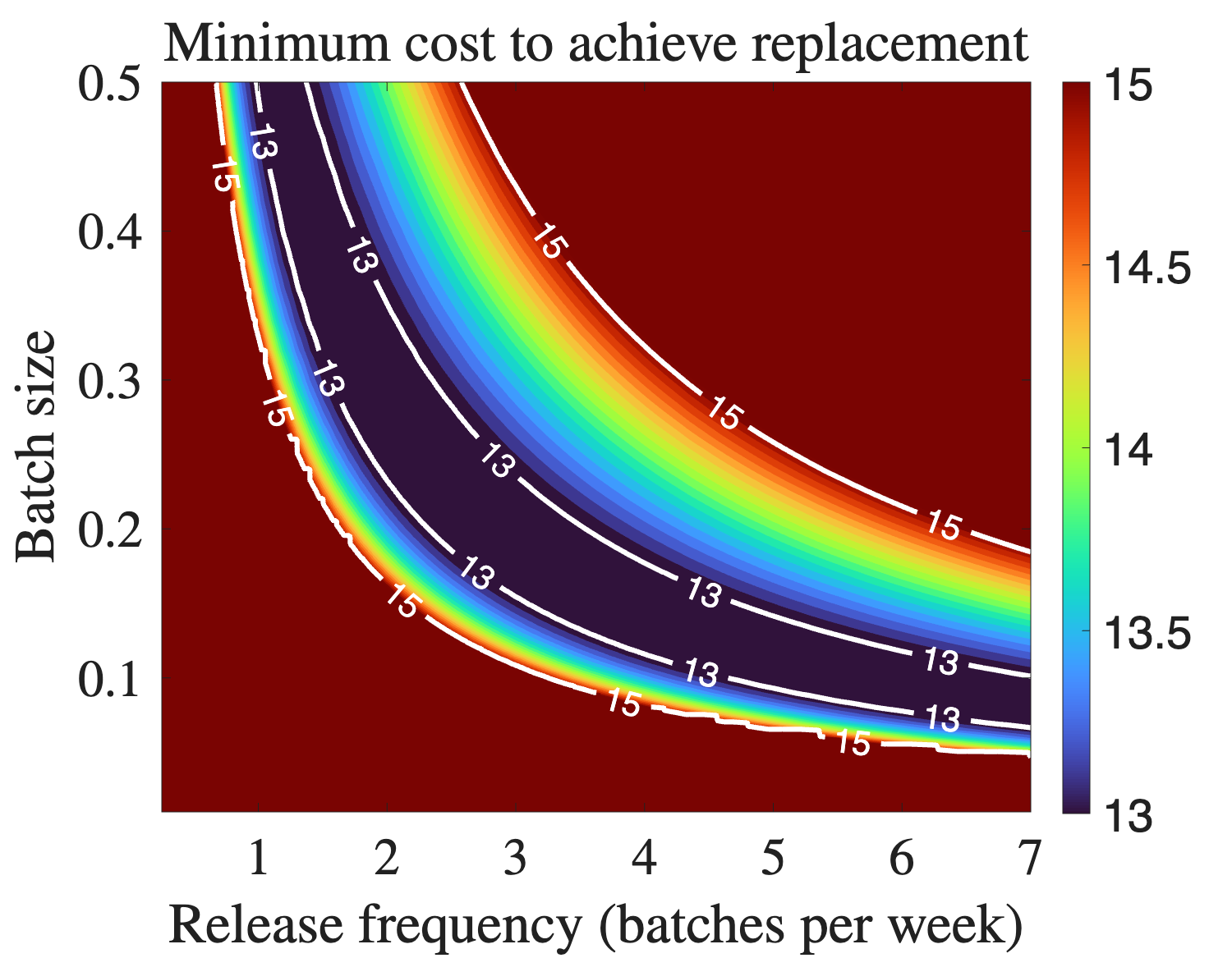}\includegraphics[width=0.33\linewidth]{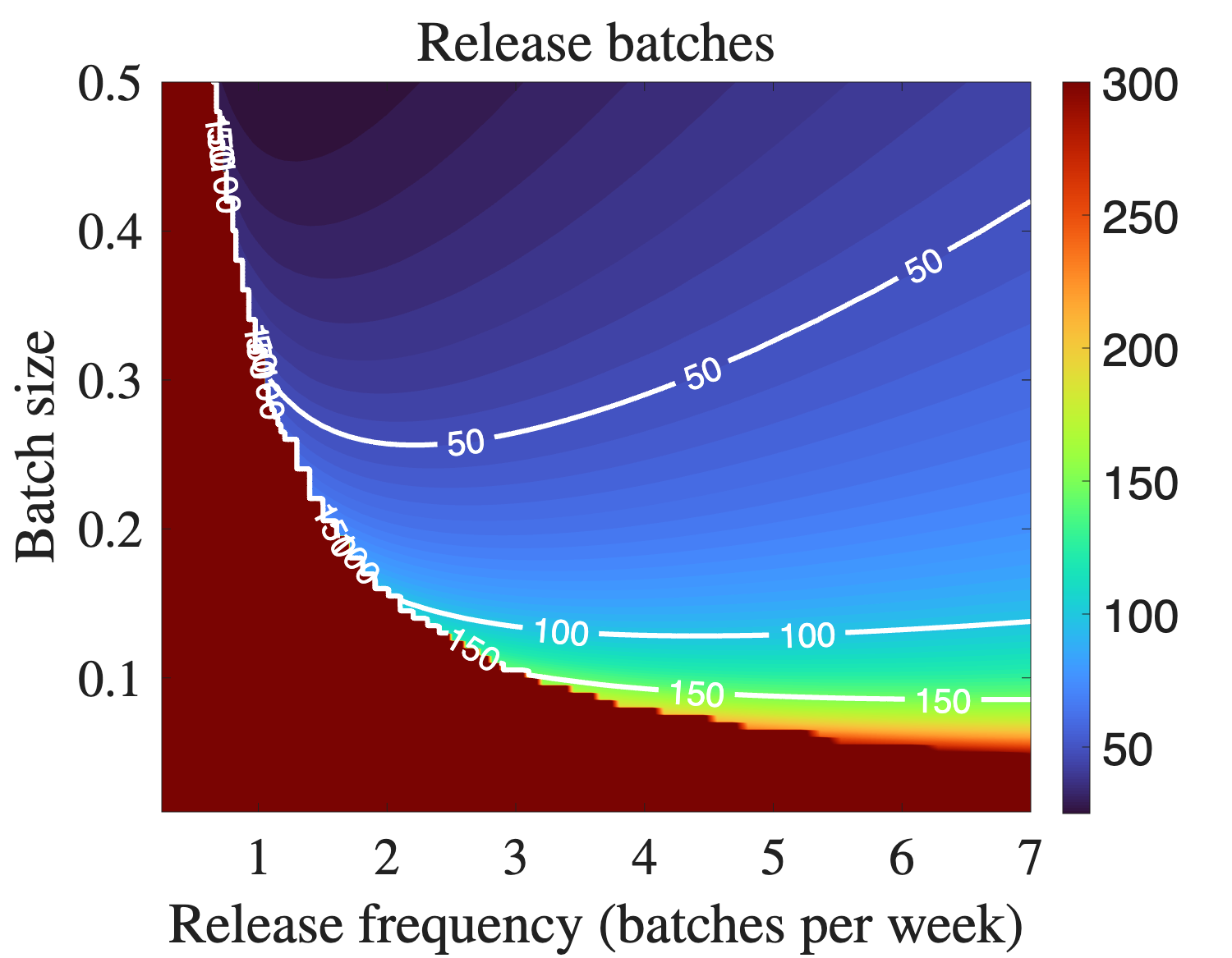}\includegraphics[width=0.33\linewidth]{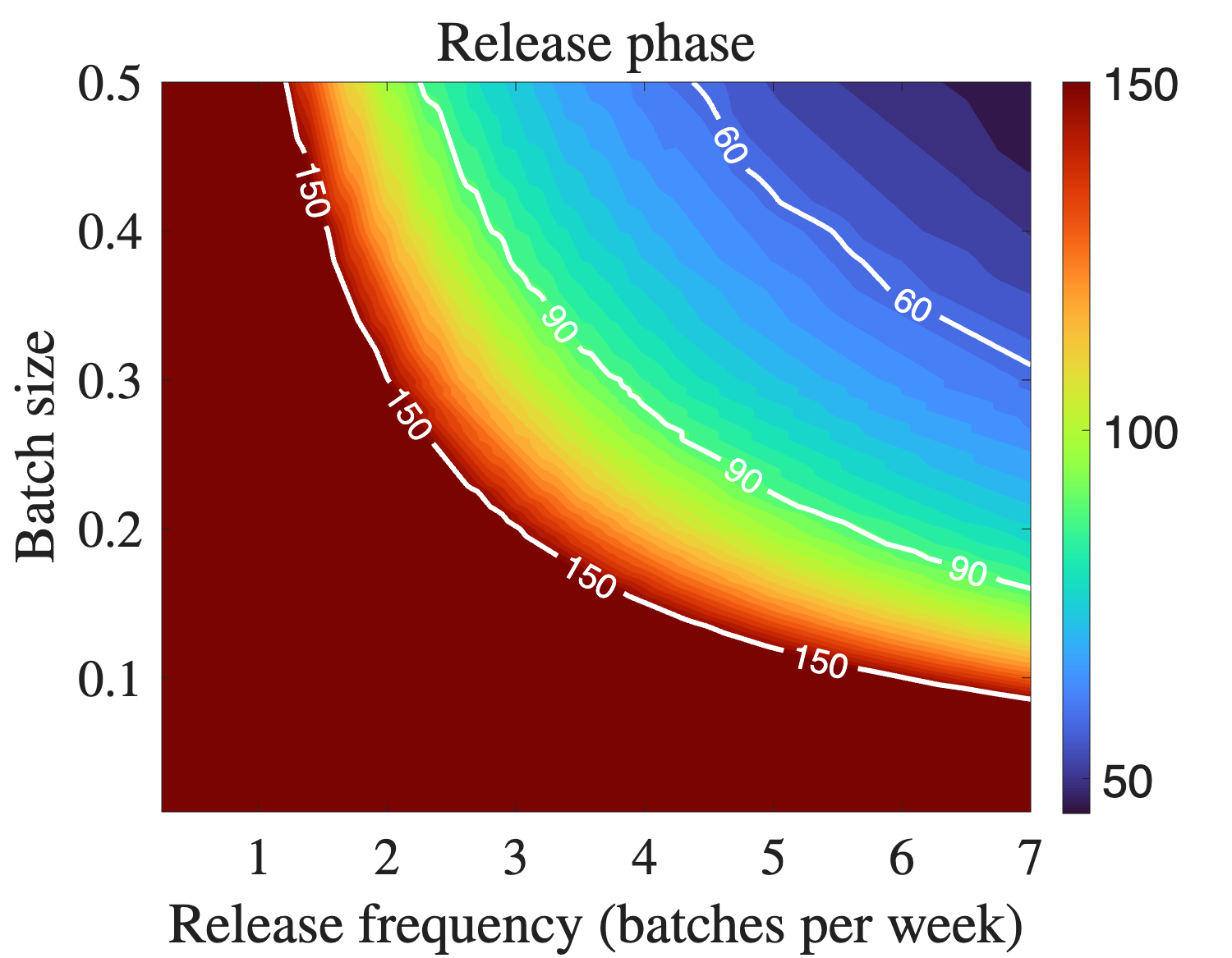}
\caption{Release optimization heatmaps. Same as \cref{fig:heatmaps} but for the reverse process via releasing wild-type females.}
\label{fig:heatmaps_reverse}
\end{figure}

\paragraph{Release wild-type females to reverse the replacement}
\Cref{fig:heatmaps_reverse} repeats the heatmap analysis of \cref{fig:heatmaps} for the reverse process. Reversal requires roughly $15\times$ as many wild-type females as the corresponding forward invasion requires \Mns-homozygous males, though the dependence on batch size and release frequency remains qualitatively similar.

Two factors contribute to this asymmetry: First, the population replacement threshold, corresponding to the coexistence equilibrium, sits at $(E_0,E_1,E_2) = (6481, 1800, 111)$ at baseline ($c=0.9$), corresponding to $(E_1+E_2)/(E_0+E_1+E_2) \approx 23\%$ \Mns-carrying fraction. Forward invasion needs only push the population from $0\%$ up to $23\%$, whereas reversal must drive it from near $100\%$ all the way back down through $23\%$, a considerably longer distance in the state space. 
Second, wild-type production is severely limited for the reversal at the starting equilibrium. A released wild-type female produces wild-type offspring only when she mates with a resident M+ male (cross \circled{2}), since mating with an MM male (cross \circled{3}) yields exclusively M+ offspring. At baseline \Mns-carrying equilibrium $(E_0,E_1,E_2) = (0, 1322, 6609)$, however, M+ males constitute only about $ E_1/(E_1+E_2)\approx17\%$ of the resident male population, so the large majority of matings by released females contribute no wild-type offspring at all. This bottleneck slows the recovery of the wild-type cohort.

\begin{figure}[ht]
\centering
\includegraphics[width=0.5\linewidth]{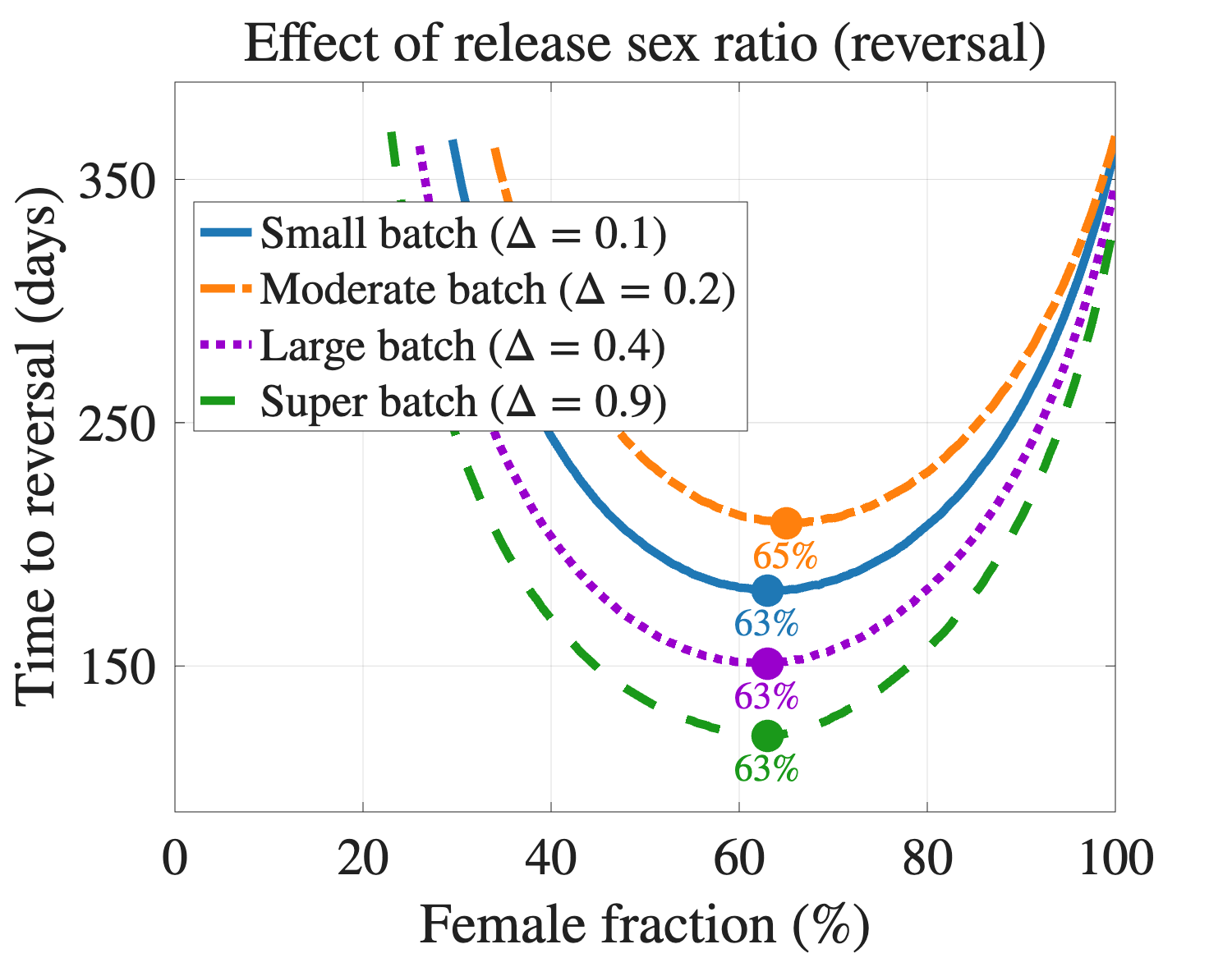}
\caption{Time to replacement for varying release sex ratio across four batch size regimes: small ($\Delta = 0.1$, $\tau = 7/7$ days, 130 batches), moderate ($\Delta = 0.2$, $\tau = 7/3$ days, 64 batches), large ($\Delta  = 0.4$, $\tau = 7/2$ days, 33.5 batches), and super-batch ($\Delta = 0.9$, $\tau = 7/1$ days, 15.5 batches). The optimal release sex ratio, giving the fastest reversal, is marked with a dot in each regime, with the corresponding female fraction indicated.}
\label{fig:release_ratio_reverse}
\end{figure}

\paragraph{Release mixed males and females}
Although releasing wild-type males alone cannot reverse the replacement, co-releasing wild-type males alongside wild-type females substantially accelerates the reversal process (\cref{fig:release_ratio_reverse}). Released ++ males enable cross \circled{1} ($F_0 \times M_0$), which produces wild-type offspring entirely, and this pathway is independent of the resident male population's composition. It therefore relaxes the bottleneck identified above, in which female-only release is constrained due to the minority of resident M+ males via cross \circled{2}. 

\Cref{fig:release_ratio_reverse} quantifies this effect by sweeping the release sex ratio across the four batch-size regimes. The optimal female fraction is remarkably stable at around 65\%, in contrast to the forward case, where the optimum shifted with batch size (\cref{fig:release_ratio}). Relative to wild-type female-only releases, mixed-sex releases at the optimal ratio accelerate reversal by 182 days (50\%), 159 days (43\%), 197 days (57\%), and 214 days (64\%) across the four regimes. These gains are considerably larger than the corresponding speedups in the forward direction (5--25\%), reflecting the fact that co-released wild-type males open a mating pathway that is otherwise almost entirely unavailable at the \Mns-carrying equilibrium, whereas in the forward case both sexes already have abundant wild-type mating partners.

\section{Conclusion} \label{sec:discussion}
We have developed and analyzed a continuous-time compartmental model of \M gene drive dynamics in \textit{Aedes aegypti}, tracking mosquito abundance jointly by life stage and by genotype at the \M locus. Two features distinguish this framework from prior theoretical treatments of \Mns. First, the model is formulated in terms of population abundance in continuous time, rather than allele frequency across discrete generations as in the population genetic models of \cite{ward2011medea} and related work; this allows demographic processes, density-dependent recruitment, and impulsive release programs to be represented explicitly, at the cost of a higher-dimensional state space. Second, the drive mechanism is generalized to admit both imperfect maternal-effect killing, through the leakage parameter $\mu_0$, and imperfect zygotic rescue, through the efficiency parameter $\mu_1$. We find both alter the equilibrium structure qualitatively rather than merely quantitatively.

Our bifurcation analysis reveals that the two imperfections act on distinct stages of the replacement process. Killing leakage $\mu_0$ governs invasion: any $\mu_0 > 0$ sustains a reservoir of wild-type offspring, eliminating the \Mns-carrying equilibrium in favor of a stable coexistence branch on which all three genotypes persist, and raising the release threshold required for establishment. Rescue efficiency $\mu_1$ instead governs the outcome once invasion has succeeded: reducing $\mu_1$ penalizes heterozygotes relative to homozygotes and stabilizes the \Mns-complete equilibrium, which is unstable for all fitness values under perfect rescue, while leaving the invasion threshold essentially unchanged. 

The global sensitivity analysis reveals a consistent two-tier structure across all quantities of interest. The \Mns-specific parameters $c$, $\mu_0$, and $\mu_1$ drive all qualitative and proportional outcomes, with the fitness coefficient dominating throughout and killing leakage, $\mu_0$, consistently second; the demographic life-history parameters influence only the absolute abundance scale and are non-significant for every fractional quantity. Predictions about invasion feasibility, threshold introduction frequency, and stable genotype composition are therefore robust to uncertainty in mosquito demography, and efforts to characterize or engineer \M constructs should prioritize minimizing drive-induced fitness cost and killing leakage. 

However, this robustness should be interpreted with the model's fitness-cost structure in mind: costs are imposed on reproduction alone in our model, so demographic parameters enter as a scaling factor on abundance rather than as a differential penalty between genotypes. Extending the fitness cost to reduce adult lifespan would couple demography directly to genotype competition and could alter this conclusion, and we regard this as a worthwhile target for further study.

The release simulations translate the threshold structure into deployment guidance. Under a fixed total budget, small-to-moderate batches released frequently establish the drive at lower total cost than infrequent large deployments, and which lever matters most depends on production capacity: at low weekly output, total cost is governed almost entirely by capacity and scheduling yields little benefit, whereas at higher output, distributing the same weekly total into smaller, more frequent batches becomes the primary efficiency gain. 

Mixed-sex releases of MM individuals outperform male-only releases, with an optimal composition near 60\% female, though the advantage is appreciable only at larger batch sizes and may not justify the added rearing complexity otherwise. Reversal is markedly more demanding than forward invasion, requiring roughly $15\times$ as many wild-type females at the respective cost-minimizing configurations, because released wild-type females can produce wild-type offspring only by mating with the scarce resident M+ males. Co-releasing wild-type males relieves this bottleneck by opening a mating pathway independent of resident male composition, accelerating reversal by 43--64\% depending on batch size, which is a substantially larger gain than the corresponding effect in the forward direction.

Several simplifications limit the scope of these conclusions. No functional synthetic \M construct has yet been demonstrated in \textit{Aedes aegypti}, so \Mns-specific parameters are drawn from \textit{Drosophila} experiments and should be regarded as indicative rather than predictive. The model omits seasonality, which is likely to matter over the year-long horizons considered in the release simulations, and it is spatially homogeneous. The latter is the more consequential omission: for a threshold-dependent drive such as \Mns, spatial structure governs both the local release density required for establishment and the potential for spatial confinement, which is among the principal practical motivations for choosing a threshold-dependent drive over an unconditional one. The present analysis is intended to establish a baseline understanding of the well-mixed dynamics, including the equilibrium structure and stability conditions that a spatial extension would inherit; incorporating dispersal is a natural next step.

\bibliographystyle{abbrv}
\bibliography{References.bib}

\begin{thebibliography}{10}

\bibitem{Achee2019}
N.~L. Achee, J.~P. Grieco, H.~Vatandoost, G.~Seixas, J.~Pinto, L.~Ching-NG,
  A.~J. Martins, W.~Juntarajumnong, V.~Corbel, L.-C. Gouagna, J.-P. David,
  J.~G. Logan, J.~Orsborne, E.~Marois, G.~J. Devine, and J.~Vontas.
\newblock Alternative strategies for mosquito-borne arbovirus control.
\newblock {\em PLOS Neglected Tropical Diseases}, 13(1):e0006822, 2019.

\bibitem{akbari2014novel}
O.~S. Akbari, C.-H. Chen, J.~M. Marshall, H.~Huang, I.~Antoshechkin, and B.~A.
  Hay.
\newblock Novel {{Synthetic Medea Selfish Genetic Elements Drive Population
  Replacement}} in {{Drosophila}}; a {{Theoretical Exploration}} of
  {{Medea-Dependent Population Suppression}}.
\newblock {\em ACS Synthetic Biology}, 3(12):915--928, Dec. 2014.

\bibitem{akbari2013synthetic}
O.~S. Akbari, K.~D. Matzen, J.~M. Marshall, H.~Huang, C.~M. Ward, and B.~A.
  Hay.
\newblock A {{Synthetic Gene Drive System}} for {{Local}}, {{Reversible
  Modification}} and {{Suppression}} of {{Insect Populations}}.
\newblock {\em Current Biology}, 23(8):671--677, Apr. 2013.

\bibitem{alphey2014genetica}
L.~Alphey.
\newblock Genetic {{Control}} of {{Mosquitoes}}.
\newblock {\em Annual Review of Entomology}, 59(1):205--224, Jan. 2014.

\bibitem{alphey2010sterileinsect}
L.~Alphey, M.~Benedict, R.~Bellini, G.~G. Clark, D.~A. Dame, M.~W. Service, and
  S.~L. Dobson.
\newblock Sterile-{{Insect Methods}} for {{Control}} of {{Mosquito-Borne
  Diseases}}: {{An Analysis}}.
\newblock {\em Vector-Borne and Zoonotic Diseases}, 10(3):295--311, Apr. 2010.

\bibitem{AMCA}
{American Mosquito Control Association}.
\newblock Vector-borne diseases.
\newblock \url{https://www.mosquito.org/vector-borne-diseases/}.

\bibitem{beeman1992maternaleffect}
R.~W. Beeman, K.~S. Friesen, and R.~E. Denell.
\newblock Maternal-{{Effect Selfish Genes}} in {{Flour Beetles}}.
\newblock {\em Science}, 256(5053):89--92, Apr. 1992.

\bibitem{bian2010endosymbiotic}
G.~Bian, Y.~Xu, P.~Lu, Y.~Xie, and Z.~Xi.
\newblock The endosymbiotic bacterium {{Wolbachia}} induces resistance to
  dengue virus in {{Aedes}} aegypti.
\newblock {\em PLoS Pathogens}, 6(4):e1000833, 2010.

\bibitem{buchman2018synthetically}
A.~Buchman, J.~M. Marshall, D.~Ostrovski, T.~Yang, and O.~S. Akbari.
\newblock Synthetically engineered {{{\emph{Medea}}}} gene drive system in the
  worldwide crop pest {{{\emph{Drosophila}}}}{\emph{ suzukii}}.
\newblock {\em Proceedings of the National Academy of Sciences},
  115(18):4725--4730, May 2018.

\bibitem{buchman2018engineered}
A.~B. Buchman, T.~Ivy, J.~M. Marshall, O.~S. Akbari, and B.~A. Hay.
\newblock Engineered {{Reciprocal Chromosome Translocations Drive High
  Threshold}}, {{Reversible Population Replacement}} in {{Drosophila}}.
\newblock {\em ACS Synthetic Biology}, 7(5):1359--1370, May 2018.

\bibitem{burt2014heritable}
A.~Burt.
\newblock Heritable strategies for controlling insect vectors of disease.
\newblock {\em Philosophical Transactions of the Royal Society B: Biological
  Sciences}, 369(1645):20130432, June 2014.

\bibitem{champer2016cheating}
J.~Champer, A.~Buchman, and O.~S. Akbari.
\newblock Cheating evolution: Engineering gene drives to manipulate the fate of
  wild populations.
\newblock {\em Nature Reviews Genetics}, 17(3):146--159, Mar. 2016.

\bibitem{chen2007synthetic}
C.-H. Chen, H.~Huang, C.~M. Ward, J.~T. Su, L.~V. Schaeffer, M.~Guo, and B.~A.
  Hay.
\newblock A {{Synthetic Maternal-Effect Selfish Genetic Element Drives
  Population Replacement}} in {{{\emph{Drosophila}}}}.
\newblock {\em Science}, 316(5824):597--600, Apr. 2007.

\bibitem{dutra2016wolbachia}
H.~L.~C. Dutra, M.~N. Rocha, F.~B.~S. Dias, S.~B. Mansur, E.~P. Caragata, and
  L.~A. Moreira.
\newblock Wolbachia blocks currently circulating {{Zika}} virus isolates in
  {{Brazilian Aedes}} aegypti mosquitoes.
\newblock {\em Cell Host Microbes}, 19:771--774, 2016.

\bibitem{dyck2005sterile}
V.~A. Dyck, J.~Hendrichs, and A.~S. Robinson, editors.
\newblock {\em Sterile Insect Technique: Principles and Practice in Area-Wide
  Integrated Pest Management}.
\newblock Springer, Dordrecht, Netherlands, 2005.

\bibitem{edgington2018population}
M.~P. Edgington and L.~S. Alphey.
\newblock Population dynamics of engineered underdominance and killer-rescue
  gene drives in the control of disease vectors.
\newblock {\em PLOS Computational Biology}, 14(3):e1006059, Mar. 2018.

\bibitem{foster2002mosquitoes}
{\relax WA}.~Foster and {\relax ED}.~Walker.
\newblock Mosquitoes ({{Culicidae}}).
\newblock In {\em Medical and Veterinary Entomology}, pages 203--262. Academic
  Press, Amsterdam, 2002.

\bibitem{hoffmann2011successful}
A.~A. Hoffmann, B.~L. Montgomery, J.~Popovici, I.~{Iturbe-Ormaetxe}, P.~H.
  Johnson, F.~Muzzi, M.~Greenfield, M.~Durkan, Y.~S. Leong, Y.~Dong, H.~Cook,
  J.~Axford, A.~G. Callahan, N.~Kenny, C.~Omodei, E.~A. McGraw, P.~A. Ryan,
  S.~A. Ritchie, M.~Turelli, and S.~L. O'Neill.
\newblock Successful establishment of {{Wolbachia}} in {{Aedes}} populations to
  suppress dengue transmission.
\newblock {\em Nature}, 476(7361):454--457, Aug. 2011.

\bibitem{HoffmannTurelli1997}
A.~A. Hoffmann and M.~Turelli.
\newblock Cytoplasmic incompatibility in insects.
\newblock In S.~L. O'Neill, J.~H. Werren, and A.~A. Hoffmann, editors, {\em
  Influential Passengers: Inherited Microorganisms and Arthropod Reproduction},
  pages 42--80. Oxford University Press, 1997.

\bibitem{indriani2020reduced}
C.~Indriani, W.~Tantowijoyo, E.~Ranc{\`e}s, B.~Andari, E.~Prabowo, D.~Yusdi,
  M.~R. Ansari, D.~S. Wardana, E.~Supriyati, I.~Nurhayati, I.~Ernesia,
  S.~Setyawan, I.~Fitriana, E.~Arguni, Y.~Amelia, R.~A. Ahmad, N.~P. Jewell,
  S.~M. Dufault, P.~A. Ryan, B.~R. Green, T.~F. McAdam, S.~L. O'Neill, S.~K.
  Tanamas, C.~P. Simmons, K.~L. Anders, and A.~Utarini.
\newblock Reduced dengue incidence following deployments of
  {{Wolbachia-infected Aedes}} aegypti in {{Yogyakarta}}, {{Indonesia}}: A
  quasi-experimental trial using controlled interrupted time series analysis.
\newblock {\em Gates Open Research}, 4:50, May 2020.

\bibitem{irvin2004assessing}
N.~Irvin, M.~S. Hoddle, D.~A. O'Brochta, B.~Carey, and P.~W. Atkinson.
\newblock Assessing fitness costs for transgenic {{{\emph{Aedes}}}}{\emph{
  aegypti}} expressing the {{GFP}} marker and transposase genes.
\newblock {\em Proceedings of the National Academy of Sciences},
  101(3):891--896, Jan. 2004.

\bibitem{james2018pathwaya}
S.~James, F.~H. Collins, P.~A. Welkhoff, C.~Emerson, H.~C.~J. Godfray,
  M.~Gottlieb, B.~Greenwood, S.~W. Lindsay, C.~M. Mbogo, F.~O. Okumu,
  H.~Quemada, M.~Savadogo, J.~A. Singh, K.~H. Tountas, and Y.~T. Tour{\'e}.
\newblock Pathway to {{Deployment}} of {{Gene Drive Mosquitoes}} as a
  {{Potential Biocontrol Tool}} for {{Elimination}} of {{Malaria}} in
  {{Sub-Saharan Africa}}: {{Recommendations}} of a {{Scientific Working Group}}
  \dag.
\newblock {\em The American Journal of Tropical Medicine and Hygiene},
  98(6\_Suppl):1--49, June 2018.

\bibitem{kamgang2008computation}
J.~C. Kamgang and G.~Sallet.
\newblock Computation of threshold conditions for epidemiological models and
  global stability of the disease-free equilibrium ({{DFE}}).
\newblock {\em Math. Biosci.}, 213(1):1--12, 2008.

\bibitem{li2020development}
M.~Li, T.~Yang, N.~P. Kandul, M.~Bui, S.~Gamez, R.~Raban, J.~Bennett, H.~M.
  S{\'a}nchez~C, G.~C. Lanzaro, H.~Schmidt, Y.~Lee, J.~M. Marshall, and O.~S.
  Akbari.
\newblock Development of a confinable gene drive system in the human disease
  vector {{Aedes}} aegypti.
\newblock {\em eLife}, 9:e51701, Jan. 2020.

\bibitem{marino2008methodology}
S.~Marino, I.~B. Hogue, C.~J. Ray, and D.~E. Kirschner.
\newblock A methodology for performing global uncertainty and sensitivity
  analysis in systems biology.
\newblock {\em Journal of Theoretical Biology}, 254(1):178--196, Sept. 2008.

\bibitem{marshall2010cartagena}
J.~M. Marshall.
\newblock The {{Cartagena Protocol}} and genetically modified mosquitoes.
\newblock {\em Nature Biotechnology}, 28(9):896--897, Sept. 2010.

\bibitem{marshall2011inverse}
J.~M. Marshall and B.~A. Hay.
\newblock Inverse {{Medea}} as a {{Novel Gene Drive System}} for {{Local
  Population Replacement}}: {{A Theoretical Analysis}}.
\newblock {\em Journal of Heredity}, 102(3):336--341, May 2011.

\bibitem{marshall2012confinement}
J.~M. Marshall and B.~A. Hay.
\newblock Confinement of gene drive systems to local populations: {{A}}
  comparative analysis.
\newblock {\em Journal of Theoretical Biology}, 294:153--171, Feb. 2012.

\bibitem{marshall2012general}
J.~M. Marshall and B.~A. Hay.
\newblock General principles of single-construct chromosomal gene drive.
\newblock {\em Evolution}, 66(7):2150--2166, July 2012.

\bibitem{nationalenvironmentagency2020aedes}
{National Environment Agency}.
\newblock Aedes {{Mosquito}}.
\newblock
  https://www.nea.gov.sg/dengue-zika/prevent-aedes-mosquito-breeding/aedes-mosquito,
  Oct. 2020.

\bibitem{oye2014regulating}
K.~A. Oye, K.~Esvelt, E.~Appleton, F.~Catteruccia, G.~Church, T.~Kuiken,
  S.~B.-Y. Lightfoot, J.~McNamara, A.~Smidler, and J.~P. Collins.
\newblock Regulating gene drives.
\newblock {\em Science}, 345(6197):626--628, Aug. 2014.

\bibitem{qu2026multistage}
Z.~Qu and T.~Wu.
\newblock Multistage {{Spatial Model}} for {{Informing Release}} of
  {{Wolbachia-infected Mosquitoes}} as {{Disease Control}}.
\newblock {\em Royal Society Open Science}, 13:251724, 2026.

\bibitem{qu2018modeling}
Z.~Qu, L.~Xue, and J.~M. Hyman.
\newblock Modeling the transmission of {{Wolbachia}} in mosquitoes for
  controlling mosquito-borne diseases.
\newblock {\em SIAM Journal on Applied Mathematics}, 78(2):826--852, Jan. 2018.

\bibitem{ryan2020establishment}
{\relax PA}.~Ryan, {\relax AP}.~Turley, G.~Wilson, {\relax TP}.~Hurst,
  K.~Retzki, J.~{Brown-Kenyon}, L.~Hodgson, N.~Kenny, H.~Cook, {\relax
  BL}.~Montgomery, {\relax CJ}.~Paton, {\relax SA}.~Ritchie, {\relax
  AA}.~Hoffmann, {\relax NP}.~Jewell, {\relax SK}.~Tanamas, {\relax
  KL}.~Anders, {\relax CP}.~Simmons, and {\relax SL}.~O?Neill.
\newblock Establishment of {{wMel Wolbachia}} in {{Aedes}} aegypti mosquitoes
  and reduction of local dengue transmission in {{Cairns}} and surrounding
  locations in northern {{Queensland}}, {{Australia}} [version 2; peer review:
  2 approved].
\newblock {\em Gates Open Research}, 3(1547), 2020.

\bibitem{SanchezMGDrivE2020}
H.~M. S{\'a}nchez~C., S.~L. Wu, J.~B. Bennett, and J.~M. Marshall.
\newblock {MGDrivE}: A modular simulation framework for the spread of gene
  drives through spatially explicit mosquito populations.
\newblock {\em Methods in Ecology and Evolution}, 11(2):229--239, 2020.

\bibitem{scott2000longitudinala}
T.~W. Scott, A.~C. Morrison, L.~H. Lorenz, G.~G. Clark, D.~Strickman,
  P.~Kittayapong, H.~Zhou, and J.~D. Edman.
\newblock Longitudinal {{Studies}} of {{{\emph{Aedes}}}}{\emph{ aegypti}}
  ({{Diptera}}: {{Culicidae}}) in {{Thailand}} and {{Puerto Rico}}:
  {{Population Dynamics}}.
\newblock {\em Journal of Medical Entomology}, 37(1):77--88, Jan. 2000.

\bibitem{soares-pinheiro2016eggs}
V.~C. {Soares-Pinheiro}, W.~{Dasso-Pinheiro}, J.~M. {Trindade-Bezerra}, and
  W.~P. Tadei.
\newblock Eggs viability of {{Aedes}} aegypti {{Linnaeus}} ({{Diptera}},
  {{Culicidae}}) under different environmental and storage conditions in
  {{Manaus}}, {{Amazonas}}, {{Brazil}}.
\newblock {\em Brazilian Journal of Biology}, 77(2):396--401, Aug. 2016.

\bibitem{styer2007mortality}
L.~M. Styer, S.~L. Minnick, A.~K. Sun, and T.~W. Scott.
\newblock Mortality and reproductive dynamics of {{Aedes}} aegypti
  ({{Diptera}}: {{Culicidae}}) fed human blood.
\newblock {\em Vector Borne Zoonotic Dis.}, 7(1):86--98, 2007.

\bibitem{van2002reproduction}
P.~{Van den Driessche} and J.~Watmough.
\newblock Reproduction numbers and sub-threshold endemic equilibria for
  compartmental models of disease transmission.
\newblock {\em Math. Biosci.}, 180(1):29--48, 2002.

\bibitem{wade1994population}
M.~J. Wade and R.~W. Beeman.
\newblock The population dynamics of maternal-effect selfish genes.
\newblock {\em Genetics}, 138(4):1309--1314, Dec. 1994.

\bibitem{wang2022symbionts}
G.-H. Wang, J.~Du, C.~Y. Chu, M.~Madhav, G.~L. Hughes, and J.~Champer.
\newblock Symbionts and gene drive: Two strategies to combat vector-borne
  disease.
\newblock {\em Trends in Genetics}, 38(7):708--723, July 2022.

\bibitem{ward2011medea}
C.~M. Ward, J.~T. Su, Y.~Huang, A.~L. Lloyd, F.~Gould, and B.~A. Hay.
\newblock \textit{Medea} selfish genetic elements as tools for altering traits
  of wild populations: {A} theoretical analysis.
\newblock {\em Evolution}, 65(4):1149--1162, 2011.

\bibitem{wu2021mgdrive}
S.~L. Wu, J.~B. Bennett, H.~M. S{\'a}nchez~C., A.~J. Dolgert, T.~M. Le{\'o}n,
  and J.~M. Marshall.
\newblock {{MGDrivE}} 2: {{A}} simulation framework for gene drive systems
  incorporating seasonality and epidemiological dynamics.
\newblock {\em PLOS Computational Biology}, 17(5):e1009030, May 2021.

\bibitem{yang2011follow}
H.~M. Yang, M.~d. L. d.~G. Macoris, K.~C. Galvani, and M.~T.~M. Andrighetti.
\newblock Follow up estimation of {{Aedes}} aegypti entomological parameters
  and mathematical modellings.
\newblock {\em Biosystems}, 103(3):360--371, Mar. 2011.

\bibitem{zettel2009yellow}
C.~Zettel and P.~Kaufman.
\newblock Yellow fever mosquito {{Aedes}} aegypti ({{Linnaeus}}) ({{Insecta}}:
  {{Diptera}}: {{Culicidae}}): {{EENY-434}}/{{IN792}}, 2/2009.
\newblock {\em EDIS}, 2009(2):1--8, Apr. 2009.

\end{thebibliography}

\section*{Acknowledgments}
ZQ was partially supported by the National Science Foundation award DMS-2316242.

\appendix
\setcounter{figure}{0} 
\setcounter{equation}{0} 
\numberwithin{figure}{section}
\renewcommand\thefigure{\thesection.\arabic{figure}} 
\renewcommand{\theHfigure}{\thesection.\arabic{figure}} 

\section{Parameterization} \label{sec:app_param}
We parameterize the model using life-history data for \textit{Aedes aegypti}, the primary mosquito vector of major mosquito-borne diseases including dengue, Zika, and yellow fever. Factory-scale production of \textit{Aedes aegypti} has been routinely achieved for SIT and \textit{Wolbachia} release programs, making it a natural candidate for gene-drive field deployment as well \cite{ward2011medea}.

\paragraph{Life span $1/\mu_{f0}$ and $1/\mu_{m0}$} Laboratory estimates report adult lifespans of 17.5 days for females and 10 days for males, a male-to-female ratio of approximately 55--60\%, appropriate as a lab-calibrated reference pair. We adopt these values as the upper bound of our parameter range, since laboratory rearing removes field mortality sources such as predation and tends to overestimate lifespan. For the baseline parameterization, representing a more realistic field scenario, we scale both lifespans down proportionally to 10 days for females and 7 days for males, adjusting the male-to-female ratio to approximately 70\%, consistent with field estimates of relatively greater male mortality.

\paragraph{Development rate $\psi$} Development proceeds through an egg-to-larvae stage lasting approximately 2 days (range 1--4 days) followed by a larvae-to-adult stage lasting approximately 10 days (range 8--30 days) \cite{soares-pinheiro2016eggs,foster2002mosquitoes}; summing these stages gives a total juvenile development time of 12 days at baseline, with a range of 9--34 days across the two stages.

\paragraph{Juvenile mortality rate $\mu_{e0}$} Reported survivorship is approximately 85\% from egg to larvae and 45\% from larvae to adult \cite{yang2011follow}, giving an overall egg-to-adult survivorship of approximately 38\%. Solving $\psi/(\psi+\mu_{e0}) = 0.38$ for the baseline development rate above yields $\mu_{e0} = 0.132$.

\paragraph{\Mns-related parameters} Since a functional synthetic \M construct has not yet been demonstrated in \textit{Aedes aegypti}, we base our \Mns-specific parameter estimates on the experimental characterization of synthetic \M elements in \textit{Drosophila} \cite{chen2007synthetic, buchman2018synthetically}, with ranges chosen to span the spectrum of efficiencies reported across these systems. The resulting baseline values, summarized in \cref{tab:paramter}, serve as a reference point for the bifurcation and sensitivity analyses.

\paragraph{Assumptions on fitness cost} We introduce two simplifying assumptions for the incorporation of \Mns-induced fitness cost. These assumptions are adopted solely for the numerical bifurcation and sensitivity analyses in the remainder of this section; the analytical results in \cref{sec:analysis} hold without these restrictions.

\noindent\underline{Equal mortality rates.} We assume that adult and juvenile mortality rates are identical across genotypes,
\begin{equation}\label{eq:assum}
\mu_{f0}=\mu_{f1}=\mu_{f2} = \mu_f,\quad \mu_{m0}=\mu_{m1}=\mu_{m2} = \mu_m,\quad \mu_{e0}=\mu_{e1}=\mu_{e2} = \mu_e.
\end{equation}
This assumption reflects empirical evidence from multiple gene-drive studies: \cite{li2020development} and \cite{irvin2004assessing} find adult survival statistically indistinguishable between transgenic and wild-type individuals. In field-relevant populations, adult mortality is dominated by extrinsic factors (predation, desiccation, temperature), against which intrinsic genotypic differences in survival are negligible; fitness costs from transgene expression are therefore assumed to manifest primarily through reduced fecundity and fertility rather than altered longevity.\\

\noindent\underline{Multiplicative fitness coefficient.} Fitness costs are encoded through a single coefficient $c \in (0,1]$, with female fecundity $v_{fi} = c^i$ and male fertility $v_{mi} = c^i$, giving the overall reproduction rate
\begin{equation}\label{eq:assum2}
\phi_{ij} = \phi_0\, c^{i+j}.
\end{equation}
Each additional copy of the \M allele imposes a multiplicative reduction in reproductive output, so heterozygotes (M+) carry a single factor of $c$ and homozygotes (MM) carry $c^2$. This multiplicative structure is standard in gene-drive modeling \cite{edgington2018population, ward2011medea} and follows the usual convention of encoding fitness costs through reproduction rather than survival. When $c=1$ there is no fitness cost; as $c$ decreases, the cost increases and \Mns-bearing genotypes become progressively less reproductively viable.

\section{Additional Bifurcation Diagrams}\label{sec:app_bifur}
\begin{figure}[ht!]
\centering
\includegraphics[width=0.33\linewidth]{pics/bifur_E0E1_mu0_0_mu1_100.png}\includegraphics[width=0.33\linewidth]{pics/bifur_E0E1_mu0_0_mu1_70.png}\includegraphics[width=0.33\linewidth]{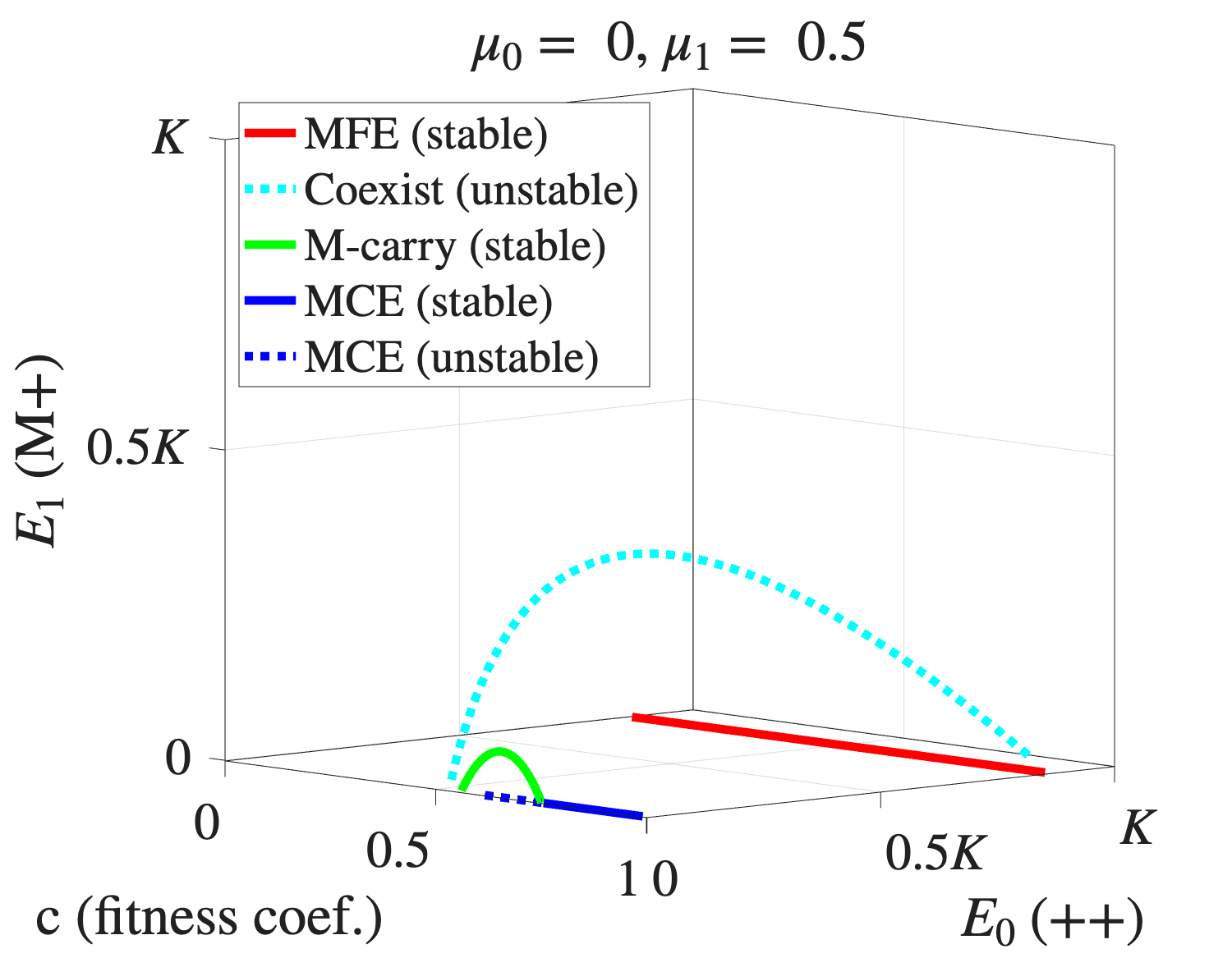}\\
\includegraphics[width=0.33\linewidth]{pics/bifur_E1E2_mu0_0_mu1_100.png}\includegraphics[width=0.33\linewidth]{pics/bifur_E1E2_mu0_0_mu1_70.png}\includegraphics[width=0.33\linewidth]{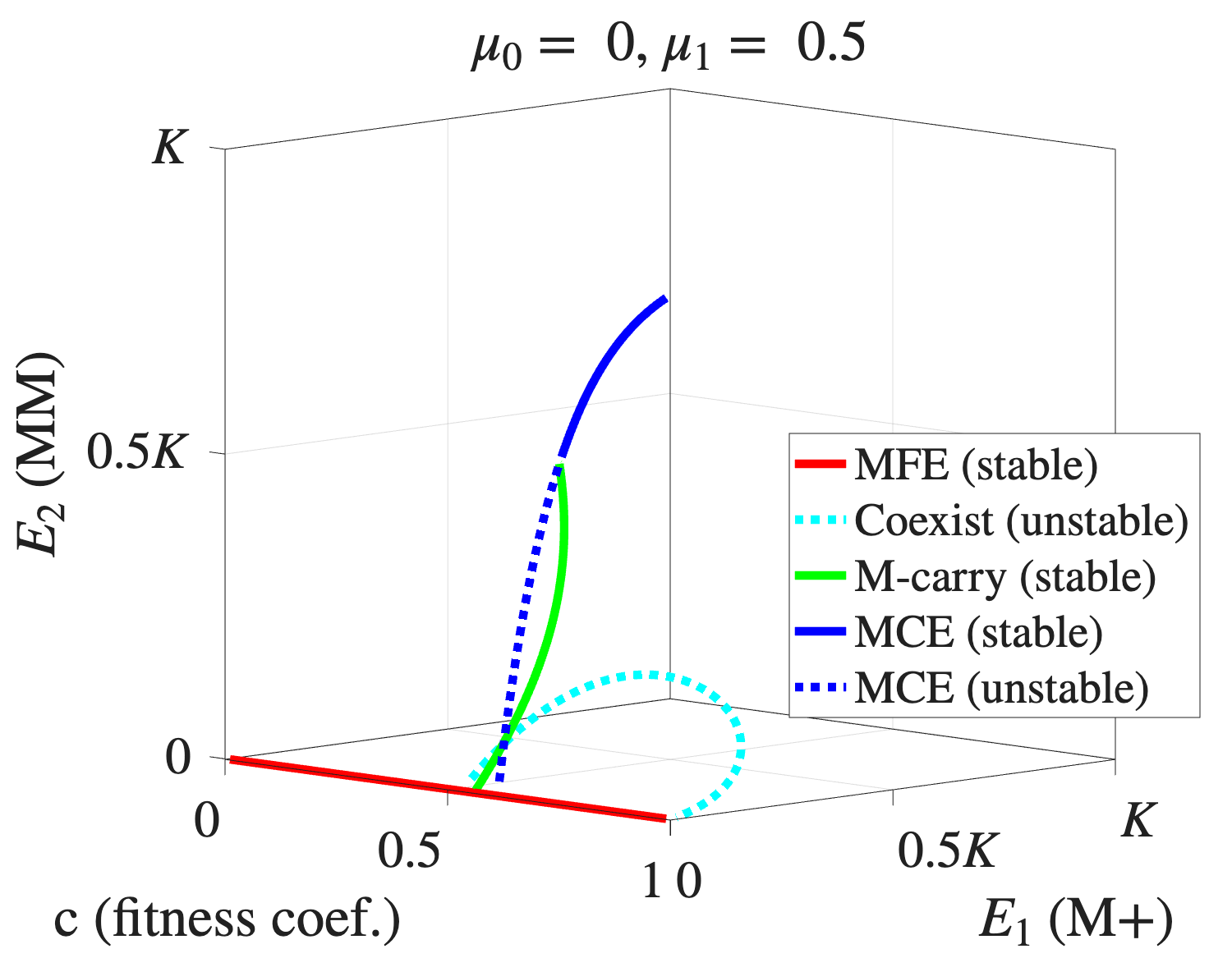}
\caption{Bifurcation diagrams for varying $\mu_1$ to supplement discussion in \cref{sec:bifur}.} 
\label{fig:bifur_mu1}
\end{figure}

\begin{figure}[ht!]
\centering
\includegraphics[width=0.33\linewidth]{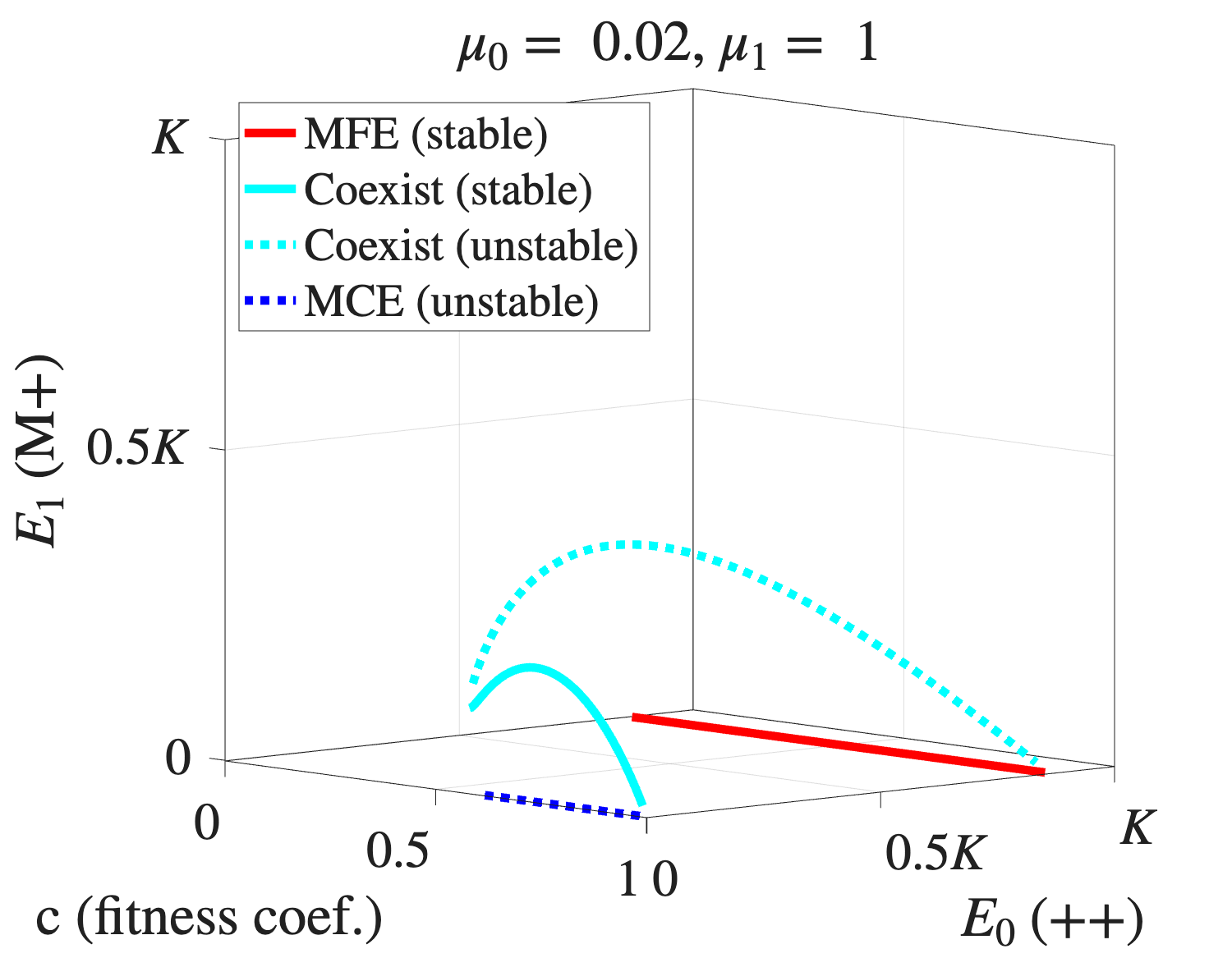}\includegraphics[width=0.33\linewidth]{pics/bifur_E0E1_mu0_10_mu1_100.png}\includegraphics[width=0.33\linewidth]{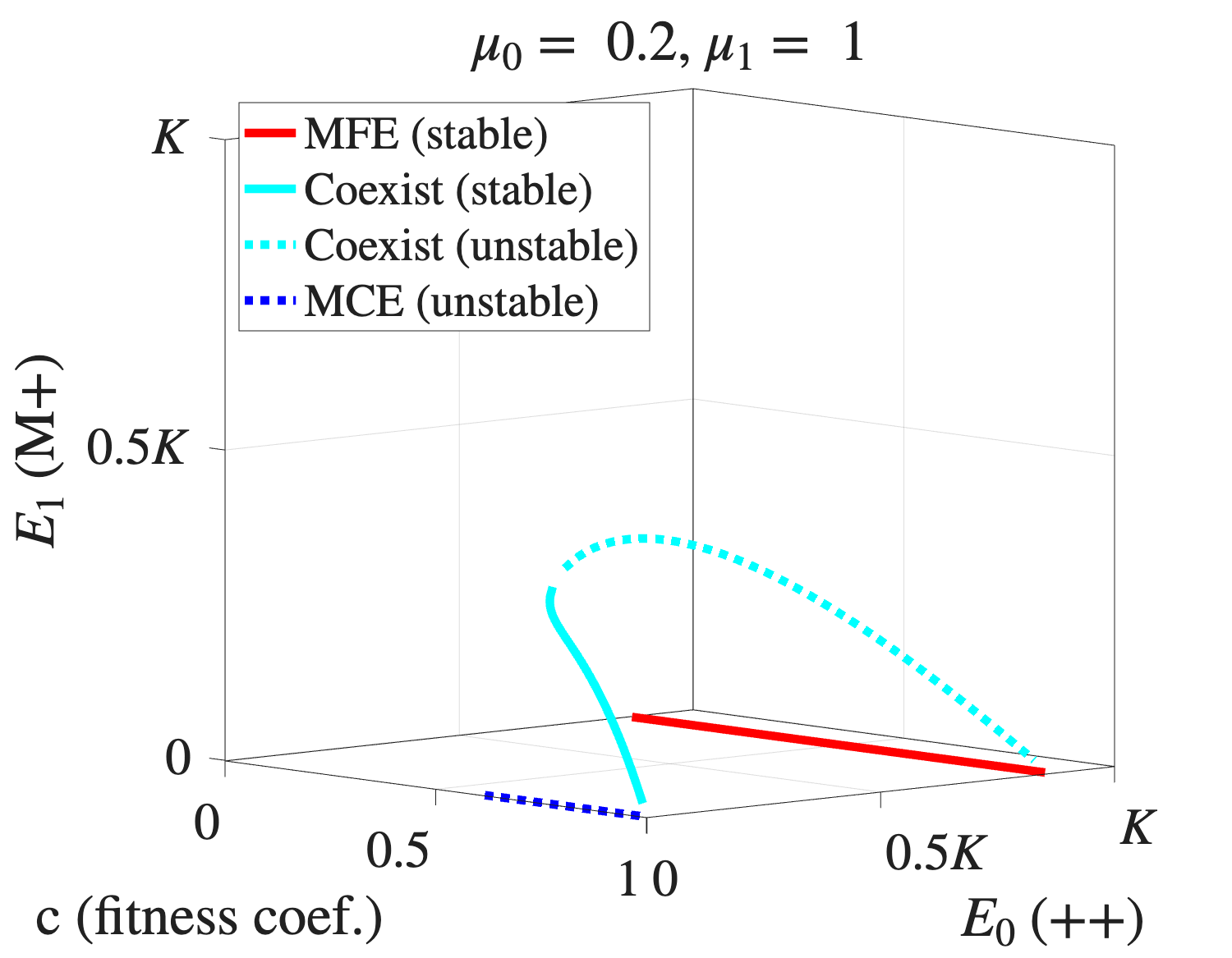}\\
\includegraphics[width=0.33\linewidth]{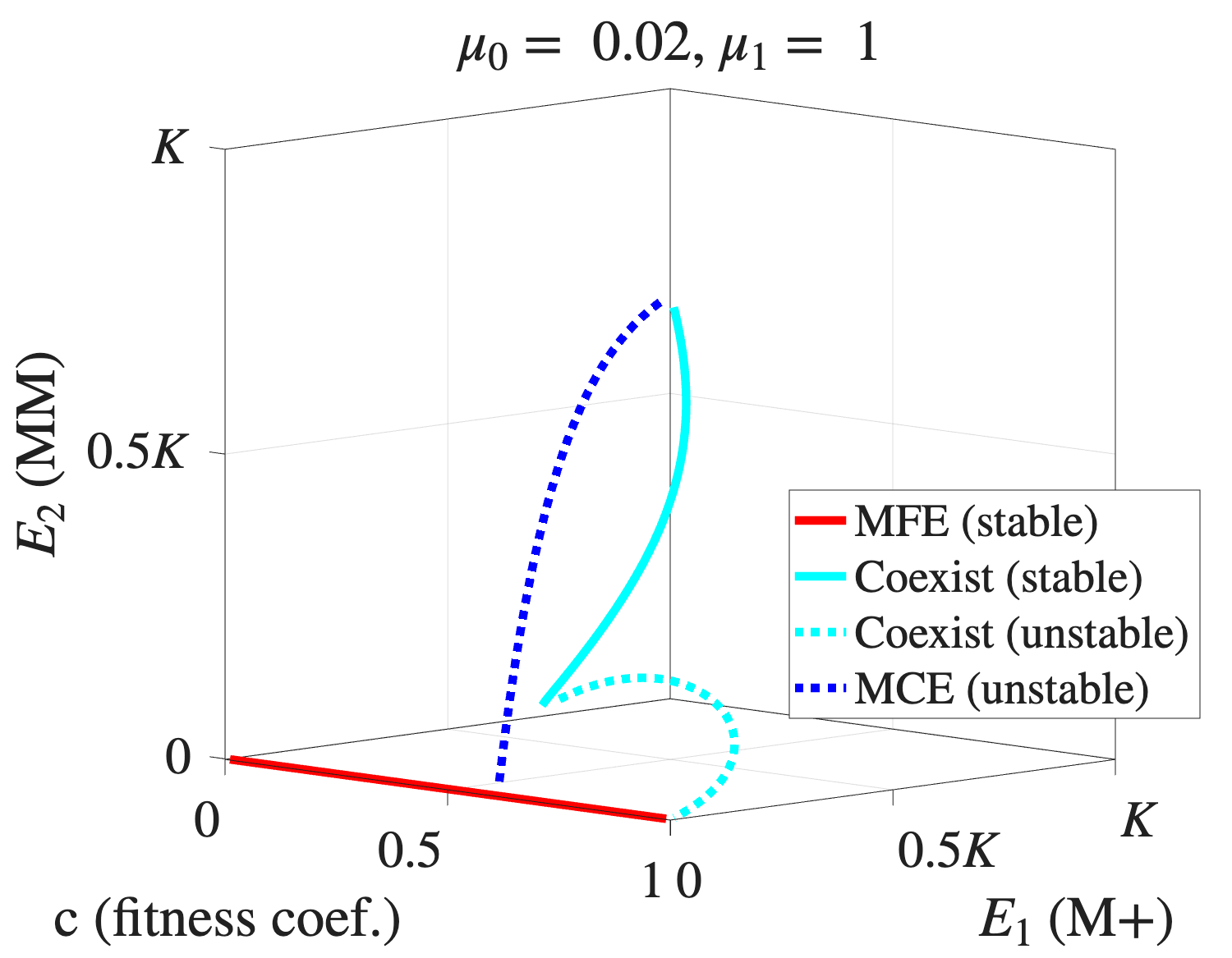}\includegraphics[width=0.33\linewidth]{pics/bifur_E1E2_mu0_10_mu1_100.png}\includegraphics[width=0.33\linewidth]{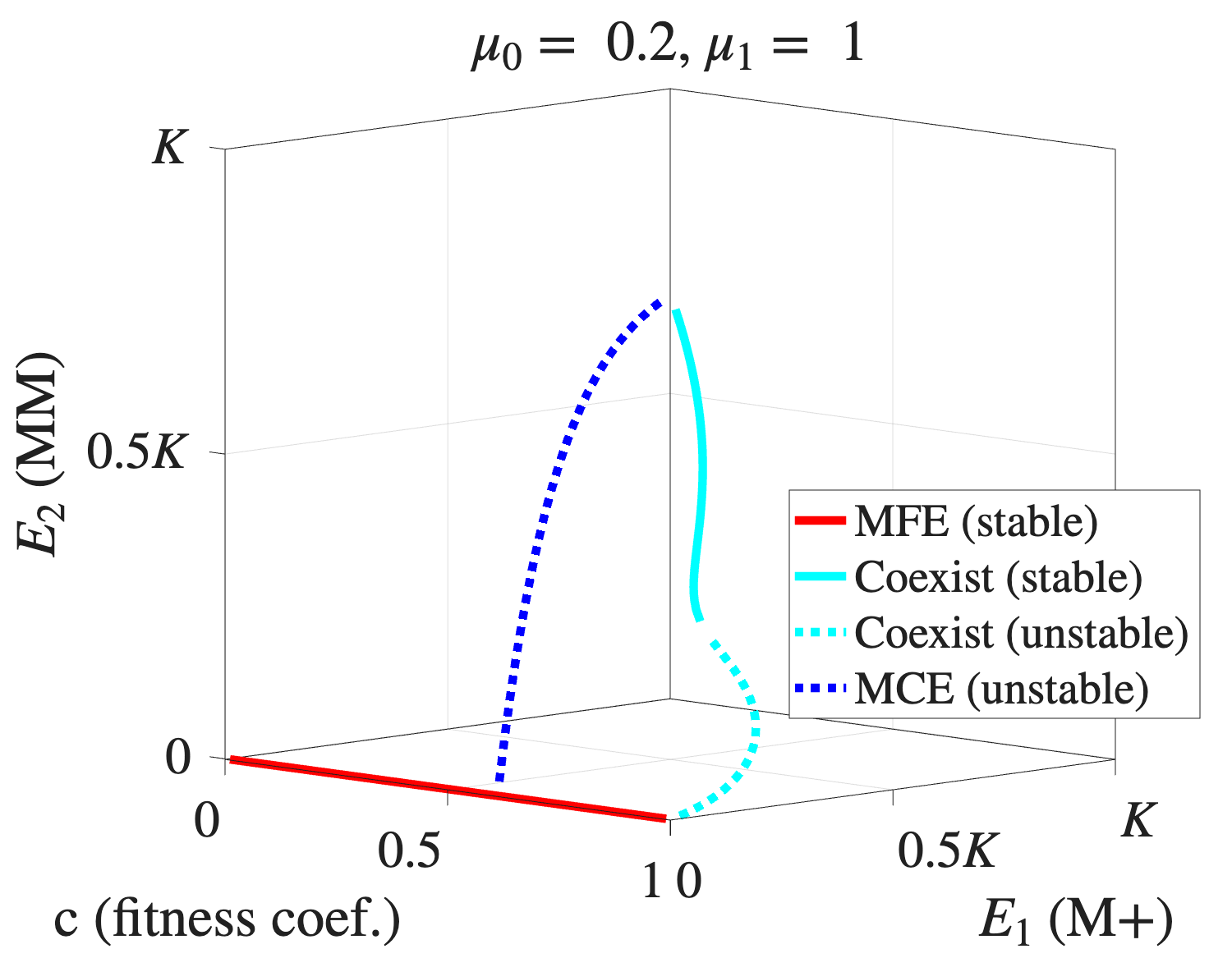}
\caption{Bifurcation diagrams for varying $\mu_0$ to supplement discussion in \cref{sec:bifur}.} 
\label{fig:bifur_mu0}
\end{figure}

This appendix presents bifurcation diagrams over a broader range of the \Mns-specific parameters to supplement the three representative scenarios shown in the main text (\cref{fig:bifur}). \Cref{fig:bifur_mu1} varies the rescue efficiency $\mu_1$ under perfect killing, and \cref{fig:bifur_mu0} varies the killing leakage $\mu_0$ under perfect rescue. More discussion on these figures can be found in the main text (\cref{sec:bifur}).

\section{Monotonicity Checks for the PRCC Analysis}\label{sec:app_SA}
\begin{figure}[ht]
\centering
\includegraphics[width=\linewidth]{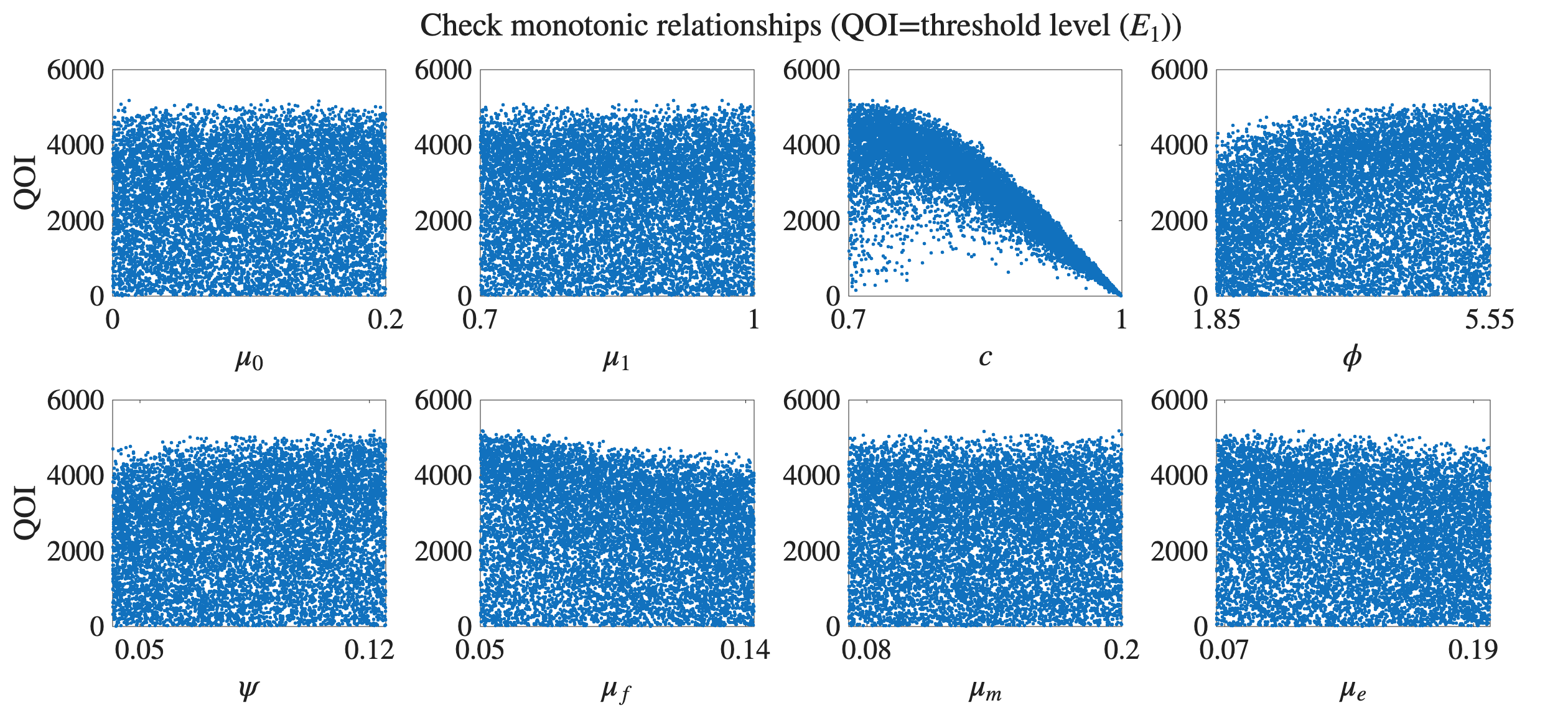}
\includegraphics[width=\linewidth]{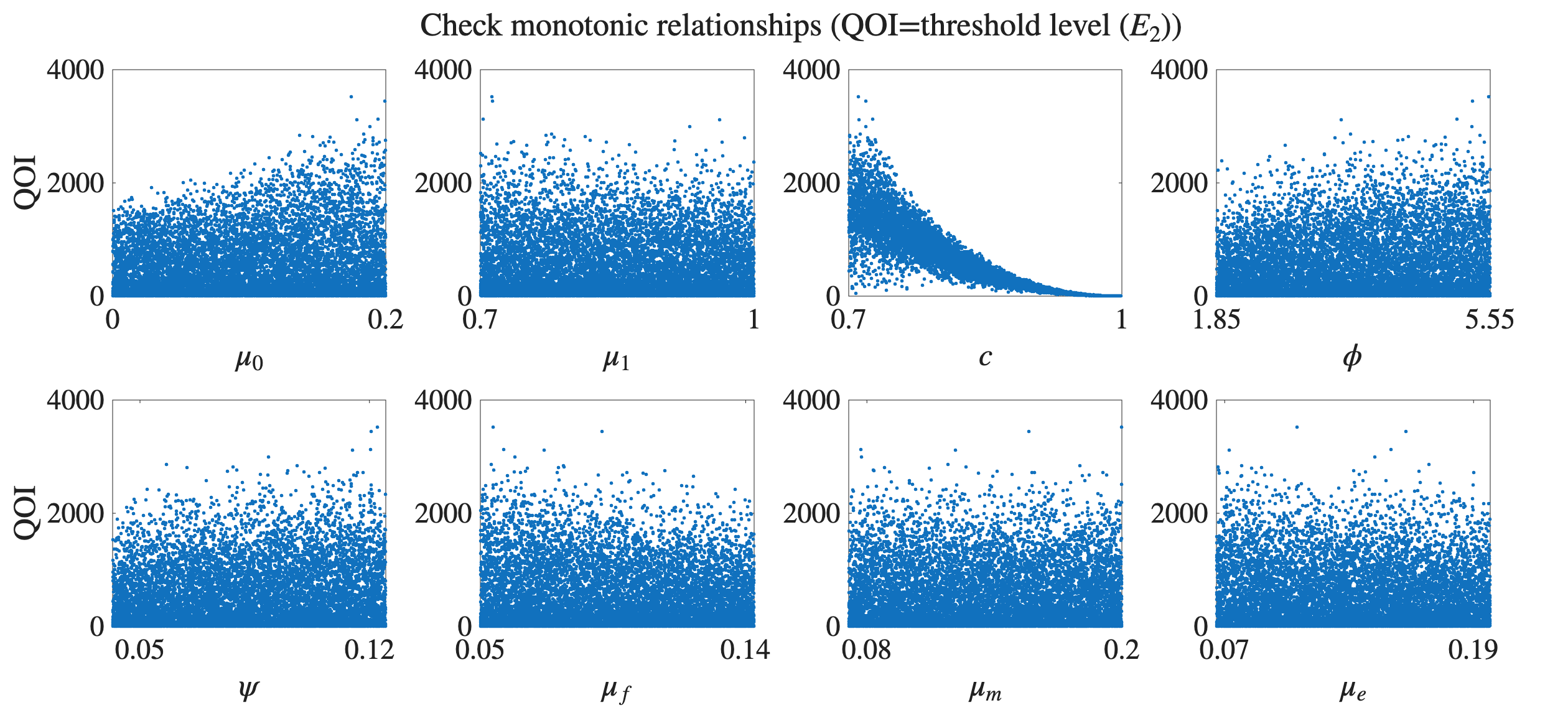}
\includegraphics[width=\linewidth]{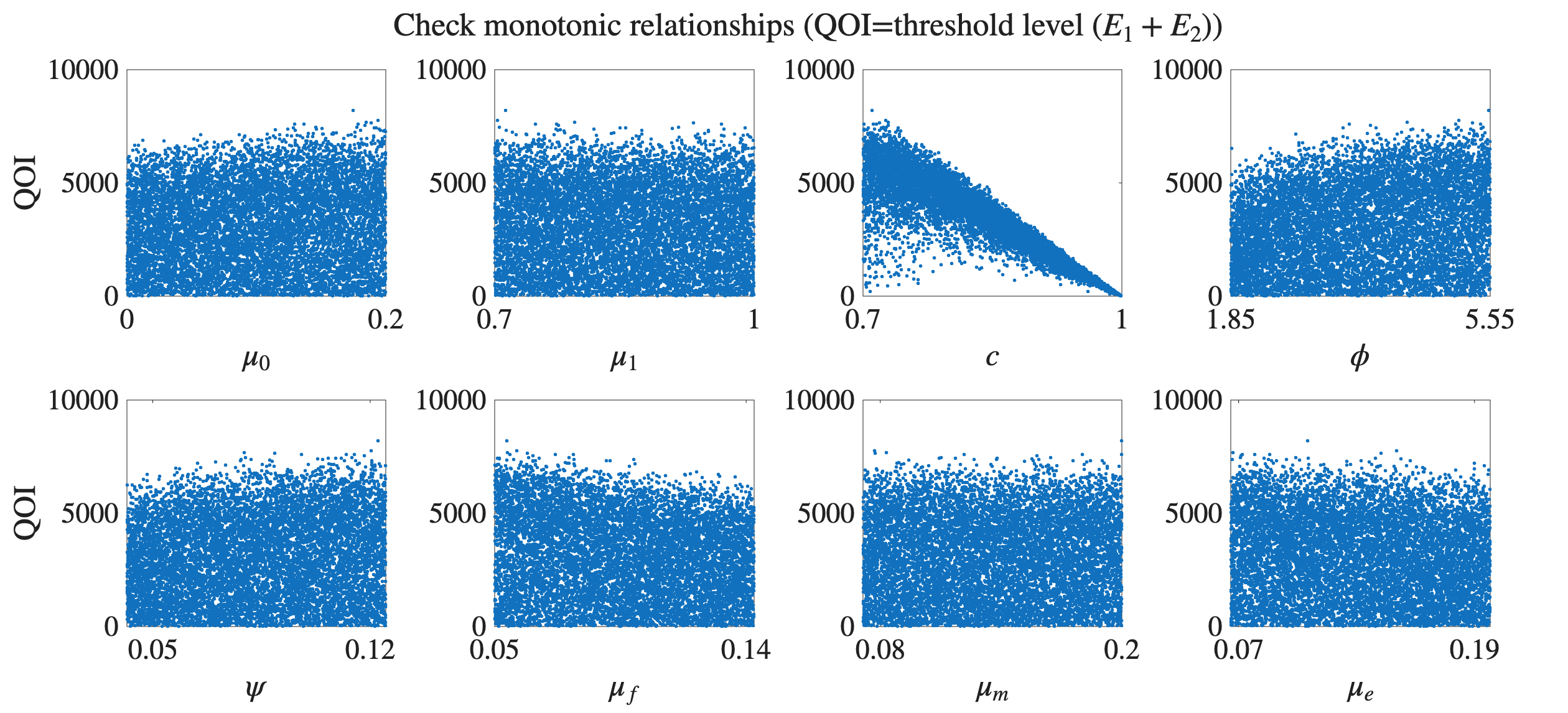}
\caption{Monotonicity check for PRCC to supplement the figures in \cref{sec:SA_threshold}.}
\label{fig:SA_PRCC_mono1}
\end{figure}

\begin{figure}[ht]
\centering
\includegraphics[width=\linewidth]{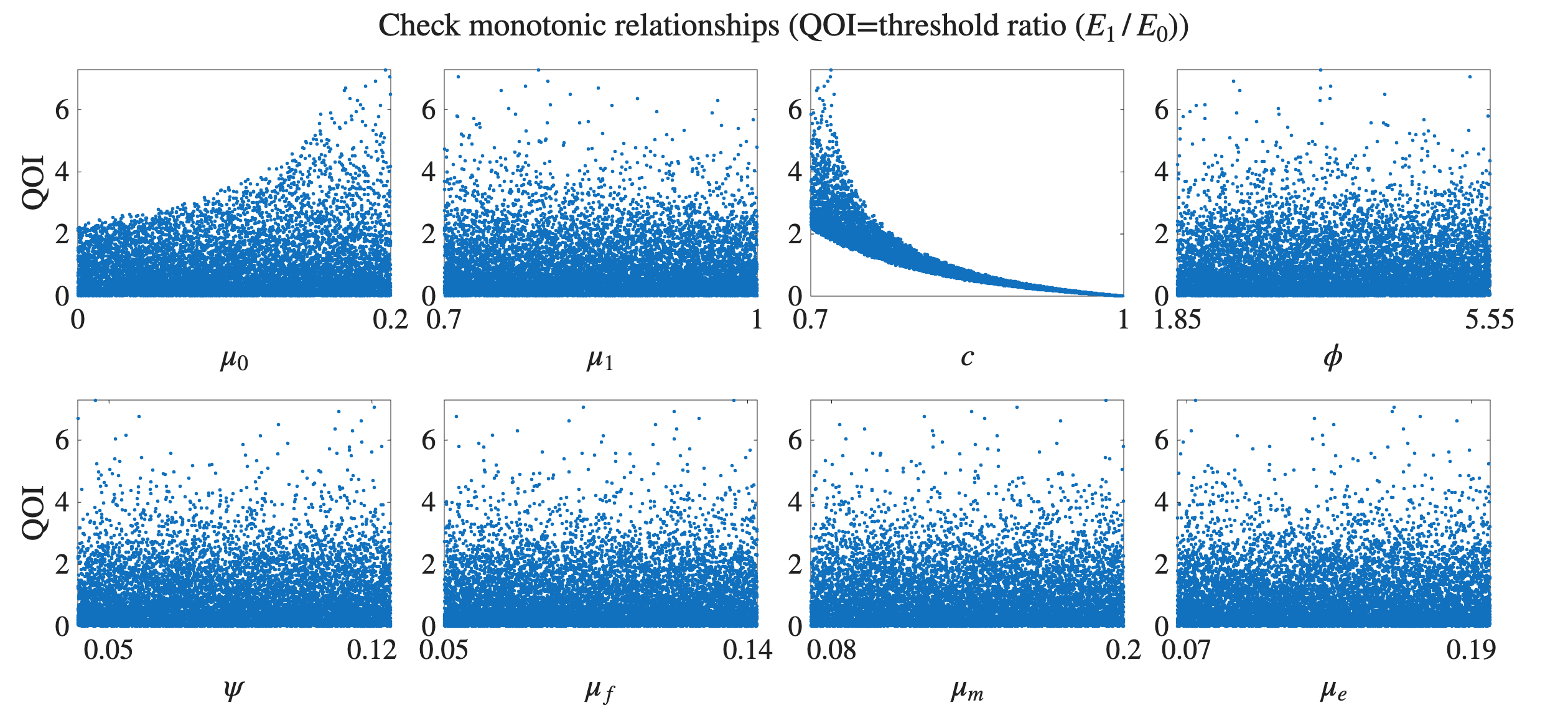}
\includegraphics[width=\linewidth]{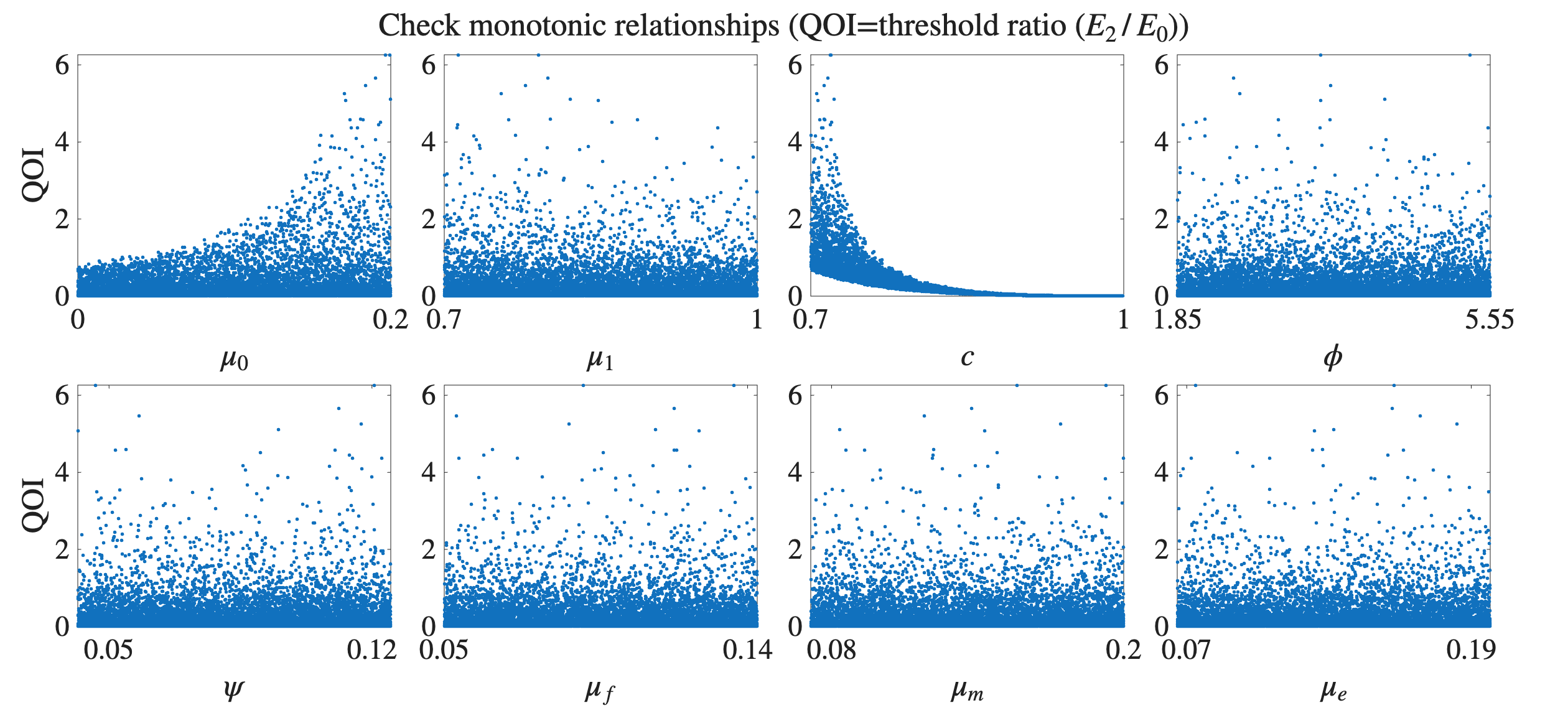}
\includegraphics[width=\linewidth]{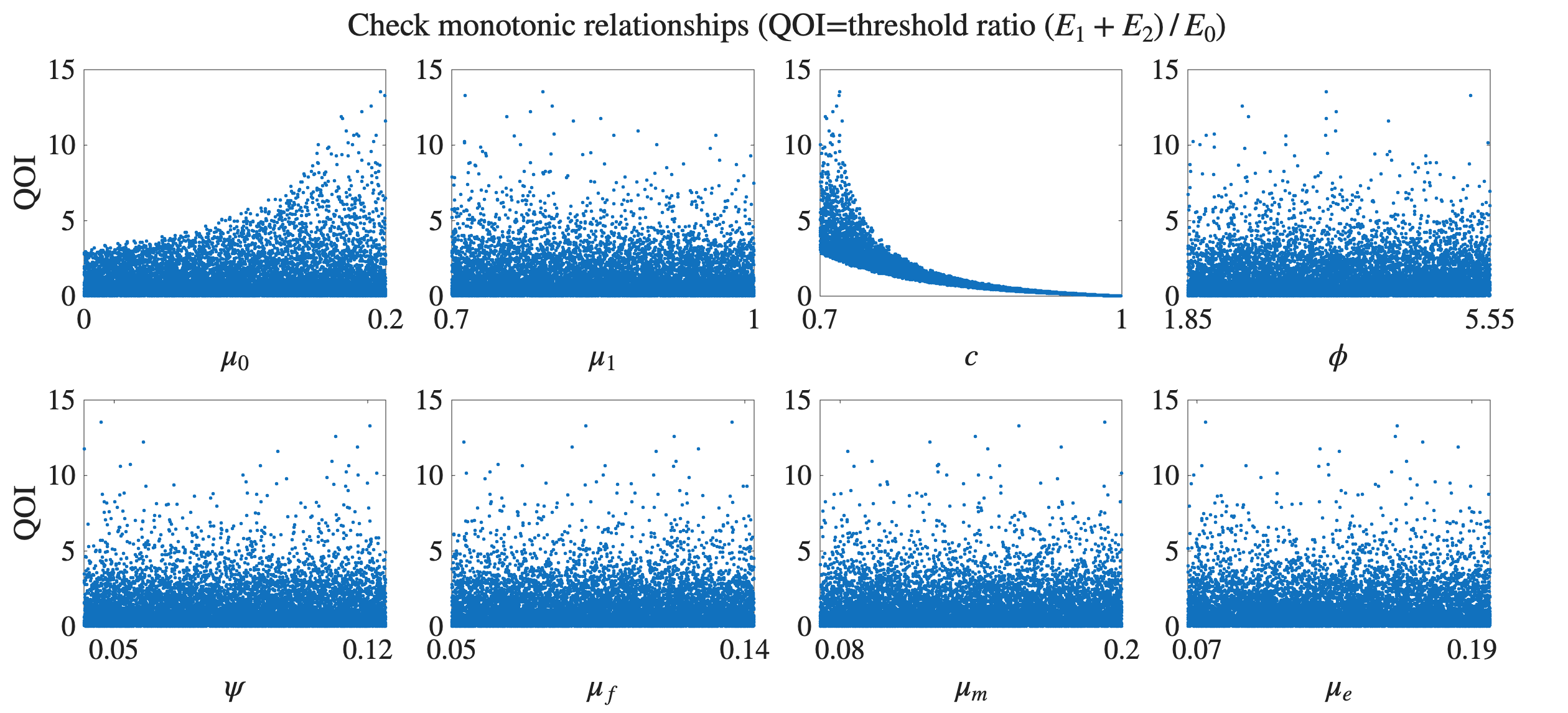}
\caption{Monotonicity check for PRCC to supplement the figures in \cref{sec:SA_threshold}.}
\label{fig:SA_PRCC_mono2}
\end{figure}

\begin{figure}[ht]
\centering
\includegraphics[width=\linewidth]{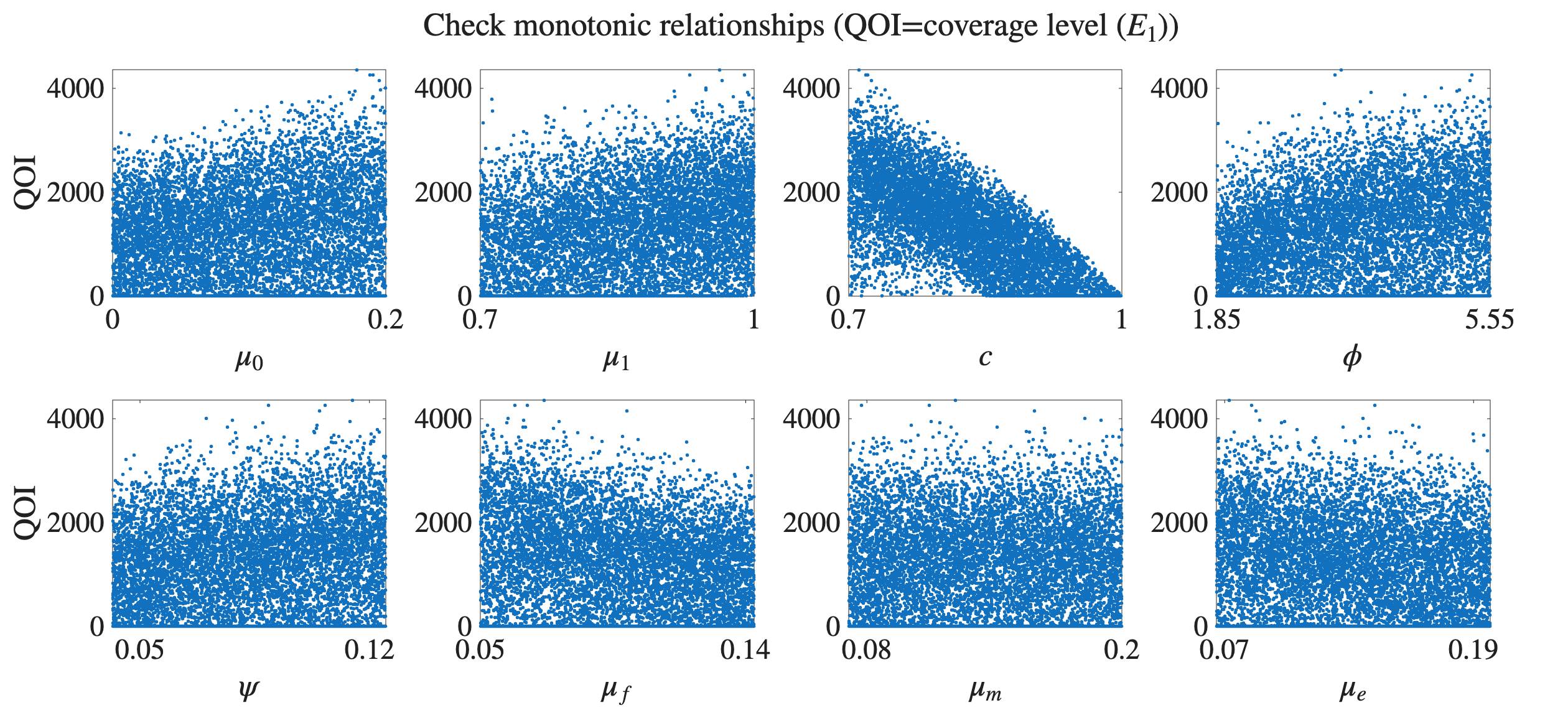}
\includegraphics[width=\linewidth]{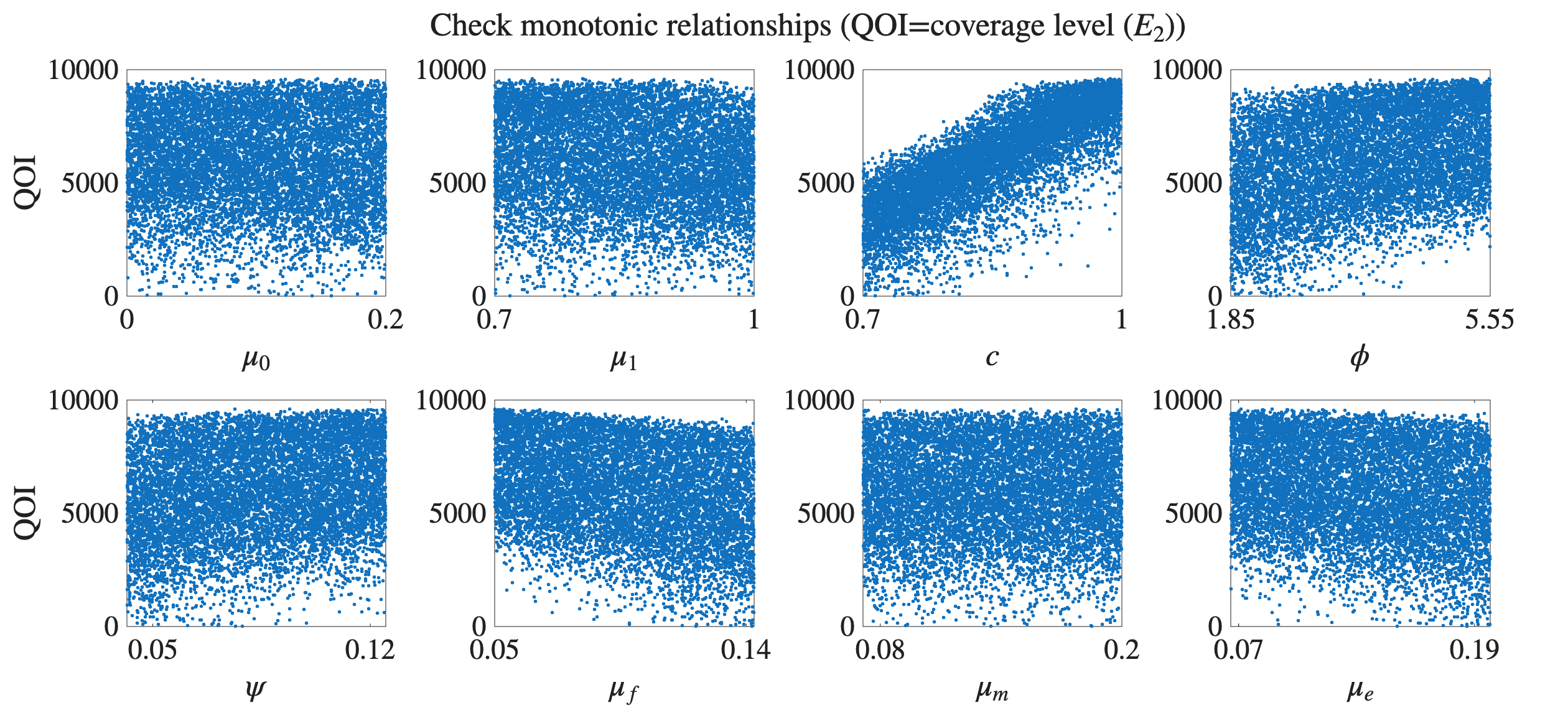}
\includegraphics[width=\linewidth]{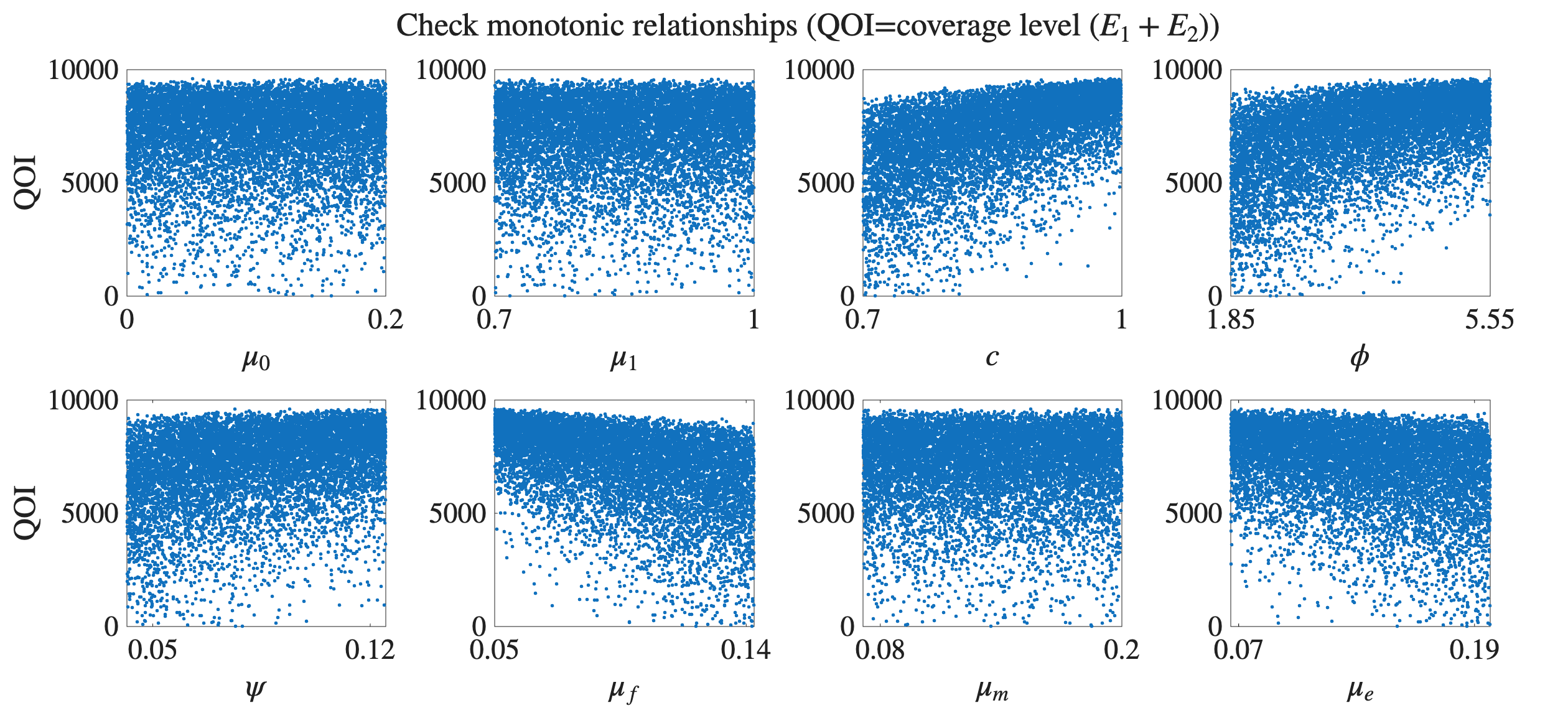}
\caption{Monotonicity check for PRCC to supplement the figures in \cref{sec:SA_coverage}.}
\label{fig:SA_PRCC_mono3}
\end{figure}

\begin{figure}[ht]
\centering
\includegraphics[width=\linewidth]{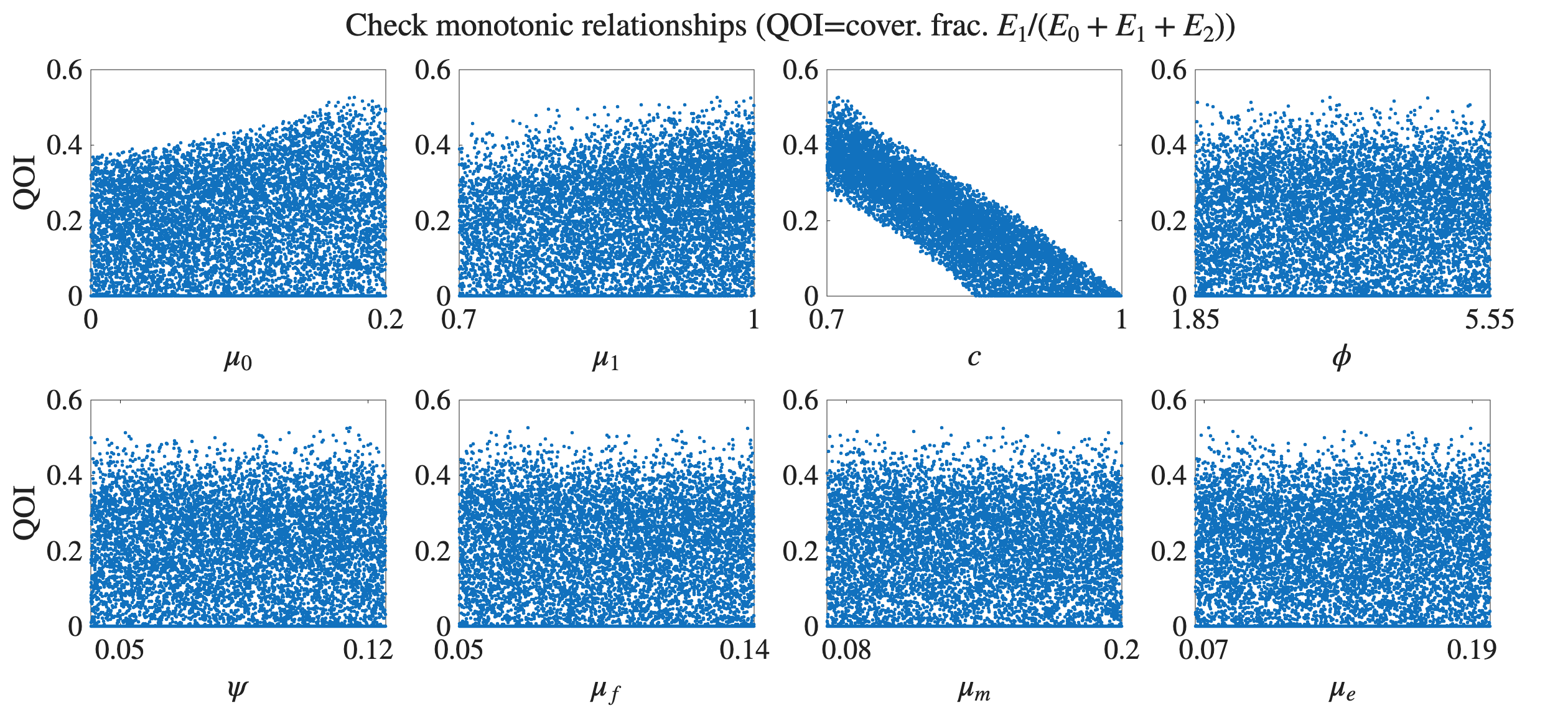}
\includegraphics[width=\linewidth]{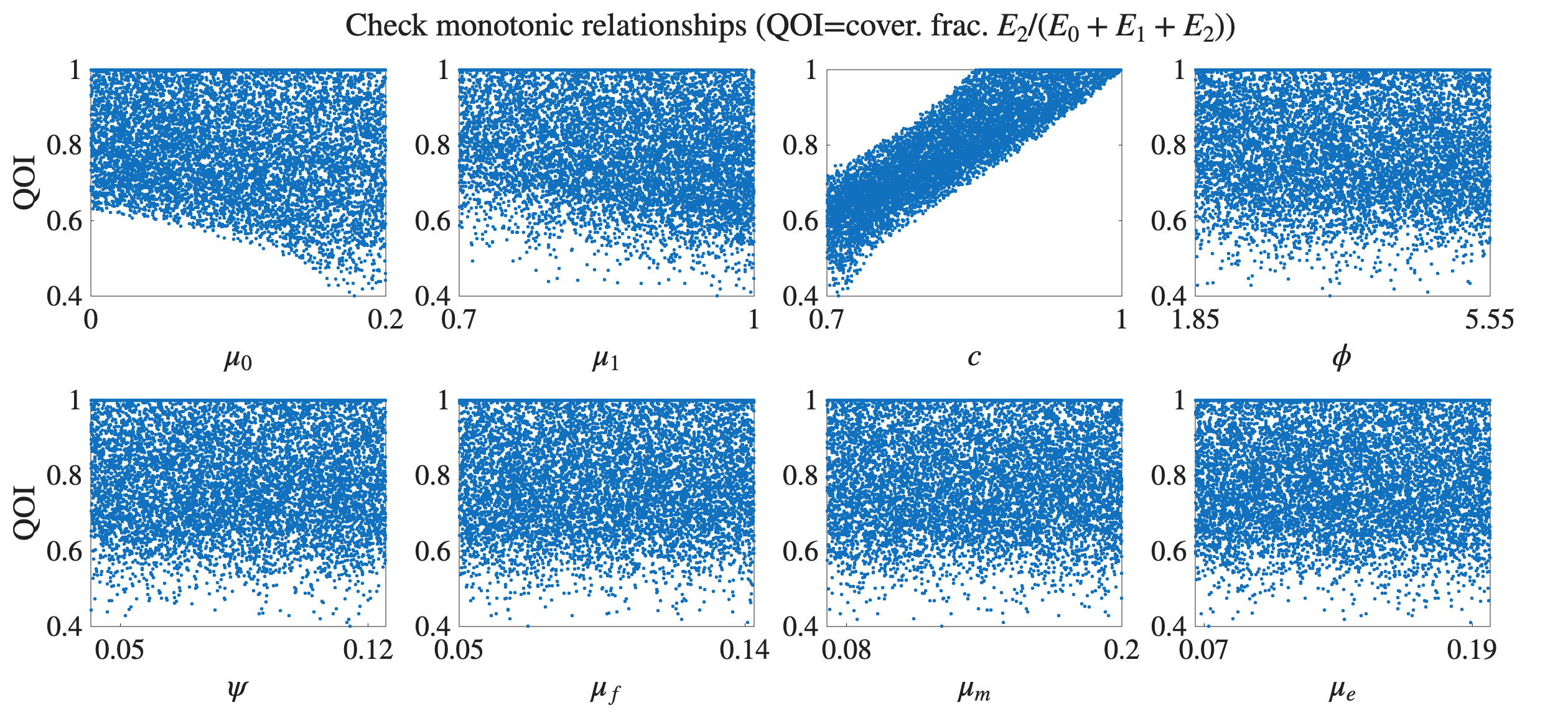}
\includegraphics[width=\linewidth]{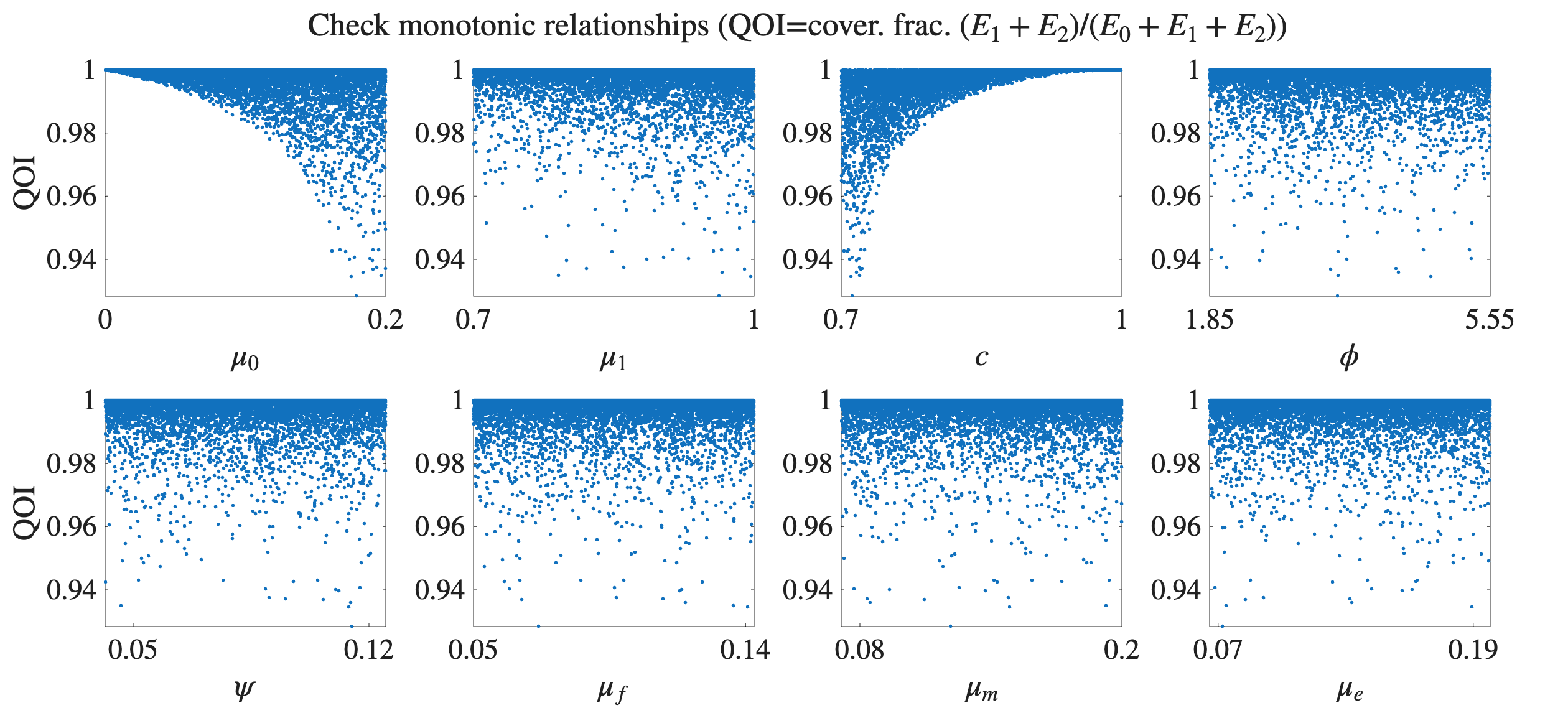}
\caption{Monotonicity check for PRCC to supplement the figures in \cref{sec:SA_coverage}.}
\label{fig:SA_PRCC_mono4}
\end{figure}

We verify the monotonicity in QOI responses as a prerequisite for interpreting the PRCC results reported in \cref{sec:SA_threshold} and \cref{sec:SA_coverage}. The scatter plots in \cref{fig:SA_PRCC_mono1,fig:SA_PRCC_mono2,fig:SA_PRCC_mono3,fig:SA_PRCC_mono4} confirm that this condition is satisfied for all QOIs considered. Note that the fitness coefficient $c$ adopts a more restricted range $c \in [0.7,1]$ specifically to exclude the non-monotonic behavior arising around $c \approx 0.6$ (results not shown).

\section{Additional Release Composition Plots}\label{sec:app_release}
\begin{figure}[ht]
\centering
\includegraphics[width=0.49\linewidth]{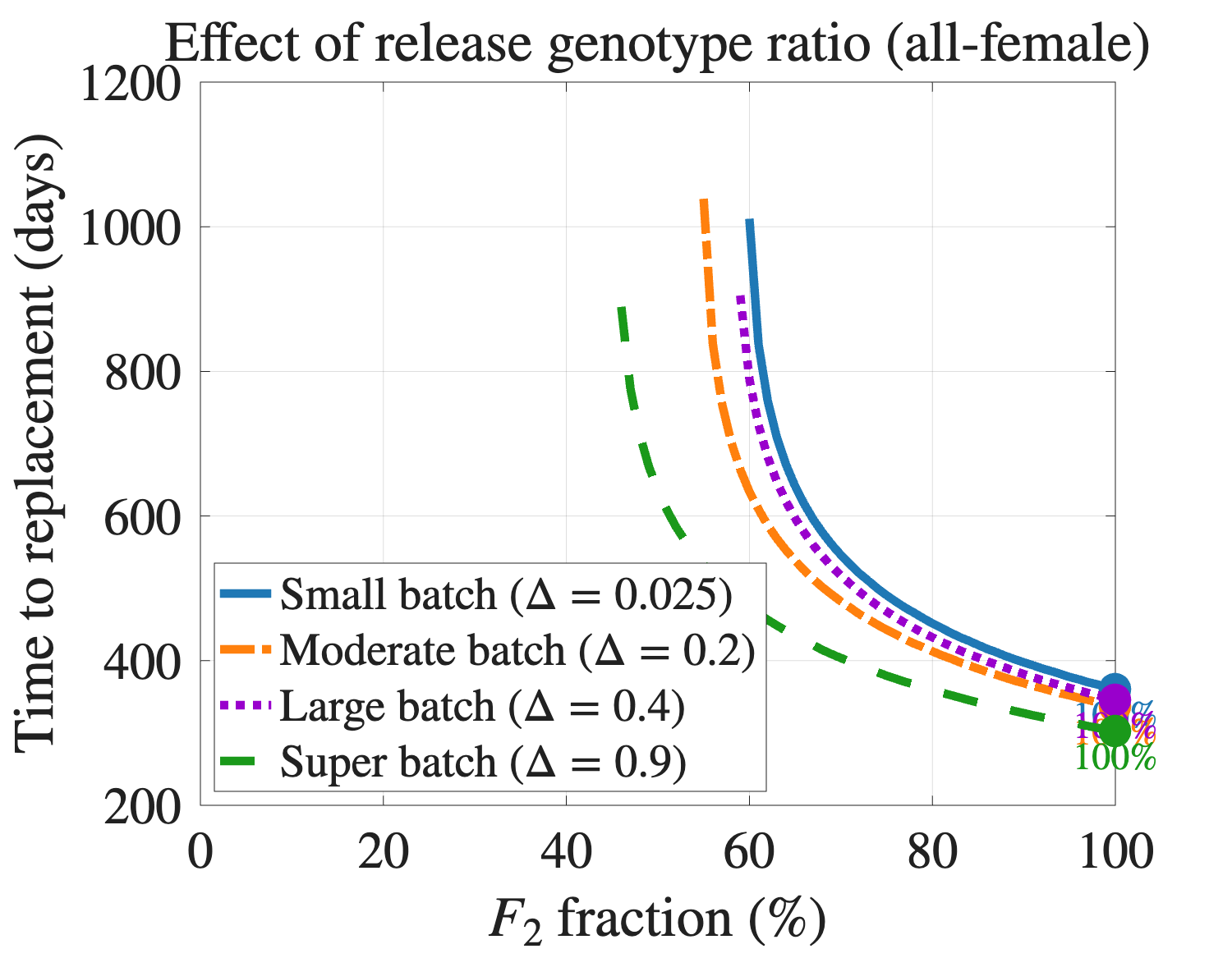}\hfill\includegraphics[width=0.49\linewidth]{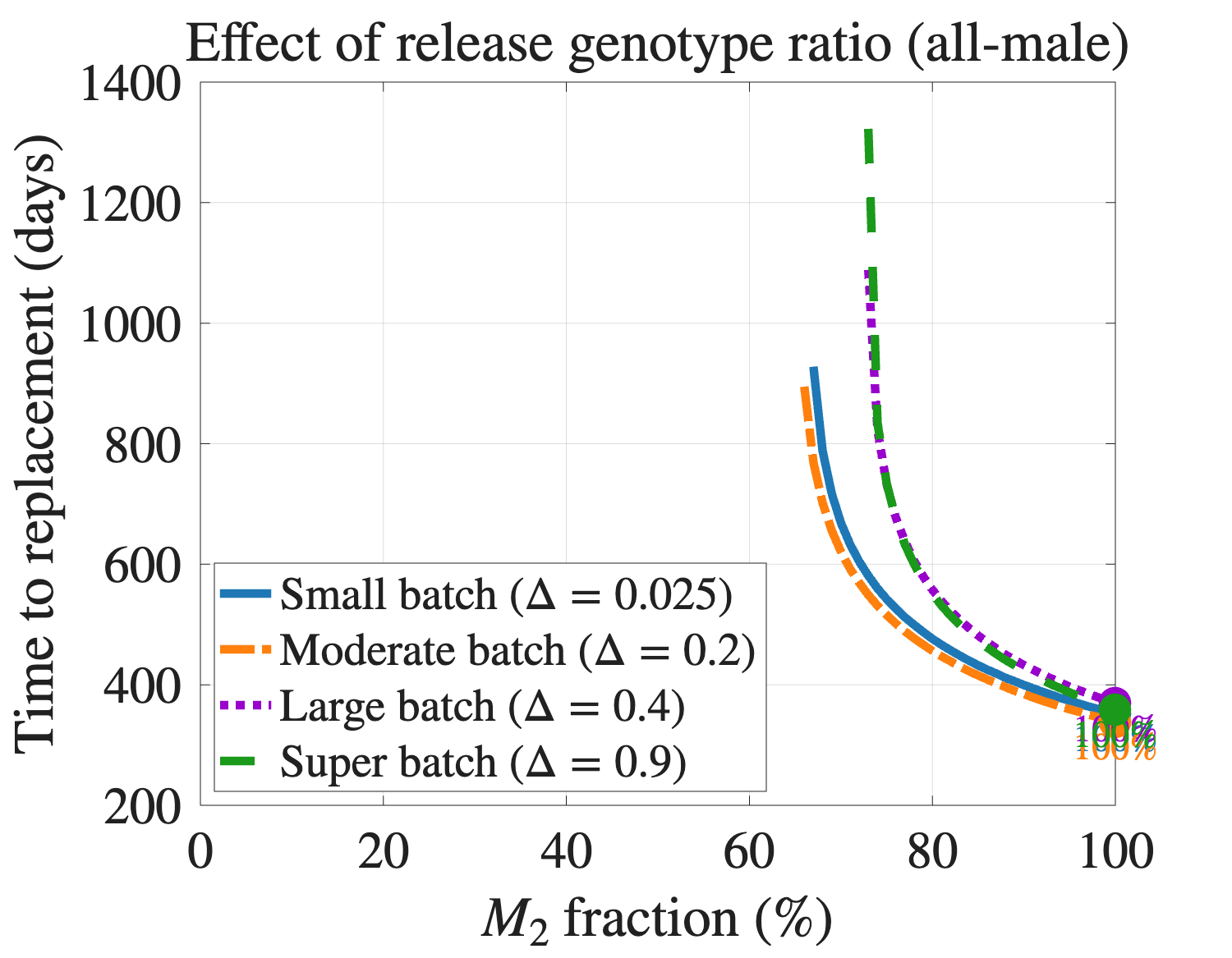}
\caption{Time to replacement for varying release genotype ratio across four batch size regimes: small ($\Delta = 0.025$, $\tau = 7/5$ days, 31 batches), moderate ($\Delta = 0.2$, $\tau = 7/0.75$ days, 4 batches), large ($\Delta  = 0.4$, $\tau = 7/0.45$ days, 2 batches), and a single super-batch ($\Delta = 0.9$)the. The optimal release ratio is 100\% MM genotypes for both male and female scenarios. Releasing mixed genotypes greatly slows down the replacement process.}
\label{fig:release_ratio_app}
\end{figure}

This appendix supplements the release composition analysis in the main text (\cref{sec:release}) by varying the genotype of the released individuals rather than their sex. Considering the same four batch size regimes as in \cref{fig:release_ratio}, we vary the fraction of the released cohort carrying the MM homozygous genotype, with the remainder released as M+ heterozygotes. \Cref{fig:release_ratio_app} shows the resulting time to replacement for female-only releases ($F_1$ and $F_2$ mix) and male-only releases ($M_1$ and $M_2$ mix). In both cases, the time to replacement decreases monotonically with the MM fraction, so a pure MM release is optimal regardless of which sex is deployed. 

\end{document}